\documentclass[a4paper,12pt]{article}

\usepackage{jheppub} 
\usepackage{hyperref}
\hypersetup{
    bookmarks=true,         
    unicode=false,          
    pdftoolbar=true,        
    pdfmenubar=true,        
    pdffitwindow=false,     
    pdfstartview={FitH},    
    pdftitle={My title},    
    pdfauthor={Author},     
    pdfsubject={Subject},   
    pdfcreator={Creator},   
    pdfproducer={Producer}, 
    pdfkeywords={keyword1} {key2} {key3}, 
    pdfnewwindow=true,      
    colorlinks=true,       
    linkcolor=red,          
    citecolor=purple,        
    filecolor=magenta,      
    urlcolor=blue,           
    linktocpage=true
}
\usepackage{graphics}
\usepackage{rotating}
\usepackage{subfigure}
\usepackage{pstricks}
\usepackage{color}
\usepackage{amsfonts}
\usepackage{mathrsfs}
\usepackage{epsfig}
\usepackage{verbatim}
\usepackage{amsmath,amssymb,amsthm,graphicx,latexsym}
\usepackage{ulem}

\newcommand{\be}{\begin{equation}}
\newcommand{\ee}{\end{equation}}
\newcommand{\bea}{\begin{eqnarray}}
\newcommand{\eea}{\end{eqnarray}}
\newcommand{\bml}{\begin{subequations}}
\newcommand{\eml}{\end{subequations}}
\newcommand{\bfig}{\begin{figure}}
\newcommand{\efig}{\end{figure}}

\newcommand{\slashed}{\hspace{-1.1ex}/}

\begin{document}
$~~~~~~~~~~~~~~~~~~~~~~~~~~~~~~~~~~~~~~~~~~~~~~~~~~~~~~~~~~~~~~~~~~~~~~~~~~~~~~~~~~~~$\textcolor{red}{\bf TIFR/TH/15-30}
\title{\textsc{\fontsize{35}{90}\selectfont \sffamily \bfseries Effective Field Theory of Dark Matter from membrane inflationary paradigm}}

\author[a]{Sayantan Choudhury}
\author[b]{Arnab Dasgupta}

\affiliation[a]{Department of Theoretical Physics, Tata Institute of Fundamental Research, Homi Bhabha Road, Colaba, Mumbai - 400005, India
\footnote{\textcolor{purple}{\bf Presently working as a Visiting (Post-Doctoral) fellow at DTP, TIFR, Mumbai, \\$~~~~~$Alternative
 E-mail: sayanphysicsisi@gmail.com}. ${}^{}$}}
\affiliation[b]{Institute of Physics, Sachivalaya Marg, Bhubaneswar, Odisha - 751005, India
}

\emailAdd{sayantan@theory.tifr.res.in, arnab.d@iopb.res.in  }

\abstract{In this article, we have studied the cosmological and particle physics constraints on dark matter relic abundance from effective field 
theory of inflation
from tensor-to-scalar ratio ($r$), in case of Randall-Sundrum single membrane (RSII) paradigm. Using semi-analytical approach we establish
a direct connection between the dark matter relic abundance ($\Omega_{DM}h^2$) and primordial gravity waves
($r$), which establishes a precise connection between inflation and generation of dark matter within the framework of effective field theory 
in RSII membrane. Further assuming the UV completeness of the effective field theory perfectly holds good in the prescribed
framework, we have explicitly shown that the membrane tension, $\sigma\leq {\cal O}(10^{-9})~M^4_p,$
bulk mass scale $M_5\leq {\cal O}(0.04-0.05)~M_p$, and cosmological constant $\tilde{\Lambda}_{5}\geq -{\cal O}(
10^{-15})~M^5_p$, in RSII membrane plays the most significant 
role to establish the connection between dark matter and inflation, using which 
we have studied the features of various mediator mass scale suppressed 
effective field theory ``relevant operators'' induced from the localized $s$, $t$ and $u$ channel interactions in RSII membrane.
Taking a completely model independent approach, we have studied an exhaustive list of tree-level Feynman diagrams for
dark matter annihilation within the prescribed setup and to check the consistency of the obtained results, further we apply the constraints as obtained from
recently observed Planck 2015 data and Planck+BICEP2+Keck Array joint datasets. Using all of these
derived results we have shown that to satisfy the bound on, $\Omega_{DM}h^2=0.1199\pm 0.0027$, as from Planck 2015 data, 
it is possible to put further stringent constraint on $r$ within, $0.01\leq r\leq 0.12$, for 
thermally averaged annihilation cross-section of dark matter, $\langle \sigma v\rangle\approx {\cal O}(10^{-28}-10^{-27}){\rm cm^3 /s}$, which are very
useful to constrain various membrane
inflationary models. 

}

\maketitle
\flushbottom
\section{Introduction}
Identifying the nature of dark matter, which will have profound consequences in the context of cosmology and particle physics.
At present a significant research is being devoted into the search for dark matter in:
\begin{enumerate}
 \item Indirect searches where the prime objective is to detect the products of dark matter annihilations or decays around the milky way \cite{Cirelli:2012tf},
 \item Direct searches where the prime target is to detect the scattering between dark matter and heavy mesons \cite{Baudis:2012ig},
 \item Collider searches (specifically at LHC) where the main goal is to
 search for mono-jet \cite{ATLAS:2012ky,Chatrchyan:2011nd,ATLAS:2012zim,CMS:rwa} and mono-photon \cite{Aad:2012fw,Chatrchyan:2012tea,ATLAS:2012moa,CMS:2011aof}.\end{enumerate}

 Despite the unknown nature of dark matter, from theoretical point of view Weakly Interacting Massive Particle (WIMP)
 is the most studied favoured candidate, whose estimated thermal relic abundance is consistent with the present observed data. 
 Particle physics beyond the Standard Model is the mostly renowned area in this context which provides such a dark matter candidate in a model dependent way. 
 To study the signatures of the dark matter candidate in a model independent way one of the powerful
 approaches is Effective Field Theory framework \cite{Busoni:2013lha,Busoni:2014sya,Busoni:2014haa}~\footnote{See ref.~\cite{Burgess:2007pt} for the review 
 on effective field theory techniques. See also ref.~\cite{Sundrum:1998sj} for the effective field theory construction
 on Randall-Sundrum (RSII) 3 membrane. To know more about effective field theory of inflation see ref.~\cite{Cheung:2007st,Weinberg:2008hq,Senatore:2010wk,LopezNacir:2011kk}.} in theoretical 
 physics, using which the various $s$, $t$ and $u$ channel interactions of dark matter candidate and the known Standard Model field contents are parametrized by a set of 
 mediator mass scale suppressed effective non-renormalizable Wilsonian operators, usually generated from integrating out the heavy mediator
 from the theory \cite{Beltran:2010ww,Duch:2014xda,Duch:2014yma,Krauss:2013wfa,Bell:2013wua,DeSimone:2013gj,Chae:2012bq,Abdallah:2015ter,Busoni:2014gta,Goodman:2010yf,Bai:2010hh,Goodman:2010ku,Rajaraman:2011wf,Fox:2011pm}. Also the effective field theory framework is a very sophisticated theoretical way to describe 
 physical phenomena occurring at a specified energy scale in terms of all possible allowed interactions. The effective field theory framework suggests that the effective operators in principle can explain 
 the direct, indirect detection and collider search of WIMPs, which require an interaction of WIMPs with Standard Model sector in the effective theory.
 Most importantly, the effective field theory framework is very advantageous as it can be testable through various experiments
 ~\footnote{In the present day research, the effective field theory framework is one of the important branches at LHC experiment for the dark matter searches
 in the mono-jet and mono-photon channels \cite{Busoni:2013lha,Busoni:2014sya,Busoni:2014haa}.}. 
 
\begin{figure}[t]
{\centerline{\includegraphics[width=16.5cm, height=7.2cm] {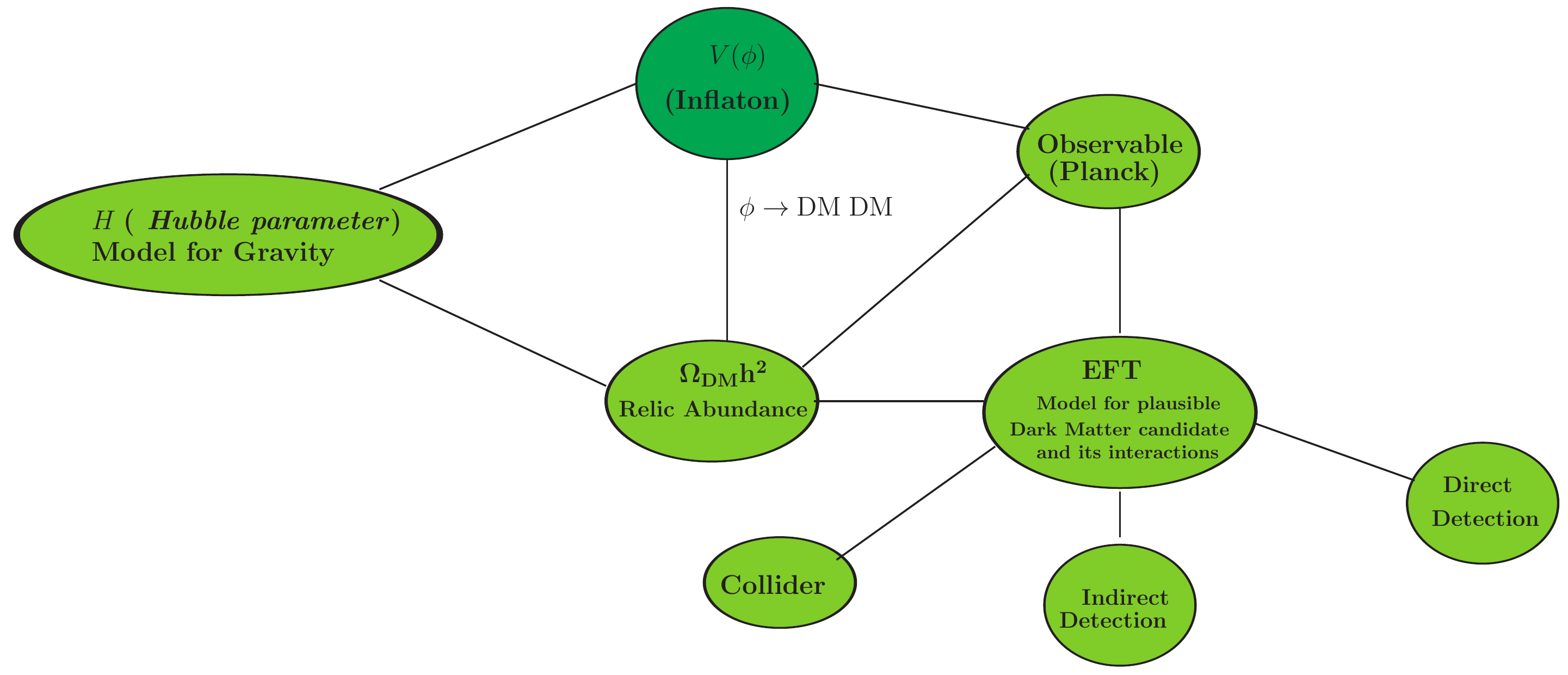}}}
\caption{Schematic representation of effective field theory setup in membrane paradigm which shows the complete algorithm of the described 
methodology in this paper.} \label{fig01}
\end{figure}

 Apart from the huge success of the effective field theory prescription, it is important to mention here that
 the justifiability of the framework depends on the separation between the characteristic 
 energy scale of the dark matter annihilation process and scale 
 of underlying microscopic interactions. In the context of indirect dark matter 
 searches, the annihilation of the non-relativistic dark matter particle contents in the galaxy occur with momentum transfer comparable to the 
 dark matter mass. On the other hand, in case of direct searches the momentum transfer process involved in the
 scattering of dark matter with 
 heavy nuclei are comparable to keV scale \cite{Busoni:2013lha,Busoni:2014sya,Busoni:2014haa}.
 In both of the cases, it is possible to implement the effective field theory technique, provided 
 the UV cut-off of the prescribed theory is larger than the typical momentum transfer. However, in the context of collider searches, the associated energy scale 
 involved is very high and the Wilsonian effective operators exist at the scale beyond the validity of effective field theory framework itself.
 Thus before applying the effective field theory prescription it is important to know about full UV complete theory, the range of applicability and consistency of framework
 in the specific context~\footnote{ In the matter sector incorporating the effects of quantum correction through the interaction between heavy and light field sector 
and finally integrating out the heavy degrees of freedom from the 4D Effective Field Theory picture the matter action \cite{Choudhury:2014sua,Baumann:2014nda}, which 
admits a systematic expansion within the light sector can be written as:
\be\begin{array}{llll}\label{mod2a}
    \displaystyle S_{matter}[\chi,\Psi]=\int d^{4}x \sqrt{-^{(4)\!}g}
  \left[{\cal L}_{DM}[\chi]+{\cal L}_{heavy}[\Psi]+{\cal L}_{int}[\chi,\Psi]\right]\\~~\displaystyle \underrightarrow{Remove~~\Psi~}~~e^{i S_{DM}[\chi]}=\int [{\cal D}\Psi] e^{iS_{matter}[\chi,\Psi]}\\
\displaystyle S_{DM}[\chi]=\int d^{4}x \sqrt{-^{(4)\!}g}
  \left[{\cal L}_{DM}[\chi]+\sum_{\alpha}J_{\alpha}(g)\frac{{\cal O}_{\alpha}[\chi]}{M^{\Delta_{\alpha}-4}_{p}}\right]
   \end{array}\ee
where $J_{\alpha}(g)$ are dimensionless Wilson coefficients that depend on the couplings g of the UV theory, and ${\cal O}_{\alpha}[\chi]$ are
local operators of dimension $\Delta_{\alpha}$.  This procedure typically generates all possible effective operators ${\cal O}_{\alpha}[\chi]$ consistent with
the symmetries of the UV theory. Also ${\cal L}_{DM}[\chi]$ and ${\cal L}_{heavy}[\Psi]$ describe the part of total Lagrangian density ${\cal L}$ involving only the light and heavy fields, 
and ${\cal L}_{int}[\chi,\Psi]$ includes all
possible interactions involving both sets of fields within Effective Field Theory prescription. After removal of heavy degrees of freedom the effective action is decomposed into a renormalizable part
 and a sum of non-renormalizable
corrections appearing through the Wilsonian operators ${\cal O}_{\alpha}[\chi]$. Such operators of dimensions less than four 
are called ``relevant operators'' \cite{Baumann:2014nda,Assassi:2013gxa}. They dominate in the IR and become small in the UV. In 4D Effective Field Theory the operators
of dimensions greater than four are called irrelevant operators. These operators become small in
the IR regime, but dominate in the UV end.
However such corrections are extremely hard to compute and at the same 
time the theoretical origin of all such corrections is not at all clear till now as it completely
belongs to the hidden sector of the theory.
One of the possibilities of the origin of such hidden sector heavy field is higher dimensional
Superstring Theory or its low energy supergravity version.
 Such a higher dimension setups dimensionally reduced to the 4D Effective Field Theory version via various compactifications.
}. In fig.~(\ref{fig01}), we have explicitly shown the schematic representation of effective field theory setup in membrane paradigm which shows the complete algorithm of the described 
methodology in this paper.

In the present context the effective field theory in four dimension is originated from varieties of string
theories, specifically the 10-dimensional $E_{8}\otimes E_{8}$ heterotic string theory is one of the strongest candidates, which
contain the standard model of particle physics and is related to an 11-dimensional superstring theory written on the orbifold
$R^{10} \otimes S^{1}/Z_{2}$. Within this UV complete field theoretic setup, the standard model particle species are confined to the 4-dimensional
space-time which is the sub-manifold of $R^{4} \otimes S^{1}/Z_{2}$. On the other hand, the graviton degrees of freedom propagate
in the bulk space-time. To visualize in a more simplified way one can think about a situation where we consider a 5-dimensional problem 
in which the matter fields are confined to the 4-dimensional space-time while gravity acts in 5 dimensional bulk space-time.
Randall-Sundrum (RSII) single membrane is one of the frameworks in which our observable universe in embedded on 3-membrane,
where effective field theory prescription is valid \cite{Choudhury:2014sua}. In field theory language one can interpret this membrane to be 
the boundary of 5-dimensional anti-de Sitter ($AdS_{5}$) bulk space-time and the corresponding effective field theory at the 
boundary is equivalent to the four dimensional Conformal Field Theory ($CFT_{4}$). In this paper, we consider 
RSII membrane paradigm as a theoretical probe using which we have studied the validity and consequences from 
effective field theory framework to explain inflation and dark matter in the light of Planck 2015 data. 

The prime objective of this paper is
to establish
a theoretical constraint to explicitly show a direct connection between the dark matter relic abundance ($\Omega_{DM}h^2$) and primordial gravity waves
($r$), which establish a precise connection between inflation and generation of dark matter within the framework of effective field theory 
in RSII membrane paradigm.

Throughout the analysis of the paper
 we assume:
 \begin{enumerate}
  \item Inflaton 
 field $\phi$ and all the effective field theory interactions are localized in the membrane of RSII set up and also minimally coupled to the gravity sector at the single membrane. 
 \item Here in the RSII membrane paradigm the extra dimension ``y'' is non-compact for which the covariant formalism is applicable.
 Also in our discussion we use the bulk and membrane parameters: 1) $M_5$ represents the 5D
quantum gravity cut-off scale, 2) $\Lambda_5$ represents the 5D bulk cosmological constant, 3) 3 membrane 
has a positive membrane tension $\sigma$ and it is localized at the
position of orbifold point $y=0$ \cite{Choudhury:2014sua,Choudhury:2015jaa,Choudhury:2011sq,Choudhury:2011rz,Choudhury:2012ib,Choudhury:2012yh}. 
The exact connecting relationship between the bulk and membrane parameters, $M_5$, $\Lambda_5$ and $\sigma$ are explicitly mentioned in the later section 
of this article. Also for the sake of simplicity, in the RSII membrane set-up, during cosmological analysis, one can choose the following sets of parameters to be free:
\begin{itemize}
 \item Bulk cosmological constant $\Lambda_5$ is the most important parameter of RSII
 set up \cite{Choudhury:2014sua,Choudhury:2015jaa,Choudhury:2011sq,Choudhury:2011rz,Choudhury:2012ib,Choudhury:2012yh}.
 Only the upper bound of $\Lambda_5$ is fixed 
 to validate the Effective Field Theory framework within the prescribed set up. Once I choose the value of $\Lambda_5$ below its upper bound 
 value, the other two parameters- 5D quantum gravity cut-off scale $M_5$ and the membrane tension $\sigma$ is fixed from their
 connecting relationship as discussed later. In this paper, we fix the values of all of these RSII membrane parameters by using 
 Planck 2015 data and Planck+BICEP2/Keck Array joint constraints.
 \item The rest of the free parameters are explicitly appearing through the various couplings and mass of dark matter content in the effective theory. 
\end{itemize}

 \item Slow-roll prescription perfectly holds good, which is useful to estimate the tensor-to-scalar ratio $r$ from the RSII membrane inflationary paradigm.
\item Initial condition is guided via the Bunch-Davies vacuum. 
\item The effective sound speed during inflation is fixed at $c_{S}=1$.
 \end{enumerate}
                     The plan of the paper is as follows.
                     \begin{itemize}
                     \item In section \ref{a1}, we will briefly review the basic setup of RSII single membrane paradigm.
                     Here we begin our discussion with the 5D action which contains the effective action localized at
                     orbifold fixed point of the membrane. Finally, we will derive the Friedmann equation in the membrane, which is very very 
                     useful for further computation in the paper. 
                      \item In section \ref{a2}, we will establish
a theoretical constraint to explicitly show a direct connection between the dark matter relic abundance ($\Omega_{DM}h^2$) and primordial gravity waves
($r$), which establish a precise connection between inflation and generation of dark matter within the framework of effective field theory 
in RSII membrane paradigm.
                      \item In section \ref{avv11}, we will explicitly study the details of 
                      Effective Field Theory of dark matter from membrane paradigm paradigm.
\item In this paper we
use various constraints arising from Planck 2015 data on the upper bound on tenor to scalar ratio,
and the bound on dark matter abundance
within $1.5\sigma-2\sigma$ statistical CL.

\item I also mention that the GR limiting result ($\rho<<\sigma$) and the difference between the high energy
  limit result ($\rho>>\sigma$) of RSII.

\item Hence in section \ref{a3v1} and \ref{a3v2}, we have studied the $s$ and $t/u$ channel interaction of the 
effective theory of Dirac dark matter with spin-0 mediator. Further, in section \ref{a4v1} and \ref{a4v2}, we have studied the $s$ and $t/u$ channel interaction of the 
effective theory of Dirac dark matter with spin-1 mediator. Then, in section \ref{a5v1} and \ref{a5v2}, we have studied the $s$ and $t/u$ channel interaction of the 
effective theory of Majorana dark matter with spin-1 mediator. After that, in section \ref{a6}, we have studied the consequences from 
s-channel interaction of complex and real scalar dark matter with spin-0 mediator. Next, in section \ref{a8}, we have studied the consequences from 
s-channel interaction of complex scalar and real vector dark matter with spin-1 mediator. Further, in section \ref{a10} and \ref{a12}, we have studied the consequences from 
s-channel interaction of complex and real vector dark matter with spin-0 and spin-1 mediator respectively. Finally, in section \ref{a14}, \ref{a15} and \ref{a16}, we have studied the consequences from 
t/u-channel interaction of complex and real scalar and complex and real vector dark matter with spin-1/2 mediator.

\item At the end, in section \ref{a18}, we summarize our obtained results point-wise. 
                     \end{itemize}
   
\section{An overview of membrane paradigm}
\label{a1}
Let us start our discussion with a very brief introduction to membrane paradigm based on Randall-Sundrum single membrane (RSII) setup.
 The RSII single membrane setup and its most generalized version from a Minkowski membrane to a Friedmann-
Robertson-Walker (FRW) membrane were derived as solutions
in specific choice of coordinates of the 5D Einstein equations in the bulk, along with the junction conditions, which are applied at
the ${\bf Z}_{2}$ -symmetric single membrane. A broader perspective, with non-compact dimensions, 
can be obtained via the well known covariant Shiromizu-Maeda-Sasaki approach, in
which the membrane and bulk metrics take its generalized structure \cite{Shiromizu:1999wj}. The key point is to use the Gauss-Codazzi
equations to project the 5D bulk curvature along the membrane using the covariant formalism. 
Here we start with the well known 5D Rundall-Sundrum (RSII) single membrane model action given by \cite{Randall:1999vf}:

\begin{eqnarray}
 S_{RS}&=& \int d^{5}x\sqrt{-{}^{(5)}\!g}\left[\frac{M^{3}_{5}}{2}~{}^{(5)}\!R-2\Lambda_{5}+{\cal L}_{bulk}+\left({\cal L}_{membrane}-\sigma\right)\delta(y)\right],
\end{eqnarray}
where the extra dimension ``y'' is non-compact for which the covariant formalism is applicable. Here $M_5$ be the 5D quantum gravity cut-off scale, $\Lambda_5$ be the 5D bulk cosmological constant, ${\cal L}_{bulk}$ be the
bulk field Lagrangian density, ${\cal L}_{membrane}$ signifies the Lagrangian density for the membrane field contents. 
It is important to mention the the scalar inflaton degree of 
freedom is embedded on the 3 membrane which has a positive membrane tension $\sigma$ and it is localized at the position of orbifold point $y=0$ in case of single membrane.
The 5D field
equations in the bulk, including explicitly the contribution of the RS single membrane is given by \cite{Maartens:2010ar}:

\begin{equation}
  ^{(5)}\!G_{AB}=\frac{1}{M_5^3}\left[-\Lambda_5 \, {}^{(5)}\!g_{AB}+
  {}^{(5)}T_{AB}+ T_{\mu\nu}^\mathrm{membrane}\delta^{\mu}_{A}\delta^{\nu}_{B}\delta(y)\right]
  \label{febfg}
\end{equation}
where $^{(5)}T_{AB}$ characterizes any 5D energy-momentum tensor of the
gravitational sector within bulk spece-time. On the other hand the total energy-momentum tensor on the membrane
is given by:

\begin{equation}
  T_{\mu\nu}^\mathrm{membrane} =T_{\mu\nu}-\sigma g_{\mu\nu},
\end{equation}

where $T_{\mu\nu}$ is the energy-momentum tensor of particles and
fields confined to the single membrane so that the constraint condition, \be T_{AB}n^B=0\ee is valid in the present context.
Let us consider further $y$ be a Gaussian normal coordinate, which is orthogonal to the single membrane,
chosen to be placed at the orbifold point $y=0$ without loss of generality, so that
an infinitesimal change along the direction of extra dimensional coordinate is represented by, \be n_AdX^A=dy,\ee
where $n^A$ be the unit normal. In case of RS single membrane setup 
the 5D metric in terms
of the induced metric on the family of $y=\mathrm{const.}$ hyper-surfaces is locally
represented by \cite{Maartens:2010ar}:

\begin{eqnarray}
  {}^{(5)}\!g_{AB}&=&g_{AB}+ n_An_B,
  \\
  {}^{(5)\!}ds^2&=&g_{\mu\nu}(x^\alpha,y)dx^\mu dx^\nu +dy^2.
  \label{gnfg}
\end{eqnarray}
Here it is assumed that the metric $g_{\mu\nu}(x^\alpha,y)$ is non-factorizable in the most generalized
prescription. In such a situation one can 
Taylor expand of the metric about the single membrane. In Gaussian normal
coordinates after applying Taylor expansion one can write \cite{Maartens:2010ar}:
%
\begin{eqnarray}
  g_{\mu\nu}(x^\alpha,y) &=& g_{\mu\nu}(x^\alpha,0) -
  \frac{1}{M_5^3}\left[ T_{\mu\nu}+{1\over 3}(\sigma-T)g_{\mu\nu}
  \right]_{y=0+} \!\!\! |y| \nonumber\\
  && 
+ \left[ -{\cal E}_{\mu\nu} + {1\over4 M^6_5} \left(
  T_{\mu\alpha} T^\alpha{}_\nu +{2\over3} (\sigma-T)T_{\mu\nu} \right)
\right. \nonumber \\ &&~~~~~~~~~~~~~~~~~~~~~~~~~~~~~~~~~~~~~~~~~~~\left. \nonumber + {1\over6} \left( {1\over6 M^6_5} (\sigma-T)^2 - \Lambda_5
  \right) g_{\mu\nu}\right]_{y=0+} \!\!\! y^2 + \dots
  \nonumber \\
  \label{tayfg}
\end{eqnarray}
Now further integrating Equation~(\ref{febfg}) along the extra dimension within the interval
$y=-\epsilon$ to $y=+\epsilon$, and after taking the limit $\epsilon\to
0$, leads to the well known Israel-Darmois junction conditions at the membrane:

\begin{eqnarray}
  g^+_{\mu\nu}&=&g^-_{\mu\nu}, \\
  K_{\mu\nu}^{+}&=&-K_{\mu\nu}^{-}=-{1\over2 M^{3}_5} \left[T_{\mu\nu}+
  {1\over3} \left(\sigma-T\right)g_{\mu\nu} \right],
  \label{junsder}
\end{eqnarray}
where we define, $T=T^\mu{}_\mu $ and and one can
evaluate quantities on the membrane by taking the physical limit $y\to+0$.
This further implies that the effective action on the 3 membrane can be expressed as:
\be\begin{array}{llll}
 \displaystyle S_{eff}=\int d^{4}x\sqrt{-{}^{(4)}\!g}\left[\frac{M^{2}_{p}}{2}({}^{(4)}\!R-2\Lambda_{4})+{\cal L}_{membrane}
+\Delta_{\cal S}-\Delta_{\cal E}\right]
\end{array}\ee
where $\Delta_{\cal S}$ and $\Delta_{\cal E}$ are the contributions from quadratic part of the energy-momentum tensor $T_{\mu\nu}$ and Weyl tensor respectively. 
Finally one can arrive at the $4$-dimensional Einstein induced field equations on the single membrane given by \cite{Maartens:2010ar,Brax:2004xh}:

\be 
G_{\mu \nu} = - \Lambda_{4}  g_{\mu \nu} + {1 \over M_p^{2}} T_{\mu \nu} 
+ \left({8\pi \over M_{5}^3}\right)^2 {\cal S}_{\mu \nu} - {\cal E}_{\mu \nu}~,
\label{eq:gmunucvfd}
\ee
where $T_{\mu \nu}$ represents the energy-momentum on the single membrane, ${\cal S}_{\mu \nu}$ is a 
rank-2 tensor that contains  contributions that are quadratic in the energy momentum tensor
$T_{\mu \nu}$ represented by:
\begin{eqnarray}
  {\cal S}_{\mu\nu}&=& \frac{M^6_5}{16\pi^2 M^2_p \sqrt{-{}^{(4)}\!g}}\frac{\delta(\sqrt{-{}^{(4)}\!g}~\Delta_{\cal S})}{\delta g^{\mu\nu}}\nonumber\\
&=&{{1\over12}}T T_{\mu\nu}
  -{{1\over4}}T_{\mu\alpha}T^\alpha{}_\nu + {{1\over24}}g_{\mu\nu}
  \left[3 T_{\alpha\beta} T^{\alpha\beta}-T^2 \right].
\end{eqnarray}

and ${\cal E}_{\mu \nu}$ characterizes the projection of the 
5-dimensional Weyl tensor on the 3-membrane which is physically equivalent to the non-local contributions pressure and energy flux for a perfect fluid, 
given by:
\begin{eqnarray}
  {\cal E}_{\mu\nu} &=&\frac{2}{\sqrt{-{}^{(4)}\!g}}\frac{\delta(\sqrt{-{}^{(4)}\!g}~\Delta_{\cal E})}{\delta g^{\mu\nu}}\nonumber\\
  &=&{}^{(5)\!}C_{ACBD} \, n^Cn^D g_\mu{}^A g_\nu{}^B
\end{eqnarray} 
which is orthogonal to the unit normal vector $n^{B}$ represented via the following condition:
\begin{equation}
  {\cal E}_{AB}n^B= 0 ={\cal E}_{[AB]}={\cal E}_{A}{}^A,
\end{equation}

which is the outcome of the Weyl tensor symmetries.

In a cosmological framework, where the 3-membrane resembles our universe and the
 metric projected onto the membrane is an homogeneous and isotropic flat
 Friedmann-Robertson-Walker (FRW) metric, the  Friedmann
 equation becomes \cite{Maartens:2010ar,Brax:2004xh}:

\be
H^2 = {\Lambda_{4} \over 3} +  {\rho \over 3 M_p^2} 
+ \left({4 \pi \over 3 M_5^3}\right)^2 \rho^2 + {\epsilon \over a^4},
\label{eq:H2}
\ee
where $\epsilon$ is an integration constant. The four and five-dimensional
 cosmological constants are related by:

\be
\Lambda_{4} = {4 \pi \over M_5^3} \left(\Lambda_5 + {4 \pi \over 3 M_5^3}~
\sigma^2 \right)~~,
\label{eq:Lam}
\ee
where $\sigma$ is the 3-membrane tension. Within RS setup the quantum gravity cut-off scale i.e. the 
5D Planck mass and effective 4D Planck mass are connected through the visible membrane tension as:
\begin{eqnarray}\label{mass}
M^{3}_{5}=\sqrt{\frac{4\pi\sigma}{3}}M_{p}.
\end{eqnarray}

Assuming that, as required by observations, the 4D cosmological constant 
is negligible $\Lambda_{4}\approx 0$ in the early universe the localized visible membrane tension is given by:
\begin{eqnarray}\label{lam}
 \sigma&=&\sqrt{-\frac{3}{4\pi}M^{3}_{5}\Lambda_{5}}=
\sqrt{-24M^{3}_{5}\tilde{\Lambda}_{5}}>0
\end{eqnarray}
where $\tilde{\Lambda}_{5}$ be the scaled 5D bulk cosmological constant defined as:
\be\label{lamc}
\tilde{\Lambda}_{5}= \frac{\Lambda_{5}}{32\pi}<0.
\ee
 Also the last term 
in Eq. (\ref{eq:H2}) rapidly becomes redundant after inflation sets in, 
the Friedmann equation in RSII membrane becomes \cite{Maartens:2010ar,Brax:2004xh}:
\begin{eqnarray}\label{eq1}
 H^{2}&=&\frac{\rho}{3M^{2}_{p}}\left(1+\frac{\rho}{2\sigma}\right)
\end{eqnarray}
where $\sigma$ be the positive membrane tension, $\rho$ signifies the energy density of the inflaton field $\phi$ and $M_{p}=2.43\times 10^{18}~{\rm GeV}$
 be the reduced 4D Planck mass. Using Eq~(\ref{lam}) in Eq~(\ref{mass}), the 5D quantum gravity cut-off scale can be expressed in terms of 5D cosmological constant as:
\begin{eqnarray}\label{mass1}
M^{3}_{5}&=&\sqrt[3]{-\frac{4\pi\Lambda_{5}}{3}}M^{4/3}_{p}=\sqrt[3]{-\frac{128\pi^2\tilde{\Lambda}_{5}}{3}}M^{4/3}_{p}.
\end{eqnarray}
In the low energy limit $\rho<<\sigma$ in which standard GR framework can be retrieved. On the other hand, in the high energy regime $\rho>>\sigma$
as the effect of membrane correction factor is dominant which is my present focus in this paper.

\section{Inflationary constraints on dark Matter abundance from membrane paradigm}
\label{a2}
In this section we will discuss the inflationary constraints on dark matter from membrane paradigm in detail. To serve this purpose 
let us first start with the total energy density $\rho$, which is localized in the single membrane and can be expressed as:
\be \rho=\rho_r +\rho_\phi ,\ee
where $\rho_r$ and $\rho_\phi$ represent the energy density during radiation and inflation, and can be written as:
\begin{align}
\rho_r &= \frac{\pi^2}{30}g_* \frac{m^4_\chi}{\Theta^4}  \\
\rho_\phi &= g_\phi m^4_\chi x \left(\frac{x}{2\pi \Theta}\right)^{3/2}e^{-x\Theta}
\label{eq:rho_den1}
\end{align}
where the parameters $x$ and $\Theta$ are defined as:
\bea x&=&\frac{m_\phi}{m_\chi},\\ \Theta&=&\frac{m_\chi}{T},\eea and $m_\phi$ is the mass of inflaton field, $T$ is 
the temperature and $m_\chi$ is the mass of the Dark Matter candidate. It is important to mention here that, the
dynamics of the radiation and the inflaton field are governed by the following version of continuity equations as: 
\begin{align}
\frac{d \rho_r}{dt} + 4H\rho_r &= \Gamma_\phi \rho_\phi \label{d1} \\
\frac{d \rho_\phi}{dt} + 3H\rho_\phi &= -\Gamma_\phi \rho_\phi \label{d2}
\end{align}
Now further combining \eqref{d1} with \eqref{d2}, we get the following simplified expression:
\begin{align}\label{weq1}
\mathcal{F}(\Theta)=\frac{d\Theta}{dt} &= H\left[\frac{2\Theta(8e^{x\Theta}g_*\pi^{7/2}+45\sqrt{2}(x/2)^{5/2}\Theta^5)}{16e^{x\Theta}g_*\pi^{7/2}+15\sqrt{2}(x/2)^{5/2}\Theta^5(2x+3)}\right].
\end{align}
where H is the Hubble parameter, which can be expressed via Friedmann equation as:
\begin{eqnarray}\label{eqvv1}
 H^{2}&=&\frac{\left(\rho_r +\rho_{\phi}\right)}{3M^{2}_{p}}\left[1+\frac{\left(\rho_r +\rho_{\phi}\right)}{2\sigma}\right],
\end{eqnarray}
where $M_{p}$ be the reduced/ effective Planck mass, $M_{p}=2.43\times 10^{18}{\rm GeV}$. 
It is important to note that, in the present context, the scale of membrane inflation $V_{inf}$ can be expressed in terms of tensor-to-scalar ratio $r$ as: 
\bea V_{inf}&\approx& \tilde{\Delta}\times \left(\frac{r}{0.12}\right),\eea
where $\tilde{\Delta}$ is around GUT scale, which is constrained by Planck 2015 and Planck 2015+ BICEP2/Keck Array joint data set. For numerical estimations we fix, \be \tilde{\Delta}\approx 1.96 \times 10^{16}~{\rm GeV}. \ee
Now for the sake of simplicity within membrane paradigm we define the following dimensionless parameter:
\bea\label{eqvv2} \alpha &=& \frac{V_{inf}}{\sigma}\approx\frac{ \tilde{\Delta}}{\sigma}\times \left(\frac{r}{0.12}\right) . \label{eq:alpha}\eea
Further substituting Eq~(\ref{eqvv2}) in Eq~(\ref{eqvv1}) we get the following expression for the Hubble parameter in terms of the dimensionless parameter $\alpha$ as:
\begin{align}
 H^2 &= \frac{(\rho_r + \rho_\phi)}{3M^2_{p}}\left[1+\frac{0.12\alpha}{2r\tilde{\Delta}^4}(\rho_r + \rho_\phi)\right].
\label{eq:h}
\end{align}
Henceforth, we will use Eq~(\ref{eq:h}) for the computation of dark matter abundance.

Additionally, for completeness it is important to note that, to validate the effective field theory prescription within the framework of
small field models of inflation, the field excursion is sub-Planckian i.e. 
\be |\Delta \phi|=|\phi_{e}-\phi_{cmb}|<M_{p},\ee
where $\phi_{e}$ and $\phi_{cmb}$ represent the inflaton field values at the end of inflation and at horizon crossing respectively.
For detailed derivation of this important result see ref.\cite{Choudhury:2014sua}~\footnote{The similar bound we have derived in the 
context of GR in ref.~\cite{Choudhury:2015pqa,Choudhury:2014wsa,Choudhury:2014kma,Choudhury:2013iaa,Choudhury:2013woa}. Also
see ref.~\cite{Choudhury:2013jya,Choudhury:2014sxa,Choudhury:2014uxa}, where 
the numerical proof is given in the context of MSSM inflation to justify the validity of the bound on field excursion.}, where using model independent semi-analytical approach we have derived
the field excursion formula for RSII single membrane in terms of inflationary observables and the final result is perfectly consistent with the 
effective field theory prescription.
Using this result  
the model independent bound on the membrane tension, the 5D cut-off scale and 5D bulk cosmological 
constant can be computed as~\footnote{
Additionally it is important to note that, in order to recover the observational successes of general relativity (GR), the high-energy regime where significant deviations occur must
take place before nucleosynthesis. Table-top tests of Newton’s laws put the lower bound on the membrane tension and 5D Planck scale as:
$\sigma > {\cal O}(2.86 \times 10^{-86})~M^{4}_{p}$ and $ M_{5} > {\cal O}(4.11 \times 10^{-11})~M_{p}.$
But such lower bound will not be able to produce large tensor-to-scalar ratio
as required by the upper bound of Planck 2015 data.}:
\bea\label{efqzx}
\large 
\displaystyle \sigma&\leq& {\cal O}(10^{-9})~M^4_p,\\ M_5&\leq& {\cal O}(0.04-0.05)~M_p,\\ \tilde{\Lambda}_{5}&\geq& -{\cal O}(
10^{-15})~M^5_p,
\eea
which we have followed during the numerical estimation performed in this paper.

Now in the context of RSII membrane paradigm, the Bolztmann equation for the dark matter relic abundance can be expressed as :
\begin{align}\label{dmbol}
\frac{dY_{\rm DM}}{d\Theta} &= \frac{s(\Theta)\langle \sigma v \rangle}{\mathcal{F}(\Theta)}\left[1+\frac{1}{3}\frac{d\ln g_s}{d\ln x}\right]\bigg((Y_{\rm DM}^{EQ})^2-Y_{\rm DM}^2 \bigg)
\end{align}
where all the informations of RSII membrane is encoded in the generalized function ${\cal F}(\Theta)$, which is defined earlier in Eq~(\ref{weq1}). 
The Boltzmann equation and the relic abundance calculation for the standard GR case is given in appendix {\bf B}.
In Eq~(\ref{dmbol}), the dark matter relic abundance $Y_{\rm DM}$ is defined as:
\be Y_{\rm DM}=\frac{n_{\rm DM}}{s(\Theta)} ,\ee
where $n_{\rm DM}$ is the number density of dark matter content and and $s(\Theta)$ characterizes the entropy density of the universe defined as:
\begin{align}\label{sen}
s(\Theta) &= \frac{S}{a^3}=\frac{2\pi^2}{45}g_{*}m^3_{\chi}\Theta^{-3},
\end{align}
where $S$ and $a$ represent the total entropy of the universe and scale factor respectively. Also $g_{\star}$ represents the total number of 
degrees of freedom.

Additionally, it is important to note that, in Eq~(\ref{dmbol}), $\langle \sigma v\rangle$ represents the thermally averaged total annihilation cross-section times the Moller velocity, 
where we have summed over final and averaged over initial spins, $Y_{\rm DM}^{EQ}$ signifies the 
dark matter relic abundance at matter and radiation equality and also
$g_{s}$ represents the effective number of relativistic degrees of freedom defined as:
\be \sqrt{g_{*}}=\sqrt{g_{s}}\left(1+\frac{1}{3}\frac{d\ln g_{s}}{d\ln \Theta}\right),\ee
which is in principle function of the parameter $x=m_\phi/m_\chi$.
\begin{figure}[t]
{\centerline{\includegraphics[width=15.5cm, height=10.5cm] {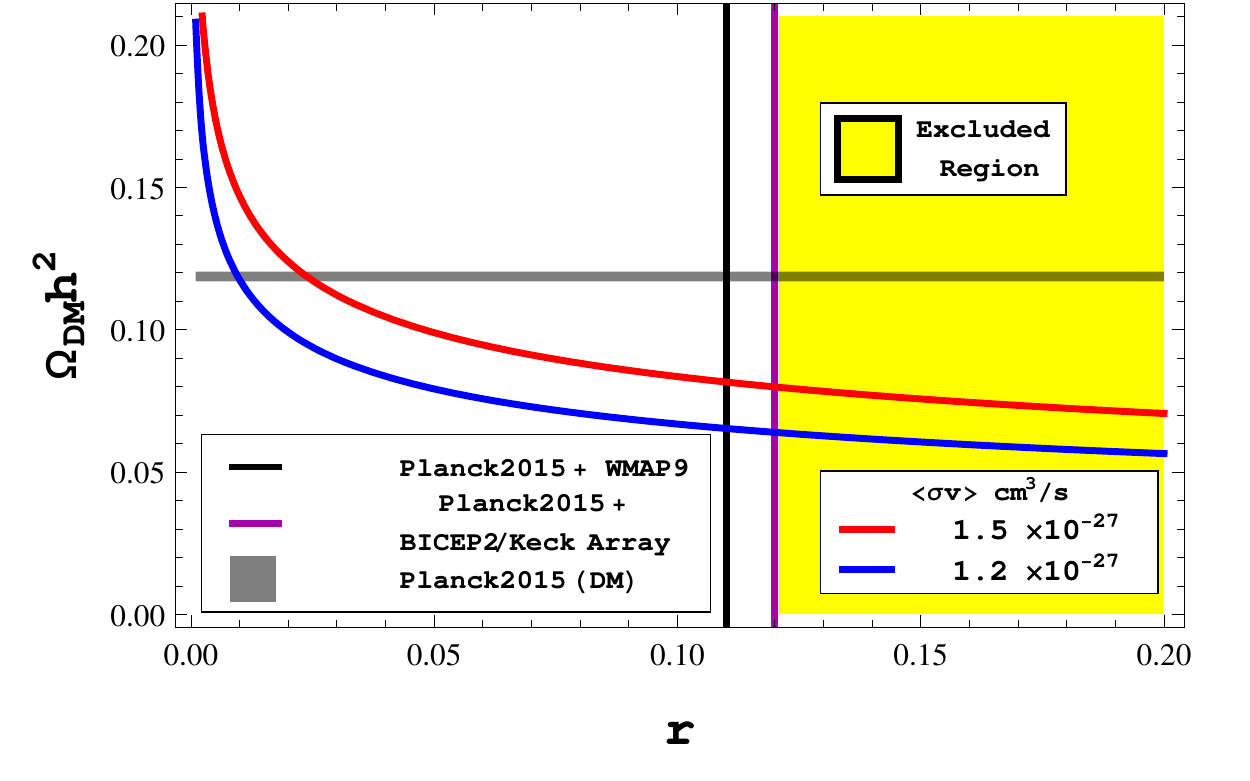}}}
\caption{The above plot shows the dependence of relic abundance $\Omega_{\rm DM} h^2$ with tensor to scalar ratio $r$ taking $\alpha = 1.77 \times 10^{64}$ for different 
thermally averaged cross-section.} \label{fig1}
\end{figure}

\begin{figure}[t]
{\centerline{\includegraphics[width=15.5cm, height=10.5cm] {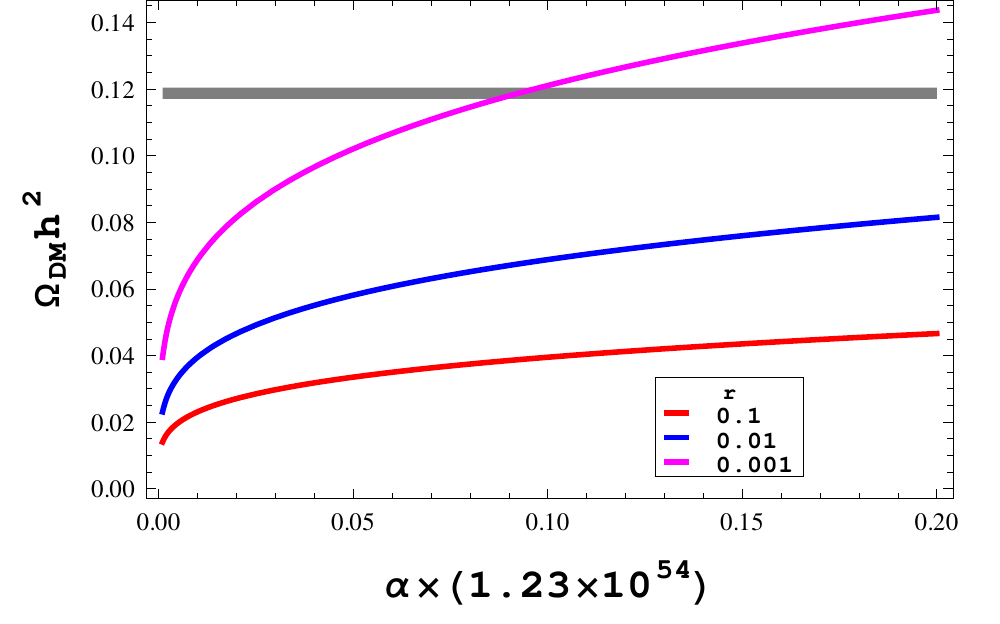}}}
\caption{The above plot shows the dependence of relic abundance $\Omega_{\rm DM} h^2$ with tensor to scalar ratio $r$ taking $\alpha = 1.77 \times 10^{64}$ for different 
thermally averaged cross-section.} \label{fig1a}
\end{figure}


Further assuming the fact that the inflaton field is very heavy compared to the dark matter content ($m_{\phi}>>m_{\chi}$), Eq~(\ref{dmbol}) can be expressed as:
\begin{align}
\frac{d Y_{\rm DM}}{d\Theta} &=
\frac{s\langle \widetilde {\sigma} v \rangle(\Theta )}{\Theta H_{GR}(\Theta)}\left[1+\frac{1}{3}\frac{d\ln g_s}{d\ln \Theta}\right]\left[(Y_{\rm DM}^{EQ})^2-Y_{\rm DM}^2 \right], 
\label{eq:beDM}
\end{align}
where we introduce three functions $\langle \widetilde{\sigma} v \rangle(\Theta)$, $H_{GR}(\Theta)$ and characteristic function in RSII membrane, $f_{membrane}(\Theta) $, which are defined as: 
\begin{align}
\langle \widetilde{\sigma} v \rangle(\Theta) &= \frac{\langle \sigma v \rangle}{f_{membrane}(\Theta)}, \label{eq:sigtld}\\
\Sigma(\Theta)&=\sqrt{\frac{g_*}{90}}\frac{\pi m^2_{\chi}}{M_p \Theta^2}, \label{eq:Sig} \\
f_{membrane}(\Theta) &= \left[1+\frac{0.12\alpha }{2r\tilde{\Delta}^4 }\left\{\rho_r(\Theta) + \rho_\phi(\Theta)\right\}\right]^{1/2}, \nonumber\\
  &= \left[1+\frac{0.12\alpha }{2r\tilde{\Delta}^4 }\left(\frac{\pi^2}{30}g_* \frac{m^4_{\rm DM}}{\Theta^4}+g_\phi m^4_\chi x \left(\frac{x}{2\pi \Theta}\right)^{3/2}e^{-x\Theta}\right)\right]^{1/2} 
\label{eq:fmem}
\end{align}
In the present context, the thermally averaged cross-section is given by the following generalized expression \cite{Gondolo:1990dk}:
\begin{align*}\label{ddq}
\langle \sigma v \rangle &= \frac{1}{8m^4TK^2_2(m/T)}\int^\infty_{4m^2}\sigma(s-4m^2)\sqrt{s}K_1(\sqrt{s}/T)ds
\end{align*}
where $K_i$'s are the modified Bessel function of the $i^{\rm th}$ order. In the non-relativistic regime Eq~(\ref{ddq}) can be recast as:
\begin{align*}
\langle \sigma v \rangle &= \frac{2\Theta^{3/2}}{\pi^{1/2}}\int^\infty_0 \sigma v \epsilon^{1/2}e^{-\Theta\epsilon}d\epsilon =\frac{\Theta^{1/2}}{\sqrt{4\pi}}\int^\infty_0 (\sigma v)v^2e^{(-xv^2/4)}dv.
\end{align*}
where we use the following expressions:
  \begin{align*}
  \epsilon &= p^2_{\rm rel}/(4m^2)= \frac{(v/2)^2}{1-(v/2)^2} ,\\
  s &= \frac{4m^2}{1-v^2/4}.
  \end{align*}
Now, the dark matter relic abundance can be found out by
\begin{align}
\Omega_{\rm DM} h^2 &=\left(\frac{ m_{\chi}sY_{\rm DM}}{3H^2 M^2_p}\right)_{today}h^2\simeq \frac{1.07\times 10^9 \textrm{GeV}^{-1}}{J(\Theta_F)g^{1/2}_* M_{p}}
\end{align} 
where using s-wave approximation the newly introduced factor $J(\Theta_F)$ is defined as:
\begin{align}\label{eqwaaa:1}
J(\Theta_F) &= \int_{\Theta_F}^\infty d\Theta  \frac{\langle \widetilde{\sigma} v \rangle(\Theta)}{\Theta^2}  \nonumber \\
 &\approx \int_{\Theta_F}^{\infty}d\Theta \int_{0}^{\infty}dv \frac{v^2 \left(a + b v^2\right)}{\sqrt{4\pi \Theta}}\frac{e^{-\Theta v^2/4}}{f_{membrane}(\Theta)} \nonumber \\
 &= \int_{\Theta_F}^{\infty} \left(a+\frac{3}{2}\frac{b}{\Theta}\right)\frac{d\Theta}{\Theta^2 f_{membrane}(\Theta)} 
\end{align}
where $\Theta_F$ is the freeze-out temperature which can be calculated by numerically solving the following transcendental equation:
\begin{align}
\Theta_F &= \ln \left(\frac{0.038g_{\rm DM} m_{\chi} M_{p}\langle \widetilde{\sigma} v\rangle (\Theta) }{g^{1/2}_*\Theta^{1/2}_F }\right) 
         = \ln \left(\frac{0.038g_{\rm DM} m_{\chi} M_{p}\langle \sigma v\rangle }{g^
{1/2}_*f_{membrane}(\Theta_F)\Theta^{1/2}_F }\right).
\end{align}
In the context of RSII membrane paradigm when the energy density of the dark matter content is very very large compared to the membrane tension i.e. $\rho_{DM}=\rho_{\chi}>>\sigma$, then one can write
the following simplified expressions for the 
memebrane characteristic function $f_{membrane}(\Theta)$ as:
\begin{align}
f_{membrane}(\Theta) &\approx \left[\frac{0.12\alpha }{2r\tilde{\Delta}^4 }\left\{\rho_r(\Theta) + \rho_\phi(\Theta)\right\}\right]^{1/2}, \nonumber\\
  &= \left[\frac{0.12\alpha }{2r\tilde{\Delta}^4 }\left(\frac{\pi^2}{30}g_* \frac{m^4_{\rm DM}}{\Theta^4}+g_\phi m^4_\chi x \left(\frac{x}{2\pi \Theta}\right)^{3/2}e^{-x\Theta}\right)\right]^{1/2},
\label{eq:ev1xxxxxc}
\end{align}
which one can use for further computation. 

In fig.~(\ref{fig1}), we have depicted the model independent inflationary constraint from the primordial gravitational waves via tensor-to-scalar ratio on the dark matter relic abundance within the framework of 
effective field theory of RSII membrane paradigm. The yellow shaded region signifies the excluded region from Planck 2015+WMAP9 and Planck 2015+BICEP2/Keck Array joint data sets. The purple and black vertical lines 
signify the upper bound on the inflationary tensor-to-scalar ratio $r=0.12$ and $r=0.11$ obtained from Planck 2015+WMAP9 and Planck 2015+BICEP2/Keck Array joint data sets respectively. Also the grey coloured horizontal 
line signifies the bound on dark matter relic abundance $\Omega_{DM}h^2=0.1199\pm 0.0027$, as obtained from Planck 2015 data. The red and blue coloured curve represents the behaviour of the dark matter relic abundance 
with respected tensor-to-scalar ratio in RSII membrane paradigm for $\langle \sigma v\rangle=1.5\times 10^{-27}{\rm cm^{3}/s}$ and $\langle \sigma v\rangle=1.2\times 10^{-27}{\rm cm^{3}/s}$ respectively. Here for both of the 
cases we fix the value of the dimensionless membrane parameter $\alpha=V_{inf}/\sigma =1.77\times 10^{64}$. From fig.~(\ref{fig1}), it is clearly observed that for $r<0.01$, various inflationary models in 
membrane paradigm are in huge tension with the Planck 2015 constraint on dark matter relic abundance. This implies that the low scale inflationary models within RSII membrane paradigm are highly disfavoured. On the 
other hand, for $0.01\leq r\leq 0.12$, various inflationary models in 
membrane paradigm are consistent with $2\sigma$ constraint on dark matter relic abundance obtained from Planck 2015 data. This implies that the high scale inflationary models within RSII membrane paradigm are favoured for 
$0.01\leq r\leq 0.12$. 

\begin{figure*}[!htb]
\centering
\subfigure[$r$ vs $\langle \sigma v\rangle$ for $m_{\chi}=100{\rm GeV}$.]{
    \includegraphics[width=14.5cm,height=8cm] {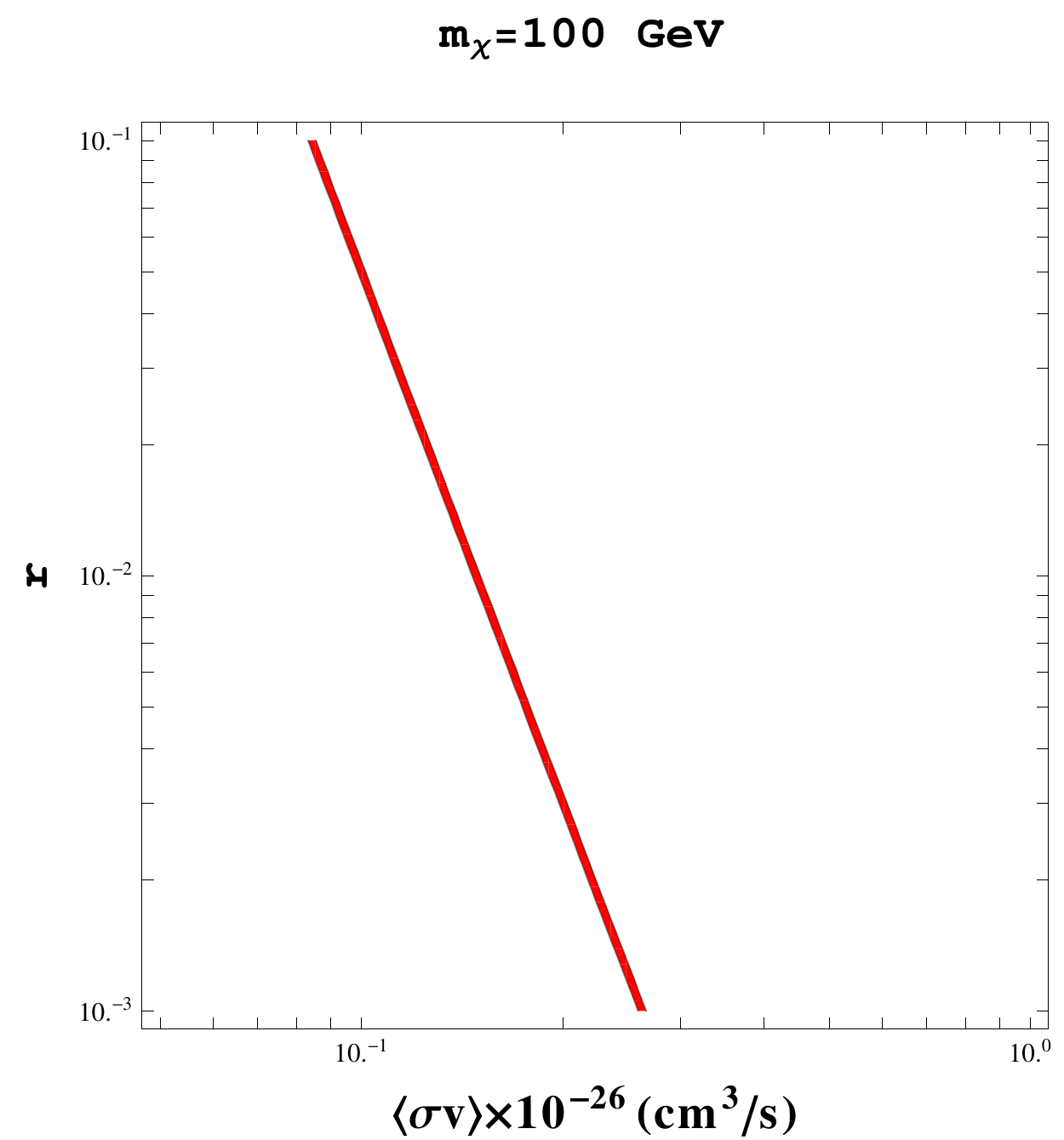}
    \label{fig2a}
}
\subfigure[$r$ vs $\langle \sigma v\rangle$ for $m_{\chi}=1{\rm TeV}$.]{
    \includegraphics[width=14.5cm,height=8cm] {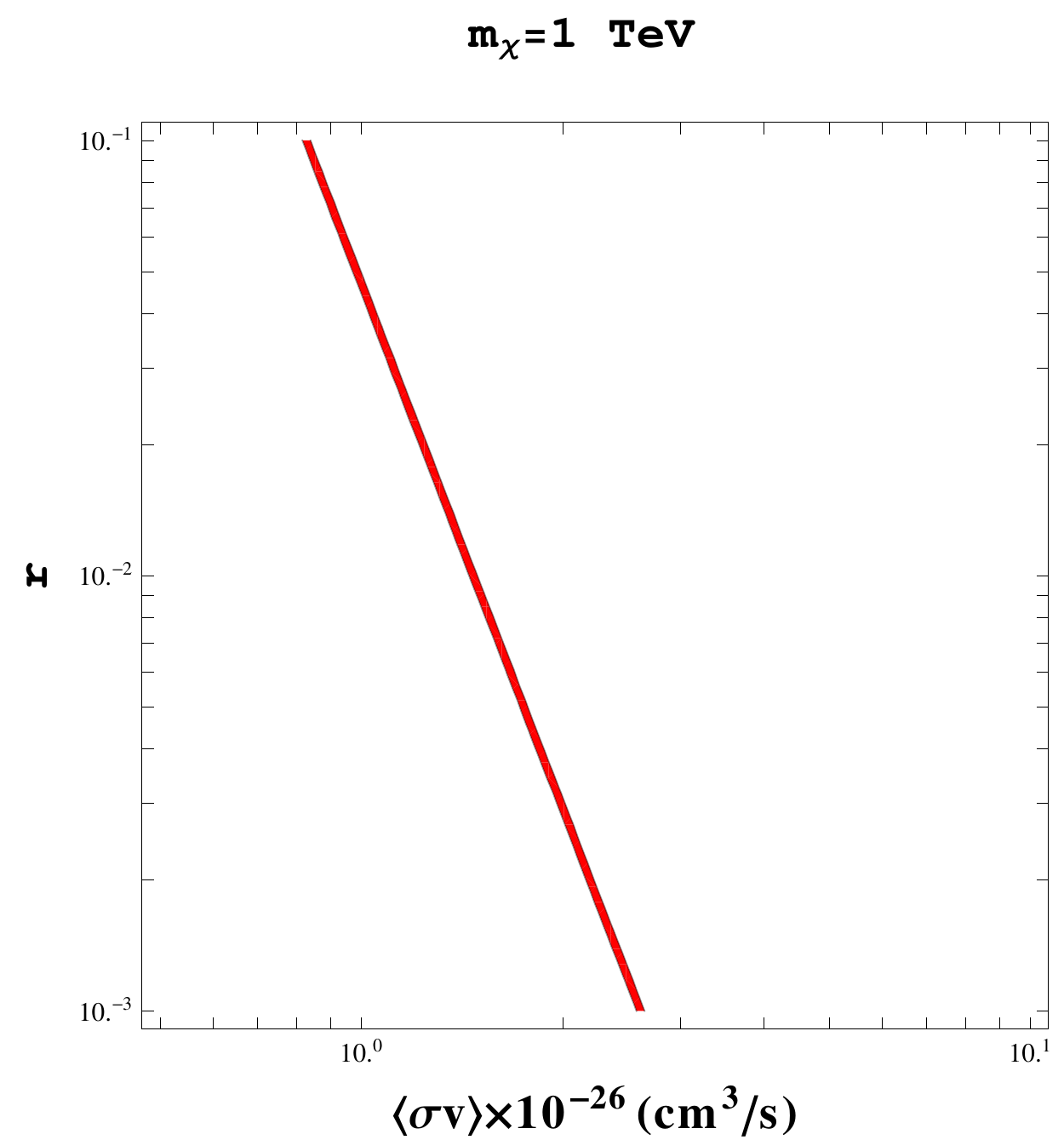}
    \label{fig2b}
}
\caption[Optional caption for list of figures]{In the above figure we have shown the allowed 
region of $\alpha$ with respect to the thermally averaged 
cross-section $\langle \sigma v \rangle $.} 
\label{fig2}
\end{figure*}

In fig.~(\ref{fig2a}) and fig.~(\ref{fig2b}), we have depicted the model independent inflationary constraint from the primordial gravitational waves via tensor-to-scalar ratio on
the thermally averaged annihilation cross-section of dark matter content for the dark matter mass, $m_{\chi}=100 {\rm GeV}$ and $m_{\chi}=1{\rm TeV}$ 
respectively, within the framework of 
effective field theory of RSII membrane paradigm. For all of these three cases from
the fig.~(\ref{fig2a}) and fig.~(\ref{fig2b}), it is clearly observed that 
the value of the inflationary tensor-to-scalar ratio decreases with the increase in
the thermally averaged annihilation cross-section of dark matter content. Also it is important to note that for of these three cases
within the allowed range of tensor-to-scalar ratio, $0.01\leq r\leq 0.12$, as constrained from
fig.~(\ref{fig1}) within RSII membrane paradigm, thermally averaged annihilation cross-section of dark matter content
is constrained within the window, $\langle \sigma v\rangle\sim{\cal O}(10^{-28}-10^{-27}){\rm cm^3/s}$,
which we will throughly follow in the rest of our analysis in this paper.

In the next section, we will explicitly study the various constraints and consequences from the effective field theory of dark matter within the framework of RSII membrane paradigm.

\section{Effective Field Theory of dark matter from membrane paradigm}
\label{avv11}
In the next couple of subsections we will explicitly study the role of the characteristic parameter $\alpha$ of RSII membrane paradigm which 
effects the allowed parameter space of effective couplings in the membrane and the mediator mass in variants of effective field theory model. More more
details effective theory see refs.~\cite{Berlin:2014tja}. For the sake of simplicity, in the present discussion we will assume that
 the couplings of dark matter with mediator and the mediator to fermions to be the 
same and we will follow this throughout the rest of the analysis performed in this paper. 
In table~(\ref{tab11}) we have explicitly mentioned the various types of dark matter, corresponding effective field theory operators, 
mediator spin and the coupling with 
mediator suppression scale.
\begin{table}[!h]
\centering
\begin{tabular}{|c|c|c|c|}
\hline\hline\hline
Type of dark matter & Effective Operator & mediator spin& coupling with \\
& & & mediator \\
\hline
\hline
Dirac Dark Matter & $\overline{\chi}\chi \overline{f}f $ &  spin-0 & $\frac{g^2}{M^2}$ \\
Majorana Dark Matter & $\overline{\chi}\chi \overline{f}f$ & spin-0 & $\frac{g^2}{M^2}$ \\
Dirac Dark Matter & $\overline{\chi}\gamma^\mu \chi \overline{f}\gamma_\mu f$ &  spin-1 & $\frac{g^2}{M^2}$ \\
Majorana Dark Matter & $\overline{\chi}\gamma^\mu \gamma_5 \chi \overline{f}\gamma_\mu f$ &  spin-1 & $\frac{g^2}{M^2}$ \\
Complex Scalar Dark Matter & $\phi^*\phi \overline{f}f$ &  spin-0 & $\frac{|\mu|^2}{M^2}$ \\
Real Scalar Dark Matter & $\phi^2 \overline{f}f$ &  spin-0 & $\frac{\mu^2}{M^2}$ \\
Complex Scalar Dark Matter & $\phi^*\partial_\mu \phi \overline{f}\gamma^\mu f$ &  spin-1 & $\frac{g^2}{M^2}$ \\
Real Scalar Dark Matter & $\phi \partial_\mu \phi \overline{f}\gamma^\mu f$ &  spin-1 & $\frac{g^2}{M^2}$ \\
Complex Vector Scalar Dark Matter & $X^\mu X^*_\mu \overline{f}f$ &  spin-0 & $\frac{|\mu_X|g}{M^2}$ \\
Real Vector Scalar Dark Matter & $X^\mu X_\mu \overline{f}f$ &  spin-0 & $\frac{|\mu_X|g}{M^2}$ \\
Complex Vector Scalar Dark Matter & $X^{*\nu} \partial_\nu X_\mu \overline{f}\gamma^\mu f$ &  spin-1 & $\frac{g_Xg}{M^2}$ \\
Real Vector Scalar Dark Matter & $X^{\nu} \partial_\nu X_\mu \overline{f}\gamma^\mu f$ &  spin-1 & $\frac{g_Xg}{M^2}$ \\
\hline\hline\hline
\end{tabular}
\caption{Tabular representation of various types of dark matter, corresponding effective field theory operators, mediator spin and the coupling with 
mediator suppression scale.}
\label{tab11}
\end{table}

\subsection{\bf Dirac dark matter: spin-0 mediator}

\subsubsection{\bf s-channel analysis}
\label{a3v1}
To start with let us consider the following localized interactions for Dirac dark matter, $\chi$ and a spin-0 mediator, $A$,
within the framework of effective field theory written in RSII membrane as:
\begin{align}
\mathcal{L}_{brane} &\supset \left[\overline{\chi} (\lambda_{\chi_S}+\lambda_{\chi_p}i\gamma^5)\chi 
+ \overline{f}(\lambda_{f_S} + \lambda_{f_p}i\gamma^5)f\right]A,
\end{align}
where $\lambda_{\chi_S},\lambda_{\chi_p}$ and $\lambda_{f_S}, \lambda_{f_p}$ are the fermionic couplings for dark matter $\chi$ and standard model (SM) fermion $f$. 
In fig.~(\ref{fig3a}) and fig.~(\ref{fig3b}), we have explicitly shown the Feynman diagrammatic representation of possible s-channel and t/u-channel processes for dirac dark matter with spin-0 mediator.
The cross-section from the above Lagrangian after taking all the coupling equal is given as: 
\begin{align}
\sigma &= \frac{1}{8\pi (s-4m^2_\chi)}\int_{t_-}^{t_+}|\mathcal{M}|^2 dt 
\nonumber \\
 &= \frac{n_cg^4}{8\pi s \left((s-m^2_A)^2 + m^2_A\Gamma^2_A\right)}\sqrt{\left[\frac{1-4m_f^2/s}{1-4m^2_\chi/s}\right]}(s-2m^2_f)(s-2m^2_\chi).
\end{align}
where $n_c=3$ for quarks and 1 for leptons, $g_A$ and $m_A$ are the respective coupling and the mass of the mediator. The matrix element for the S-matrix and the symbol $t_{\pm}$ is given by:
\bea |\mathcal{M}|^2 &=& \frac{(s-2m^2_\chi)(s-2m^2_f)}{(s-m^2_A)^2+m^2_A\Gamma_A}, \\
t_{\pm} &=& (m^2_\chi + m^2_f-\frac{s}{2}) \pm \frac{\sqrt{(s-4m^2_\chi)(s-4m^2_f)}}{2}.\eea
\begin{figure*}[!ht]
\centering
\subfigure[$\chi\bar{\chi}\rightarrow f\bar{f}$~process for s-channel.]{
    \includegraphics[width=6.5cm,height=4.2cm] {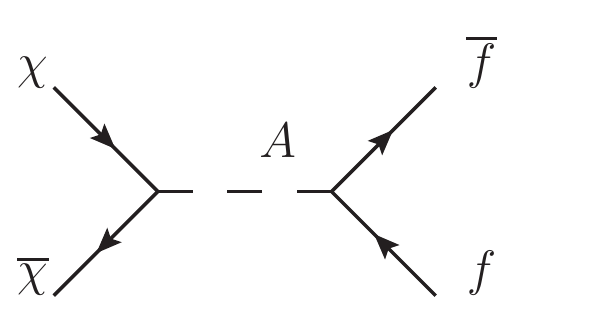}
    \label{fig3a}
}
\subfigure[$\chi{\chi}^c\rightarrow f\bar{f}$~process for s-channel.]{
    \includegraphics[width=6.5cm,height=4.2cm] {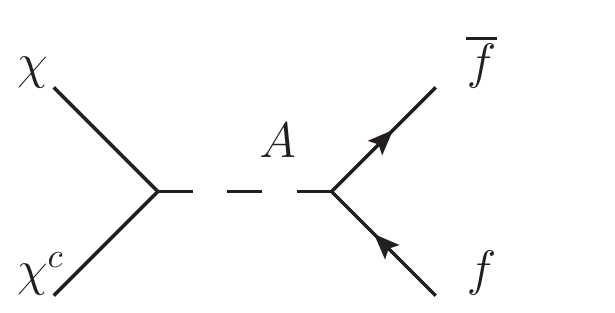}
    \label{fig3b}
}
\caption[Optional caption for list of figures]{Feynman diagrammatic representation of s-channel channel processes for dirac dark matter and Majorana with spin-0 mediator.} 
\label{fig3}
\end{figure*}
\begin{figure*}[!ht]
\centering
\subfigure[$g_{A}$ vs $m_{A}$ for s-channel with $m_{\chi}=100{\rm GeV}$.]{
   \includegraphics[width=14.5cm,height=8cm]{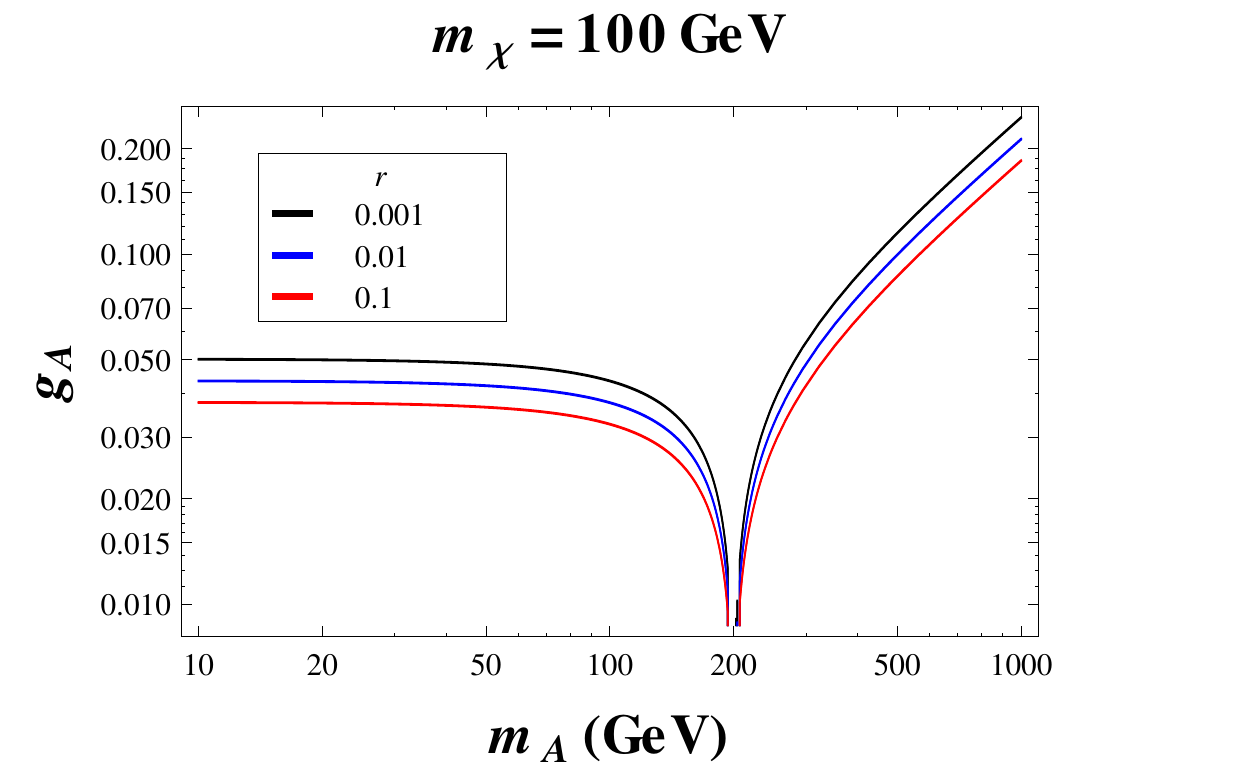}
    \label{fig4a}
}
\subfigure[$g_{A}$ vs $m_{A}$ for s-channel with $m_{\chi}=1{\rm TeV}$.]{
    \includegraphics[width=14.5cm,height=8cm]{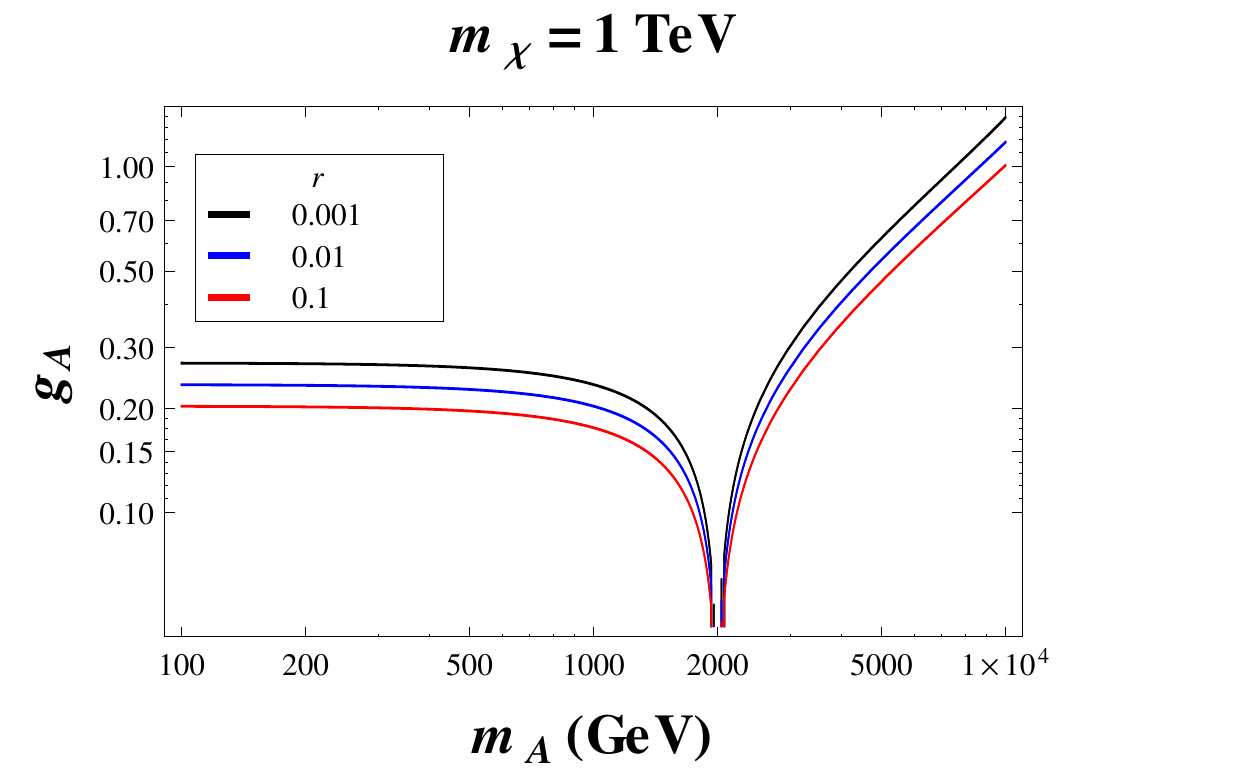}
    \label{fig4b}
}
\caption[Optional caption for list of figures]{In the above figure we have shown the allowed 
region of $g_A$ with respect to the mass of the mediator $m_A$, for the 
s-channel processes.} 
\label{fig4}
\end{figure*}

For a low velocity perturbation regime by taking the following approximation: \be s=4m^2_{\chi}{(1+v^2/4)},\ee we finally get the following simplified expression for the product of annihilation cross-section and velocity as:
\begin{align}
\sigma v \simeq a + \mathcal{O}(v^2) \end{align}
where in the present context the factor $a$ is given by:
\begin{align}
    a &= n_cg^2 \frac{2m^2_\chi - m^2_f}{2\pi(m^4-8m^2m^2_{\chi}+16m^4_\chi + m^2\Gamma^2)}\sqrt{1-\frac{m^2_f}{m^2_\chi}}\end{align}
    and the mediator's width to SM fermions is given by:
    \begin{align}
    \Gamma &=\sum_{f}\Gamma(A\rightarrow f\bar{f})= \frac{g^4n_cm}{4\pi}\left[1-4\frac{m^2_f}{m^2}\right]^{1/2}\left[1-2\frac{m^2_f}{m^2}\right].
\end{align}
 The relic abundance is given as:
\begin{align}
 \Omega_{DM} h^2 &= \frac{1.07\times 10^9}{J(x_f)g^{1/2}_*M_p}\end{align}
 where the $J(x_f)$ for this spin-0 mediator s-channel process is given by:
 \begin{align}J(x_f) &= \int_{x_f}^{\infty}n_cg^2\frac{2m^2_\chi - m^2_f}{2\pi(m^4-8m^2m^2_{\chi}+16m^4_\chi + m^2\Gamma^2)x^2f_{membrane}(x)}\sqrt{1-\frac{m^2_f}{m^2_\chi}}dx
\end{align}
where the function $f_{membrane}(x)$ is the characteristic parameter for RS single braneworld and can be expressed in terms of tensor-to-scalar ratio ($r$) which is given in Eq.\eqref{eq:ev1xxxxxc}.
Now for the GR limiting case of the $f_{membrane}(x) \rightarrow 1$ and then the relic abundance will only depend on the mass of the Dark Matter ($m_\chi$), $g_A$ the 
coupling with the spin-0 mediator and the mass of the mediator ($m_A$). 
\begin{align}
 \Omega_{DM} h^2 &= \frac{1.07\times 10^9x_f}{g^{1/2}_*M_p}\left(n_cg^2 \frac{2m^2_\chi - m^2_f}{2\pi(m^4-8m^2m^2_{\chi}+16m^4_\chi + m^2\Gamma^2)}\sqrt{1-\frac{m^2_f}{m^2_\chi}}\right)^{-1}.~~
\end{align}
In order to constrain the coupling ($g_A$) and mass ($m_A$) we take the present data of the relic abundance ($\Omega_{DM}h^2 = 0.1199\pm0.0027$\cite{Ade:2015xua}) and constrain the function 
$J(x_f)$, which in turn constrain the coupling $g_A$ and $m_A$ for a particular tensor-to-scalar ratio ($r$). We have not shown the GR limiting case as it has been extensively been explored in 
\cite{Berlin:2014tja}.
In fig.~(\ref{fig4a}) and fig.~(\ref{fig4b}), we have depicted the behaviour of the effective coupling of spin-0 mediator $g_{A}$ with the varying mass the spin-0 mediator $m_{A}$ 
for s-channel process with three distinct value of the tensor-to-scalar ratio $r=0.001$, $r=0.01$ and $r=0.1$ 
respectively in RSII membrane. We also consider two different values of the dark matter mass $m_{\chi}=100 {\rm GeV}$ and $m_{\chi}=1\;{\rm TeV}$
 for the s-channel analysis. From fig.~(\ref{fig4a}) and fig.~(\ref{fig4b}), it is clearly observed
that the behaviour of the effective coupling of spin-0 mediator $g_{A}$ with the varying mass the spin-0 mediator $m_{A}$ are similar in both of the cases and also 
sensitive in the vicinity of $m_{A}=2\times 10^{2}\;{\rm GeV}$ and $m_{A}=2\times 10^{3}\;{\rm GeV}$ respectively as it has a resonance 
(i.e $2m_X = m_A$).
Most importantly, in both the sides of $m_{A}=2\times 10^{2}\;{\rm GeV}$ and $m_{A}=2\times 10^{3}\;{\rm GeV}$ the coupling of spin-0 mediator $g_{A}$ behave in completely opposite manner.

\begin{figure*}[!htb]
\centering
\subfigure[$g_{A}$ vs $m_{A}$ for t/u-channel with $m_{\chi}=100{\rm GeV}$.]{
    \includegraphics[width=14.5cm,height=8cm]{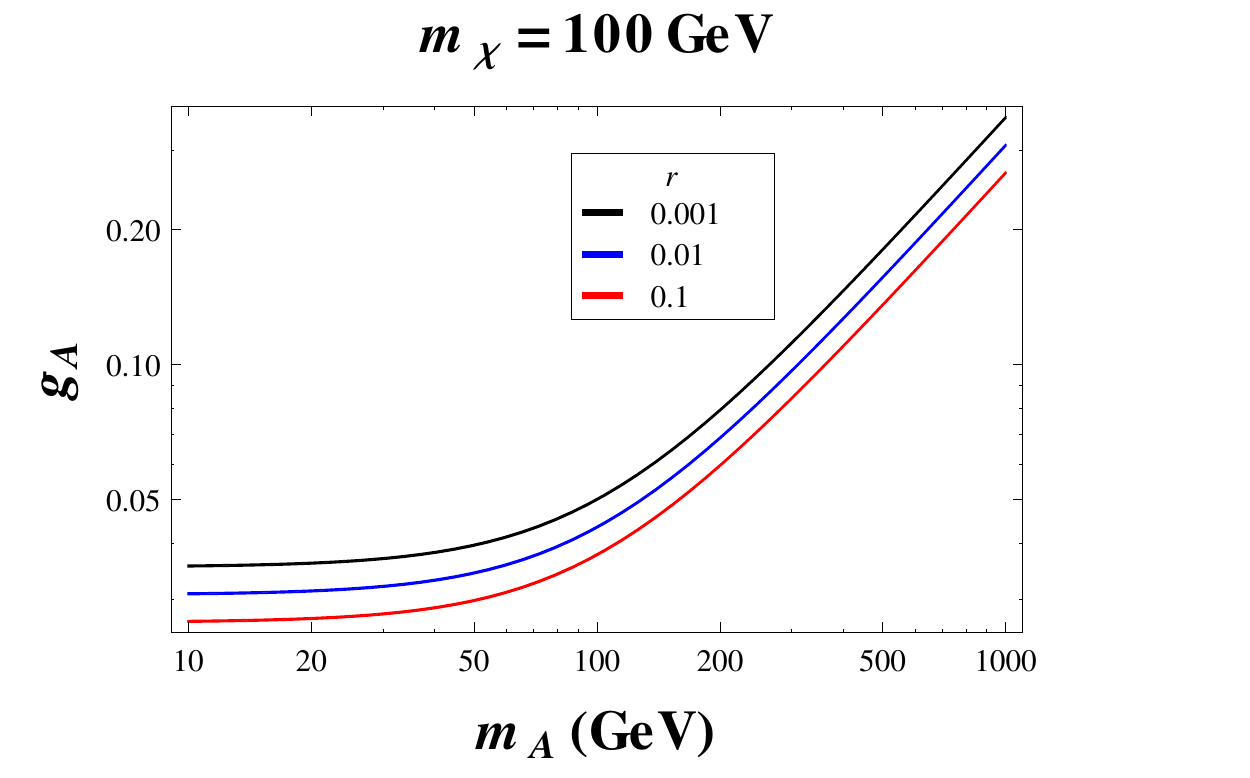}
    \label{fig5a}
}
\subfigure[$g_{A}$ vs $m_{A}$ for t/u-channel with $m_{\chi}=1{\rm TeV}$.]{
    \includegraphics[width=14.5cm,height=8cm]{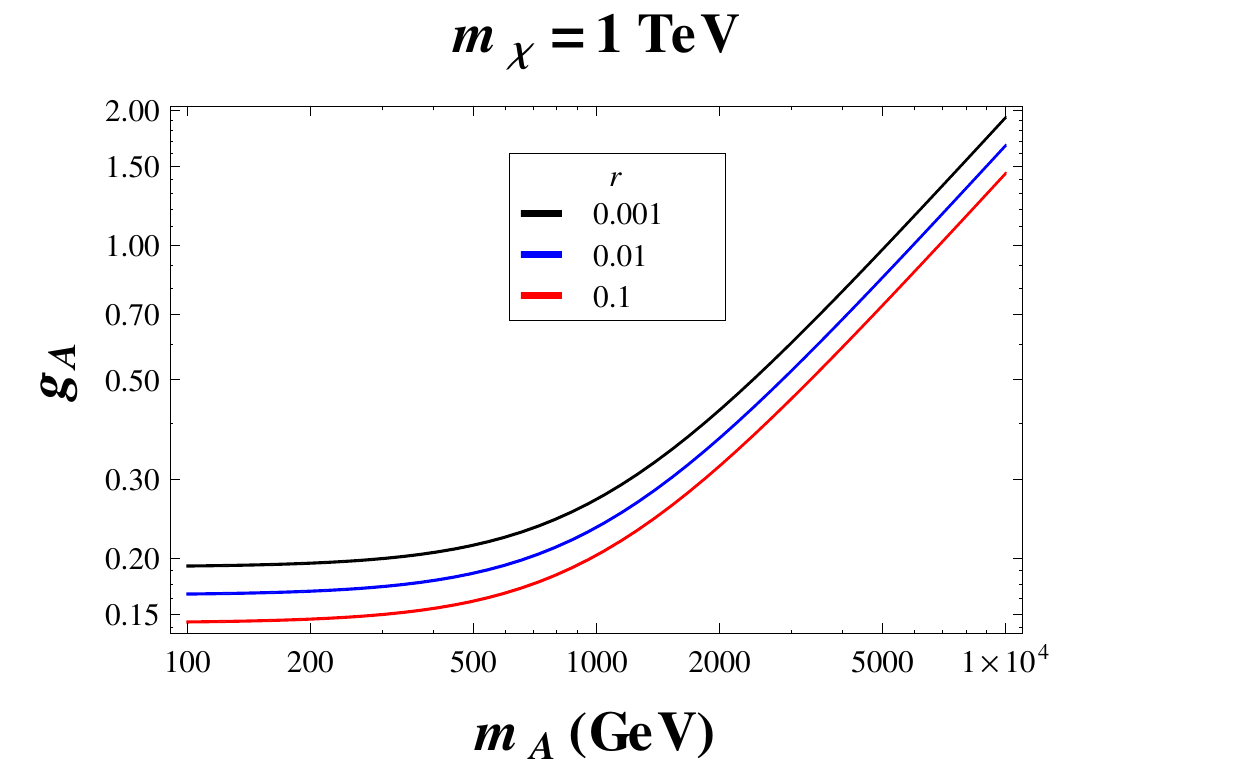}
    \label{fig5b}
}
\caption[Optional caption for list of figures]{In the above figure we have shown the allowed 
region of $g_A$ with respect to the mass of the mediator $m_A$, for the t/u - channel processes.} 
\label{fig5}
\end{figure*}
\subsubsection{\bf t/u- channel analysis}
\label{a3v2}
Next we consider the following localized interactions for Dirac dark matter, $\chi$ and a spin-0 mediator, $A$,
within the framework of effective field theory written in RSII membrane as:
\begin{align}
\mathcal{L}_{membrane} &\supset \overline{\chi} (\lambda_{\chi_S}+\lambda_{\chi_p}i\gamma^5)fA 
+ \overline{f}(\lambda_{f_S} + \lambda_{f_p}i\gamma^5)\chi A^\dag
\end{align}
Now, considering all the couplings to be of the same order the thermally averaged cross-section becomes:
$\lambda_{f_S} \thicksim \lambda_{f_p} \thicksim \lambda_{\chi_S} \thicksim \lambda_{\chi_p} \thicksim g_A$ the amplitude is given as 
\begin{align}
\sigma &= \frac{1}{8\pi (s-4m^2_\chi)}\int_{t_-}^{t_+}|\mathcal{M}|^2 dt,
\end{align}
where the matrix element for the S-matrix and $t_{\pm}$ is defined as:
\bea |\mathcal{M}|^2 &=& g^4_A n_c \frac{(m^2_f+m^2_\chi -t)^2}{(t - m^2_A)^2}, \\
t_{\pm} &=& (m^2_\chi + m^2_f-\frac{s}{2}) \pm \frac{\sqrt{(s-4m^2_\chi)(s-4m^2_f)}}{2}. \eea 
where $n_c=3$ for quarks and 1 for leptons, $g_A$ and $m_A$ are the respective coupling and the mass of the mediator. Now taking the following approximation: \be s=4m^2_\chi/\left(1-v^2/4\right)\ee 
we finally get the following simplified expression for the product of annihilation cross-section and velocity as:
\begin{align}\sigma v = a + \mathcal{O}(v^2) \end{align} where in the present context, the factor $a$ is given by: 
\begin{align}
a &\approx \frac{n_c  g^4 \sqrt{1-m^2_f/m^2_\chi}}{4 \pi \left(m^2 - m^2_f + m^2_\chi \right)^2}
 m^2_\chi. 
\label{eq:dfs0}
\end{align}
%
\begin{figure*}[!ht]
\centering
\subfigure[$\chi\bar{\chi}\rightarrow f\bar{f}$~process for t/u-channel.]{
    \includegraphics[width=6.5cm,height=4.2cm] {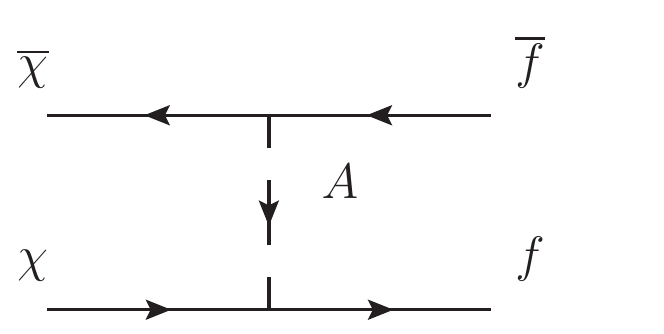}
    \label{fig3a}
}
\subfigure[$\chi{\chi}^c\rightarrow f\bar{f}$~process for t/u-channel.]{
    \includegraphics[width=6.5cm,height=4.2cm] {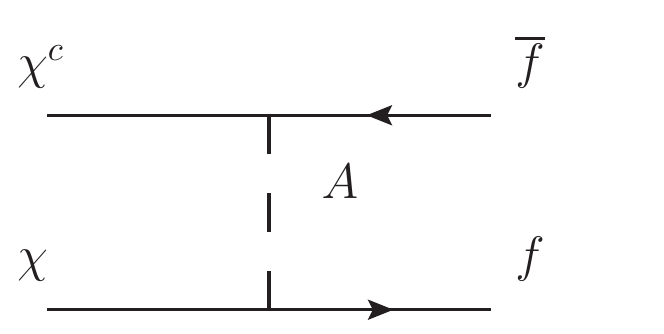}
    \label{fig3b}
}
\caption[Optional caption for list of figures]{Feynman diagrammatic representation of t/u-channel channel processes for Dirac dark matter and Majorana with spin-0 mediator.} 
\label{fig3t}
\end{figure*}
The dark matter relic abundance in this case is given as:
\begin{align}
 \Omega_{DM} h^2 &= \frac{1.07\times 10^9}{J(x_f)g^{1/2}_*M_p}\end{align}
 where $J(x_f)$ for spin-0 mediator in t/u channel is given by:
 \begin{align} J(x_f) &= \int_{x_f}^{\infty}\frac{n_c  g^4 \sqrt{1-m^2_f/m^2_\chi}m^2_{\chi}}{4 \pi \left(m^2 - m^2_f + m^2_\chi \right)^2x^2f_{membrane}(x)}dx
\end{align}
where the function $f_{membrane}(x)$ is the characteristic parameter for RS single braneworld and this can be expressed in terms of tensor-to-scalar ratio ($r$) which is given in Eq.\eqref{eq:ev1xxxxxc}.
Now for the GR limiting case of the $f_{membrane}(x) \rightarrow 1$ and then the relic abundance will only depend on the mass of the Dark Matter ($m_\chi$), $g_A$ the 
coupling with the spin-0 mediator and the mass of the mediator ($m_A$). 
\begin{align}
 \Omega_{DM} h^2 &= \frac{1.07\times 10^9x_f}{g^{1/2}_*M_p}\left(\frac{n_c  g^4 \sqrt{1-m^2_f/m^2_\chi}}{4 \pi \left(m^2 - m^2_f + m^2_\chi \right)^2}
 m^2_\chi\right)^{-1}.
\end{align}
In order to constrain the coupling ($g_A$) and mass ($m_A$) we take the present data of the relic abundance ($\Omega_{DM}h^2 = 0.1199\pm0.0027$\cite{Ade:2015xua}) and constrain the function 
$J(x_f)$, which in turn constrain the coupling $g_A$ and $m_A$ for a particular tensor-to-scalar ratio ($r$). We have not shown the GR limiting case as it has been extensively been explored in 
\cite{Berlin:2014tja}.
In fig.~(\ref{fig5a}) and fig.~(\ref{fig5b}), we have depicted the behaviour of the effective coupling of spin-0 mediator $g_{A}$ with the varying mass the spin-0 mediator $m_{A}$ 
for t/u-channel process with three distinct value of the tensor-to-scalar ratio $r=0.001$, $r=0.01$ and $r=0.1$ 
respectively in RSII membrane. We also consider two different values of the dark matter mass $m_{\chi}=100 {\rm GeV}$ and $m_{\chi}=1{\rm TeV}$ for the t/u-channel analysis. From fig.~(\ref{fig5a}) and fig.~(\ref{fig5b}),
it is clearly observed
that the behaviour of the effective coupling of spin-0 mediator $g_{A}$ with the varying mass the spin-0 mediator $m_{A}$ are similar for both of the cases and also 
sensitive in the vicinity of $m_{A}=1\times 10^{2}\;{\rm GeV}$ and $m_{A}=1\times 10^{3}\;{\rm GeV}$ respectively as it has a resonance 
(i.e $2m_X = m_A$).
Most importantly, in both the sides of $m_{A}=1\times 10^{2}\;{\rm GeV}$ and $m_{A}=1\times 10^{3}\;{\rm GeV}$
the coupling of spin-0 mediator $g_{A}$ behave in completely opposite manner.

\subsection{\bf Dirac Dark matter: spin-1 mediator}

\subsubsection{\bf s-channel analysis}
\label{a4v1}
\begin{figure*}[!htb]
\centering
\subfigure[$\chi\bar{\chi}\rightarrow f\bar{f}$~process for s-channel.]{
    \includegraphics[width=6.5cm,height=4.2cm]{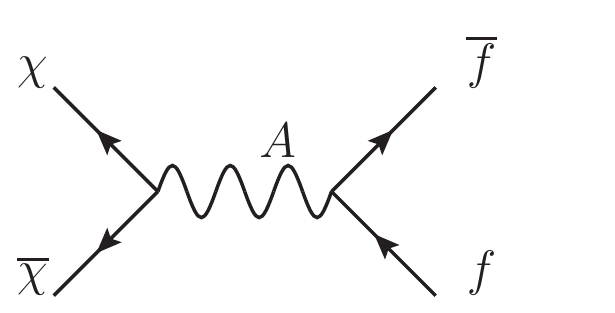}
    \label{fig6a}
}
\subfigure[$\chi{\chi}^c\rightarrow f\bar{f}$~process for s-channel.]{
    \includegraphics[width=6.5cm,height=4.2cm]{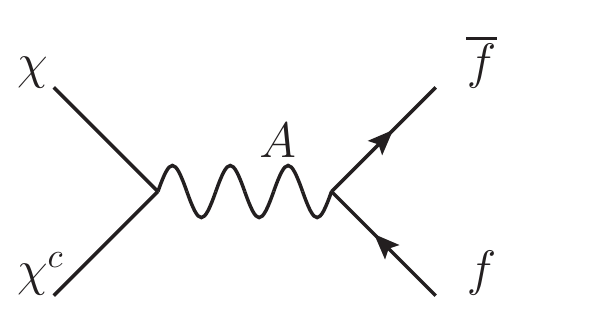}
    \label{fig6b}
}
\caption[Optional caption for list of figures]{Feynman diagrammatic representation of s-channel processes for dirac dark matter with spin-1 mediator.} 
\label{fig6}
\end{figure*}

\begin{figure*}[!htb]
\centering
\subfigure[$\chi\bar{\chi}\rightarrow f\bar{f}$~process for t/u-channel.]{
    \includegraphics[width=6.5cm,height=4.2cm]{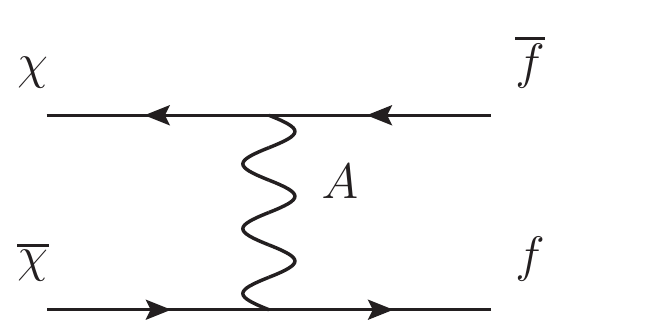}
    \label{fig7a}
}
\subfigure[$\chi{\chi}^c\rightarrow f\bar{f}$~process for t/u-channel.]{
    \includegraphics[width=6.5cm,height=4.2cm]{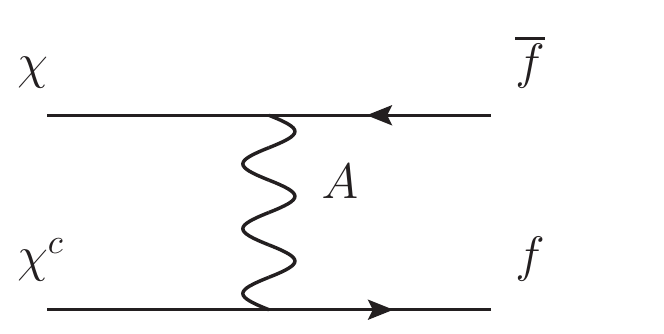}
    \label{fig7b}
}
\caption[Optional caption for list of figures]{Feynman diagrammatic representation of t/u-channel processes for Dirac dark matter with spin-1 mediator.} 
\label{fig7}
\end{figure*}

\begin{figure*}[!htb]
\centering
\subfigure[$g_{A}$ vs $m_{A}$ for s-channel with $m_{\chi}=100{\rm GeV}$.]{
    \includegraphics[width=14.5cm,height=8cm]{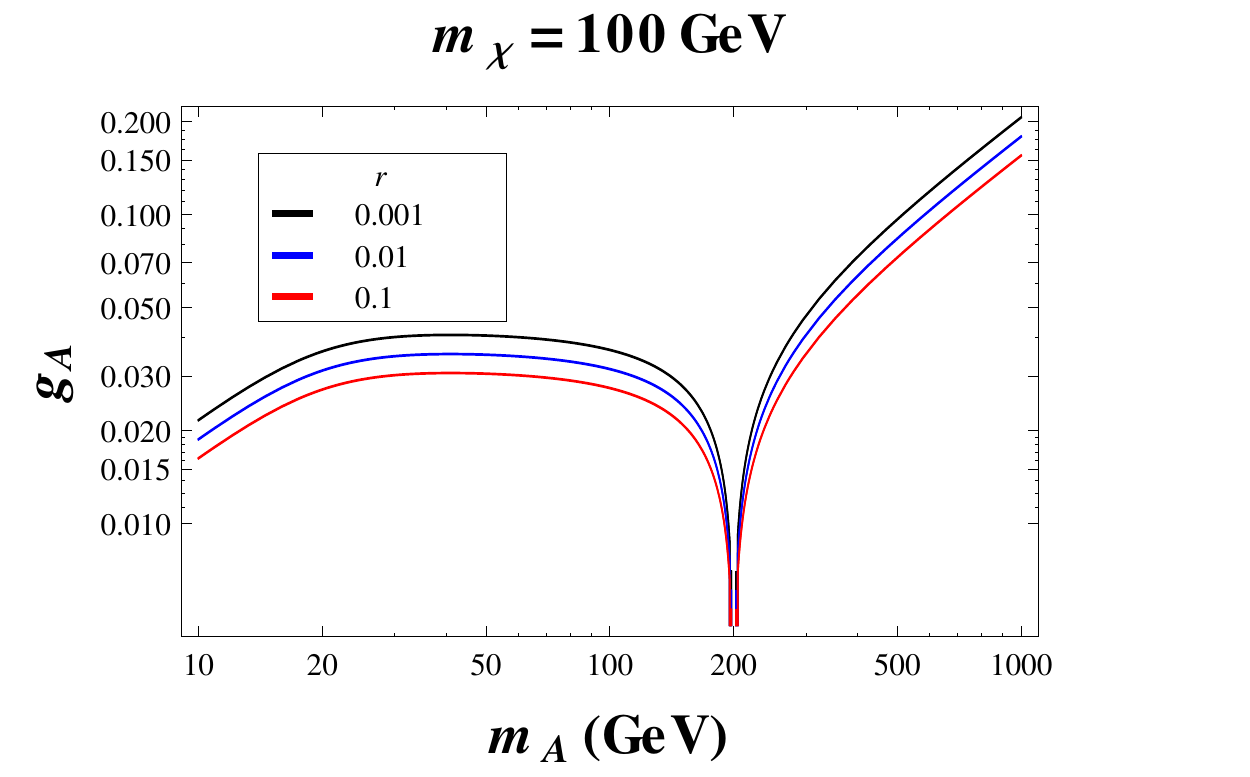}
    \label{fig8a}
}
\subfigure[$g_{A}$ vs $m_{A}$ for s-channel with $m_{\chi}=1{\rm TeV}$.]{
    \includegraphics[width=14.5cm,height=8cm]{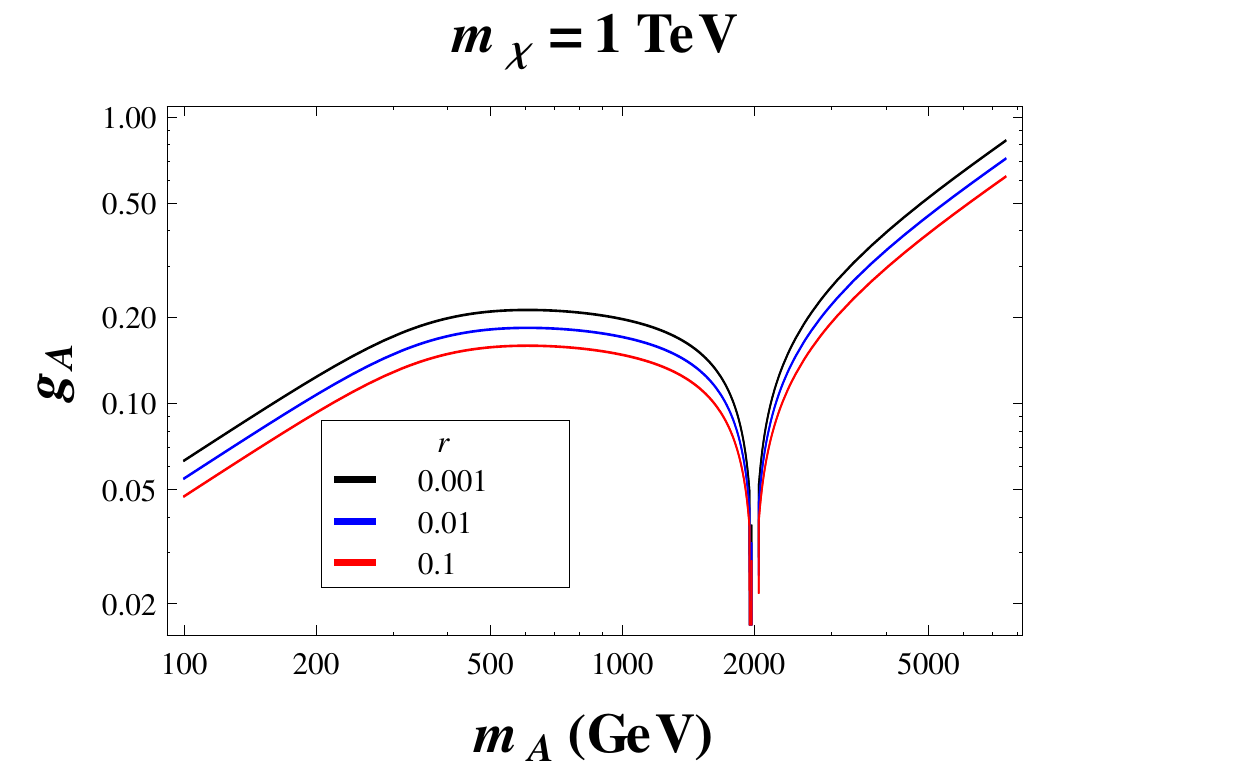}
    \label{fig8b}
}
\caption[Optional caption for list of figures]{In the above figure we have shown the allowed 
region of $g_A$ with respect to the mass of the mediator $m_A$, the upper panel is for the 
s-channel processes.} 
\label{fig8}
\end{figure*}

Next we consider the following localized interactions for a Dirac dark matter particle, $\chi$, and a spin-1 mediator, $V_\mu$,
within the framework of effective field theory written in RSII membrane as:
\begin{align}
\mathcal{L}_{membrane} &\supset \left[\overline{\chi}\gamma^\mu(g_{\chi_V} + g_{\chi_a}\gamma_5)\chi 
+ \overline{f}\gamma^\mu(g_{f_V} + g_{f_a})f \right]V_\mu. 
\end{align}
In fig.~(\ref{fig6a}), fig.~(\ref{fig6b}), fig.~(\ref{fig7a}) and fig.~(\ref{fig7b}), we have explicitly shown the
Feynman diagrammatic representation of possible s-channel and t/u-channel processes for Dirac dark matter with spin-1 mediator respectively.
After taking the equal coupling approximation the cross-section from the above Lagrangian is computed as: 
\begin{align}
\sigma &= \frac{1}{8\pi (s-4m^2_\chi)}\int_{t_-}^{t_+}|{\cal M}|^2 dt 
\end{align}
where the matrix element for the S-matrix and the symbol $t_{\pm}$ is given by:
\begin{align}|\mathcal{M}|^2 &= \frac{2n_cg^4}{3\pi s \left((s-m^2_A)^2 + m^2\Gamma^2 \right)}\left\{4m^2_\chi m^2_f\left(7-6\frac{s}{m^2}+\frac{3s^2}{m^4}\right)-4m^2_\chi s \right. \nonumber \\
    &\left.~~~~~~~~~~~~+ s(s-4m^2_f) + (s-4m^2_f)(s+2m^2_\chi) 
  + 2(s+2m^2_f)(s-m^2_\chi)\right\},\\
  t_{\pm} &= (m^2_\chi + m^2_f-\frac{s}{2}) \pm \frac{\sqrt{(s-4m^2_\chi)(s-4m^2_f)}}{2}.\end{align}
where $n_c=3$ for quarks and 1 for leptons, $g_A$ and $m_A$ are the respective coupling and the mass of the mediator. For a low velocity perturbation by taking the following approximation: \be s=4m^2_{\chi}{(1+v^2/4)} \ee
we finally get the following simplified expression for the product of annihilation cross-section and velocity as: 
\begin{align}
\sigma v &\simeq a + \mathcal{O}(v^2)\end{align}
 where in the present context, the factor $a$ is given by:
 \begin{align}
    a &= 2n_cg^4 \frac{m^2_\chi}{\pi m^4}\sqrt{1-\frac{m^2_f}{m^2_\chi}}
    \frac{1 - 2\frac{m^2_f}{m^2} + 4 \frac{m^2_\chi m^2_f}{m^4}}{2\pi(1-8\frac{m^2_{\chi}}{m^2}+16\frac{m^4_\chi}{m^4} + \frac{\Gamma^2}{m^2})}\end{align}
    and the mediator's width to SM fermions is given by:
    \begin{align}
    \Gamma &=\sum_{f}\Gamma(A\rightarrow f\bar{f})= \frac{g^2n_cm}{6\pi}\left[1-4\frac{m^2_f}{m^2}\right]^{1/2}\left[1-2\frac{m^2_f}{m^2}\right] \nonumber
\end{align}
The drak matter relic abundance in this case is given as:
\begin{align}
 \Omega_{DM} h^2 &= \frac{1.07\times 10^9}{J(x_f)g^{1/2}_*M_p}\end{align}
 where $J(x_f)$ for spin-1 mediator s-channel is give by:
 \begin{align}J(x_f) &= \int_{x_f}^{\infty}2n_cg^4 \frac{m^2_\chi}{\pi m^4}\sqrt{1-\frac{m^2_f}{m^2_\chi}}
    \frac{1 - 2\frac{m^2_f}{m^2} + 4 \frac{m^2_\chi m^2_f}{m^4}}{2\pi(1-8\frac{m^2_{\chi}}{m^2}+16\frac{m^4_\chi}{m^4} + \frac{\Gamma^2}{m^2})^2x^2f_{membrane}(x)}dx
\end{align}
where the function $f_{membrane}(x)$ is the characteristic parameter for RS single braneworld and this can be expressed in terms of tensor-to-scalar ratio ($r$) which is given in Eq.\eqref{eq:ev1xxxxxc}. 
Now for the GR limiting case of the $f_{membrane}(x) \rightarrow 1$ and then the relic abundance will only depend on the mass of the Dark Matter ($m_\chi$), $g_A$ the 
coupling with the spin-0 mediator and the mass of the mediator ($m_A$). 
\begin{align}
 \Omega_{DM} h^2 &= \frac{1.07\times 10^9x_f}{g^{1/2}_*M_p}\left(2n_cg^4 \frac{m^2_\chi}{\pi m^4}\sqrt{1-\frac{m^2_f}{m^2_\chi}}
    \frac{1 - 2\frac{m^2_f}{m^2} + 4 \frac{m^2_\chi m^2_f}{m^4}}{2\pi(1-8\frac{m^2_{\chi}}{m^2}+16\frac{m^4_\chi}{m^4} + \frac{\Gamma^2}{m^2})}\right)^{-1}.
\end{align}
In order to constrain the coupling ($g_A$) and mass ($m_A$) we take the present data of the relic abundance ($\Omega_{DM}h^2 = 0.1199\pm0.0027$\cite{Ade:2015xua}) and constrain the function 
$J(x_f)$, which in turn constrain the coupling $g_A$ and $m_A$ for a particular tensor-to-scalar ratio ($r$). We have not shown the GR limiting case as it has been extensively been explored in 
\cite{Berlin:2014tja}.
In fig.~(\ref{fig8a}) and fig.~(\ref{fig8b}), we have depicted the behaviour of the effective coupling of spin-1 mediator $g_{A}$ with the varying mass the spin-1 mediator $m_{A}$ 
for s-channel process with three distinct value of the tensor-to-scalar ratio $r=0.001$, $r=0.01$and $r=0.1$ 
respectively in RSII membrane. We also consider two different values of the dark matter mass $m_{\chi}=100 \;{\rm GeV}$ and $m_{\chi}=1\;{\rm TeV}$
 for the s-channel analysis. From fig.~(\ref{fig8a}) and fig.~(\ref{fig8b}), it is clearly observed
that the behaviour of the effective coupling of spin-1 mediator $g_{A}$ with the varying mass the spin-1 mediator $m_{A}$ are similar for both of the cases and also 
sensitive in the vicinity of $m_{A}=2\times 10^{2}\;{\rm GeV}$ and $m_{A}=2\times 10^{3}\;{\rm GeV}$ respectively as it has a resonance 
(i.e $2m_X = m_A$).
Most importantly, in both the sides of $m_{A}=2\times 10^{2}\;{\rm GeV}$ and $m_{A}=2\times 10^{3}\;{\rm GeV}$
the coupling of spin-1 mediator $g_{A}$ behave in completely opposite manner.

\begin{figure*}[!htb]
\centering
\subfigure[$g_{A}$ vs $m_{A}$ for t/u-channel with $m_{\chi}=100{\rm GeV}$.]{
   \includegraphics[width=14.5cm,height=8cm]{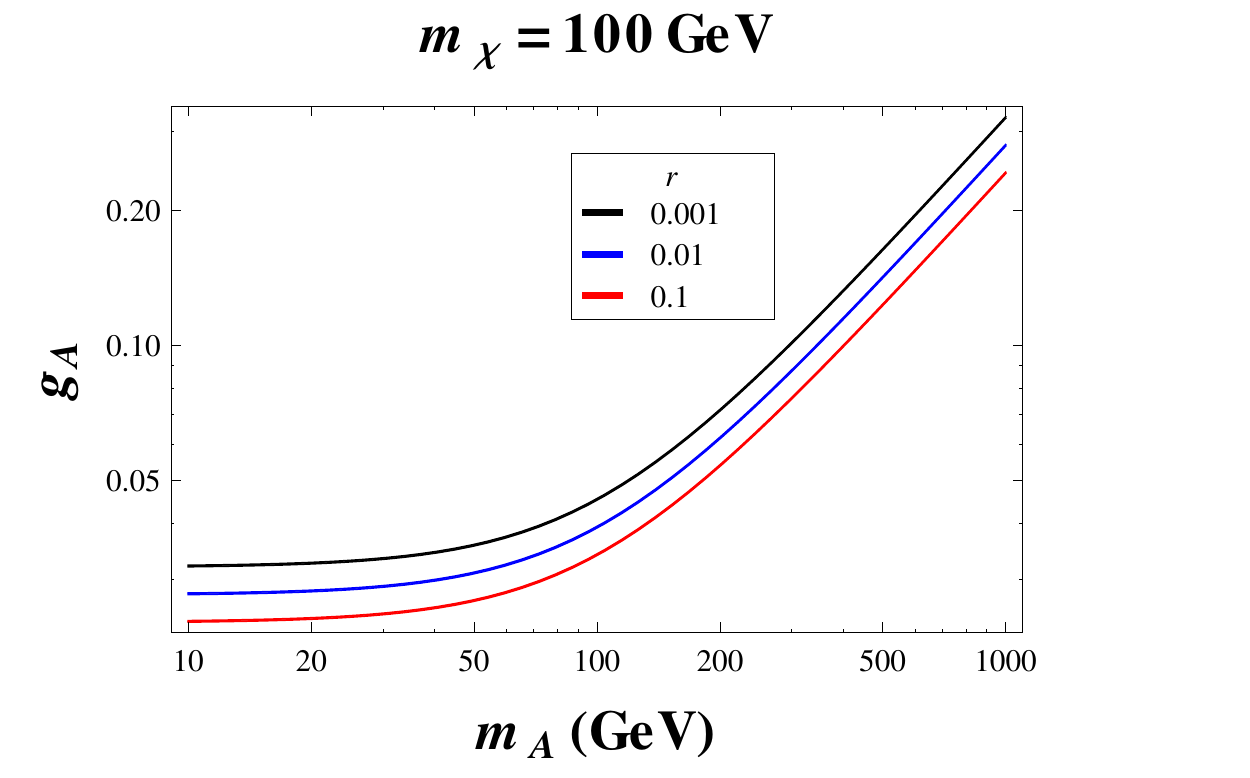}
    \label{fig9a}
}
\subfigure[$g_{A}$ vs $m_{A}$ for t/u-channel with $m_{\chi}=1{\rm TeV}$.]{
    \includegraphics[width=14.5cm,height=8cm]{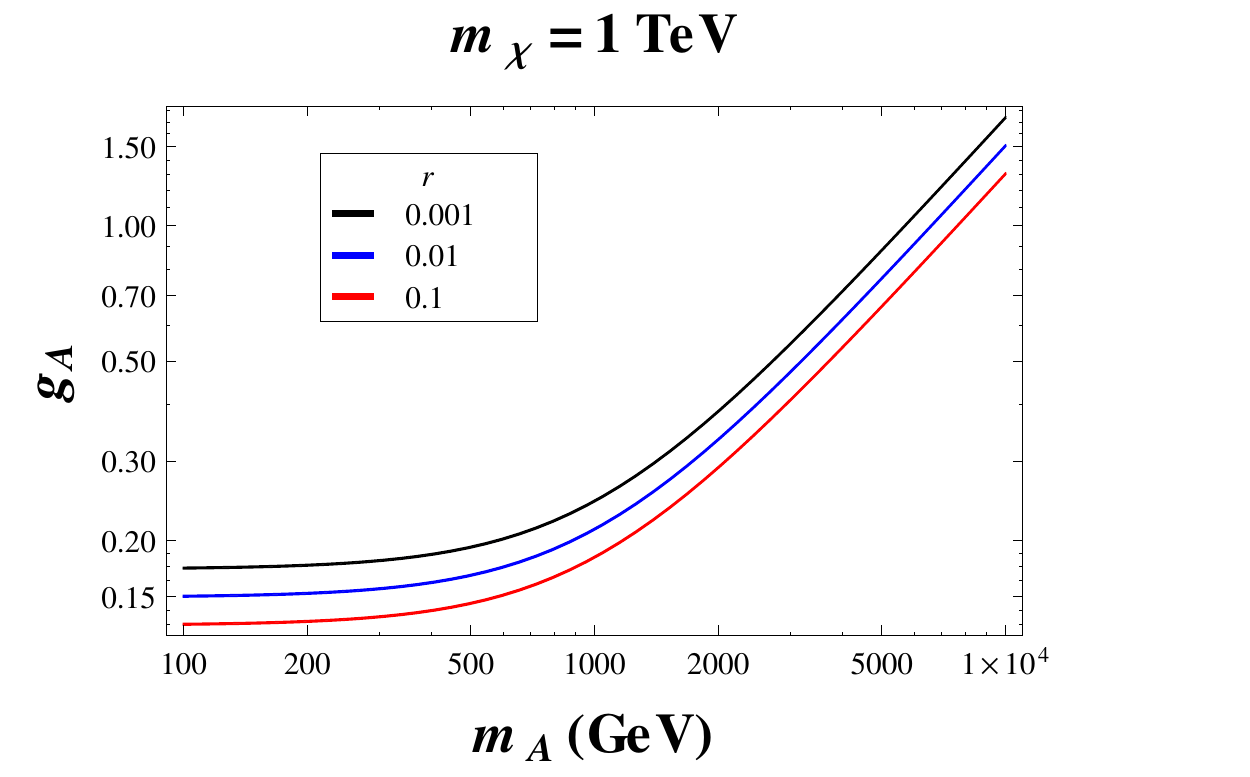}
    \label{fig9b}
}
\caption[Optional caption for list of figures]{In the above figure we have shown the allowed 
region of $g_A$ with respect to the mass of the mediator $m_A$, for the t/u - channel processes.} 
\label{fig9}
\end{figure*}
\subsubsection{\bf t/u- channel analysis}
\label{a4v2}
Next we consider the following localized interactions for a Dirac dark matter particle, $\chi$, and a spin-1 mediator, $V_\mu$,
within the framework of effective field theory written in RSII membrane as:
\begin{align}
\mathcal{L}_{membrane} &\supset \overline{\chi}\gamma^\mu(g_{\chi_V} + g_{\chi_a}\gamma_5)f V_\mu 
+ \overline{f}\gamma^\mu(g_{f_V} + g_{f_a})\chi V^\dag_\mu 
\end{align}

Now, taking all the couplings to be the same the cross-section becomes
\begin{align}
\sigma &= \frac{1}{8\pi (s-4m^2_\chi)}\int_{t_-}^{t_+}|{\cal M}|^2 dt
\end{align}
where the matrix element for the S-matrix and the symbol $t_{\pm}$ is given by:
\begin{align}
 |\mathcal{M}|^2 &= \frac{g_A^4 n_c}{4\pi s \left(t - m_A^2\right)^2}\left((s-2m^2_\chi)(s-2m^2_f) + (m^2_\chi + m^2_f -t)^2\right),\\
 t_{\pm} &= (m^2_\chi + m^2_f-\frac{s}{2}) \pm \frac{\sqrt{(s-4m^2_\chi)(s-4m^2_f)}}{2}. 
\end{align}
where $n_c=3$ for quarks and 1 for leptons, $g_A$ and $m_A$ are the respective coupling and the mass of the mediator. 
Now taking the following approximation: \be s=4m^2_\chi/\left(1-v^2/4\right)\ee we finally get the following simplified expression for the product of annihilation cross-section and velocity as:
\begin{align} \sigma v = a + \mathcal{O}(v^2) \end{align} where the factor $a$ is given by:
\begin{align}
a &\approx \frac{n_c  g^4 \sqrt{2m^2_\chi -m^2_f}}{8 \pi \left(m^2 - m^2_f + m^2_\chi \right)^2}
\left[1-\frac{m^2_f}{m^2_\chi}\right]^{1/2}.
\label{eq:dfs0}
\end{align}
 The dark matter relic abundance in this case is given as:
\begin{align}
 \Omega_{DM} h^2 &= \frac{1.07\times 10^9}{J(x_f)g^{1/2}_*M_p}\end{align}
 where $J(x_f)$ for spin-1 mediator $t/u$ channel is given by:
 \begin{align}J(x_f) &= \int_{x_f}^{\infty}\frac{n_c  g^4 \sqrt{2m^2_\chi -m^2_f}}{8 \pi \left(m^2 - m^2_f + m^2_\chi \right)^2x^2f_{membrane}(x)}
\left[1-\frac{m^2_f}{m^2_\chi}\right]^{1/2}dx
\end{align}
where the function $f_{membrane}(x)$ is the characteristic parameter for RS single braneworld and this can be expressed in terms of tensor-to-scalar ratio ($r$) which is given in Eq.\eqref{eq:ev1xxxxxc}.
Now for the GR limiting case of the $f_{membrane}(x) \rightarrow 1$ and then the relic abundance will only depend on the mass of the Dark Matter ($m_\chi$), $g_A$ the 
coupling with the spin-0 mediator and the mass of the mediator ($m_A$). 
\begin{align}
 \Omega_{DM} h^2 &= \frac{1.07\times 10^9x_f}{g^{1/2}_*M_p}\left(\frac{n_c  g^4 \sqrt{2m^2_\chi -m^2_f}}{8 \pi \left(m^2 - m^2_f + m^2_\chi \right)^2}
\left[1-\frac{m^2_f}{m^2_\chi}\right]^{1/2}\right)^{-1}.
\end{align}
In order to constrain the coupling ($g_A$) and mass ($m_A$) we take the present data of the relic abundance ($\Omega_{DM}h^2 = 0.1199\pm0.0027$\cite{Ade:2015xua}) and constrain the function 
$J(x_f)$, which in turn constrain the coupling $g_A$ and $m_A$ for a particular tensor-to-scalar ratio ($r$). We have not shown the GR limiting case as it has been extensively been explored in 
\cite{Berlin:2014tja}.
In fig.~(\ref{fig9a}) and fig.~(\ref{fig9b}), we have depicted the behaviour of the effective coupling of spin-1 mediator $g_{A}$ with the varying mass the spin-1 mediator $m_{A}$ 
for t/u-channel process with three distinct value of the tensor-to-scalar ratio $r=0.001$, $r=0.01$and $r=0.1$ 
respectively in RSII membrane. We also consider three different values of the dark matter mass $m_{\chi}=100 {\rm GeV}$ and $m_{\chi}=1{\rm TeV}$
for the t/u-channel analysis. From fig.~(\ref{fig9a}) and fig.~(\ref{fig9b}), it is clearly observed
that the behaviour of the effective coupling of spin-1 mediator $g_{A}$ with the varying mass the spin-1 mediator $m_{A}$ are similar for all of the three cases, where the coupling decreases with mediator mass.

\subsection{\bf Majorana dark Matter: spin-1 mediator}

\subsubsection{\bf s-channel analysis}
\label{a5v1}
\begin{figure*}[!htb]
\centering
\subfigure[$\chi\bar{\chi}\rightarrow f\bar{f}$~process for s-channel.]{
    \includegraphics[width=6.5cm,height=4.2cm] {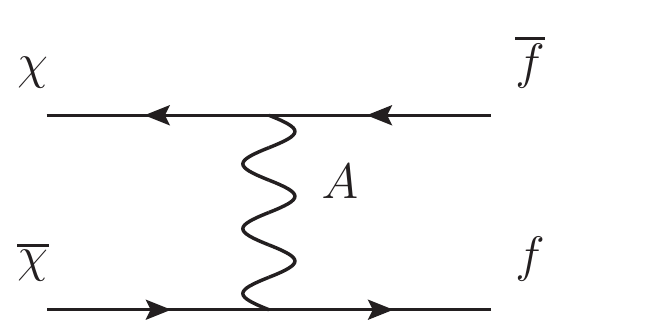}
    \label{fig10a}
}
\subfigure[$\chi{\chi}^c\rightarrow f\bar{f}$~process for t/u-channel.]{
    \includegraphics[width=6.5cm,height=4.2cm] {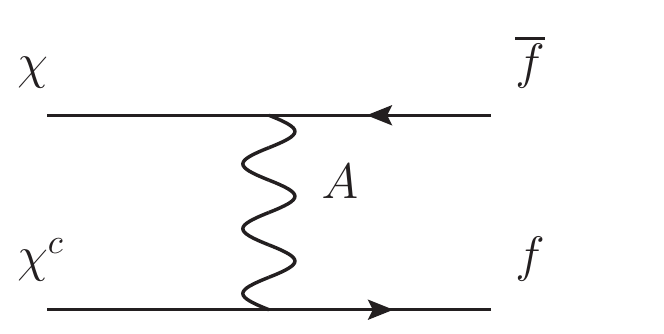}
    \label{fig10b}
}
\caption[Optional caption for list of figures]{Feynman diagrammatic representation of s-channel and t/u channel processes for Dirac dark matter with spin-1 mediator.} 
\label{fig13}
\end{figure*}

Further we consider the following Lagrangian for a Majorana dark matter particle, $\chi$, that interacts with the SM via a spin-1 
mediator, $V_\mu$, within the framework of effective field theory written in RSII membrane as:
\begin{align}
\mathcal{L}_{membrane} &\supset \left[\frac{1}{2}g_{\chi_a}\overline{\chi}\gamma^\mu \gamma^5 \chi 
+ \overline{f}(g_{f_V} + g_{f_a}\gamma^5)f\right]V_\mu 
\end{align}
In fig.~(\ref{fig10a}) and fig.~(\ref{fig10b}), we have explicitly shown the
Feynman diagrammatic representation of possible s-channel and t/u-channel processes for dirac dark matter with spin-1 mediator respectively.
Now taking equal coupling approximation for the couplings the cross-section from the above Lagrangian can be computed as:
\begin{align}
\sigma &= \frac{1}{8\pi (s-4m^2_\chi)}\int_{t_-}^{t_+}|\mathcal{M}|^2 dt
\end{align}
where the matrix element for the $S$ matrix and the symbol $t_{\pm}$ is defined as:
\begin{align}
 |\mathcal{M}|^2 &= \frac{2n_cg^4}{3\pi s \left((s-m^2)^2 + m^2\Gamma^2 \right)}\left\{4m^2_\chi m^2_f\left(7-6\frac{s}{m^2}+\frac{3s^2}{m^4}\right)-4m^2_\chi s \right.\nonumber \\
    &\left.~~~~~~~~~~~ + s(s-4m^2_f) + (s+2m^2_f)(s-4m^2_\chi)\right\},\\
    t_{\pm} &= (m^2_\chi + m^2_f-\frac{s}{2}) \pm \frac{\sqrt{(s-4m^2_\chi)(s-4m^2_f)}}{2}.
\end{align}
where $n_c=3$ for quarks and 1 for leptons, $g_A$ and $m_A$ are the respective coupling and the mass of the mediator. 
\begin{figure*}[!ht]
\centering
\subfigure[$g_{A}$ vs $m_{A}$ for s-channel with $m_{\chi}=100{\rm GeV}$.]{
    \includegraphics[width=14.5cm,height=8cm] {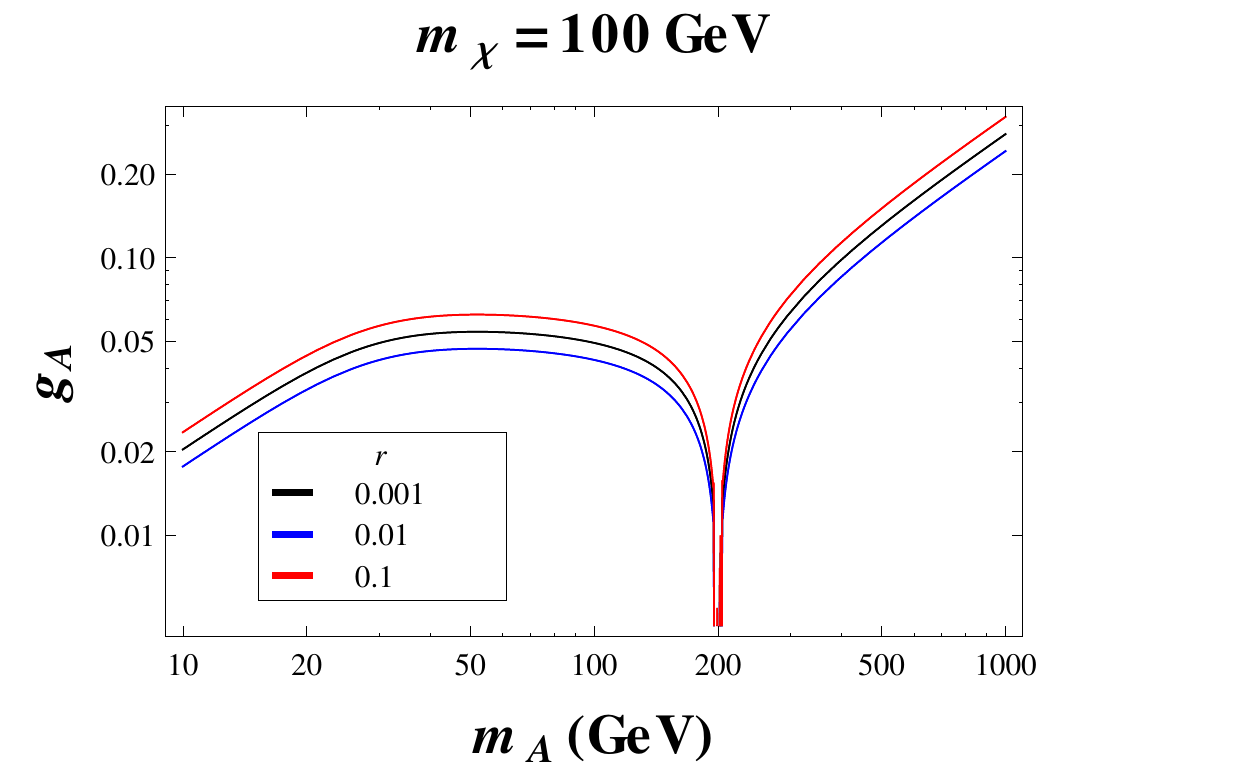}
    \label{fig11a}
}
\subfigure[$g_{A}$ vs $m_{A}$ for s-channel with $m_{\chi}=1{\rm TeV}$.]{
    \includegraphics[width=14.5cm,height=8cm] {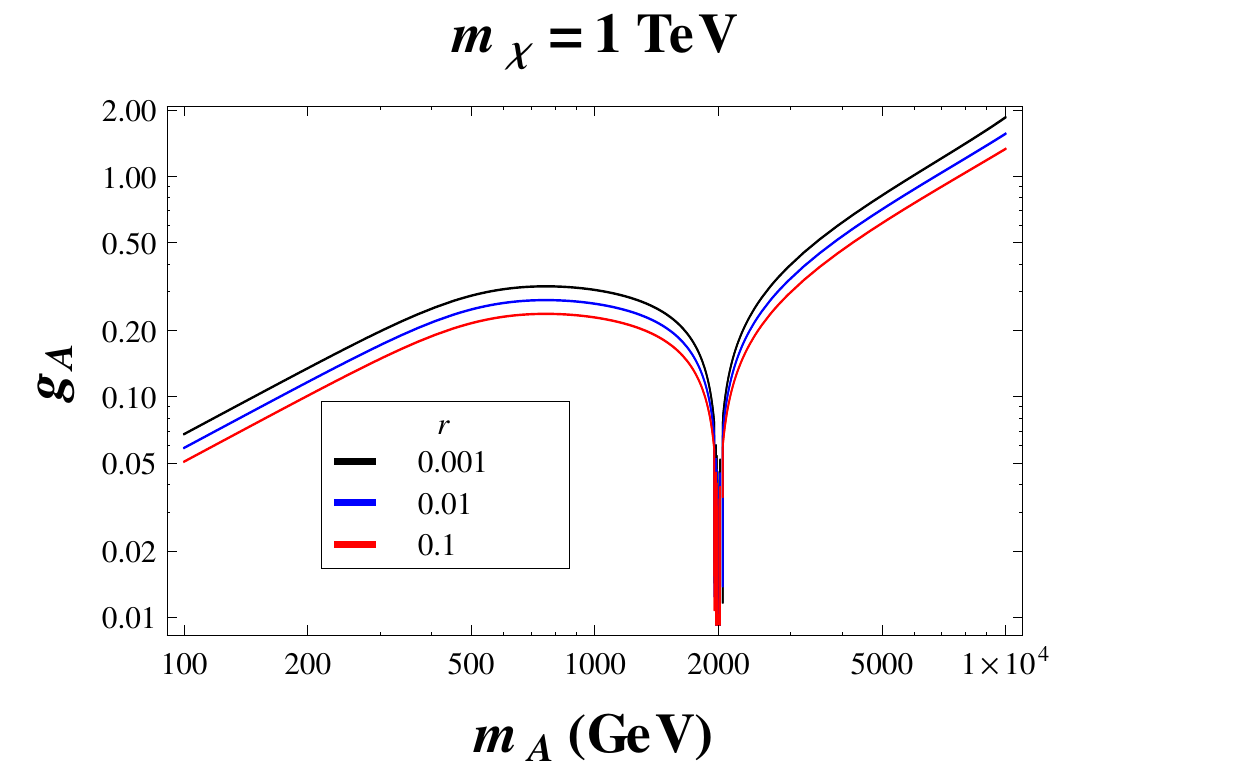}
    \label{fig11b}
}
\caption[Optional caption for list of figures]{In the above figure we have shown the allowed 
region of $g_A$ with respect to the mass of the mediator $m_A$, for the
s-channel processes.} 
\label{fig11}
\end{figure*}
Now taking the following approximation: \be s=4m^2_\chi/\left(1+v^2/4\right)\ee we finally get the following simplified expression for the product of annihilation cross-section and velocity as:
\begin{align} \sigma v = a + \mathcal{O}(v^2) \end{align} where the factor $a$ is given by:
\begin{align}
    a &= n_cg^4 \frac{m^2_f}{2\pi m^4}\sqrt{1-\frac{m^2_f}{m^2_\chi}}\frac{1-8\frac{m^2_\chi}{m^2}  
      + 16 \frac{m^4_\chi}{m^4} }{1-8\frac{m^2_{\chi}}{m^2}+16 \frac{m^4_\chi}{m^4} + \frac{\Gamma^2}{m^2}}, \end{align}
  and the mediator's width to SM fermions is given by:
    \begin{align}
    \Gamma &=\sum_{f}\Gamma(A\rightarrow f\bar{f})=\frac{g^2n_cm}{6\pi}\left[1-4\frac{m^2_f}{m^2}\right]^{1/2}\left[1-2\frac{m^2_f}{m^2}
    \right].
\end{align}
 The relic abundance is given as:
\begin{align}
 \Omega_{DM} h^2 &= \frac{1.07\times 10^9}{J(x_f)g^{1/2}_*M_p}\end{align}
 where $J(x_f)$ for spin-1 mediator s-channel process is given by:
 \begin{align}J(x_f) &= \int_{x_f}^{\infty}n_cg^4 \frac{m^2_f}{2\pi m^4}\sqrt{1-\frac{m^2_f}{m^2_\chi}}\frac{1-8\frac{m^2_\chi}{m^2}  
      + 16 \frac{m^4_\chi}{m^4} }{\left(1-8\frac{m^2_{\chi}}{m^2}+16 \frac{m^4_\chi}{m^4} + \frac{\Gamma^2}{m^2}\right)x^2f_{membrane}(x)}
\left[1-\frac{m^2_f}{m^2_\chi}\right]^{1/2}dx.
\end{align}
where the function $f_{membrane}(x)$ is the characteristic parameter for RS single braneworld and this can be expressed in terms of tensor-to-scalar ratio ($r$) which is given in Eq.\eqref{eq:ev1xxxxxc}. 
Now for the GR limiting case of the $f_{membrane}(x) \rightarrow 1$ and then the relic abundance will only depend on the mass of the Dark Matter ($m_\chi$), $g_A$ the 
coupling with the spin-0 mediator and the mass of the mediator ($m_A$). 
\begin{align}
 \Omega_{DM} h^2 &= \frac{1.07\times 10^9x_f}{g^{1/2}_*M_p}\left(n_cg^4 \frac{m^2_f}{2\pi m^4}\sqrt{1-\frac{m^2_f}{m^2_\chi}}\frac{1-8\frac{m^2_\chi}{m^2}  
      + 16 \frac{m^4_\chi}{m^4} }{1-8\frac{m^2_{\chi}}{m^2}+16 \frac{m^4_\chi}{m^4} + \frac{\Gamma^2}{m^2}}\right)^{-1}.
\end{align}
In order to constrain the coupling ($g_A$) and mass ($m_A$) we take the present data of the relic abundance ($\Omega_{DM}h^2 = 0.1199\pm0.0027$\cite{Ade:2015xua}) and constrain the function 
$J(x_f)$, which in turn constrain the coupling $g_A$ and $m_A$ for a particular tensor-to-scalar ratio ($r$). We have not shown the GR limiting case as it has been extensively been explored in 
\cite{Berlin:2014tja}.
In fig.~(\ref{fig11a}) and fig.~(\ref{fig11b}), we have depicted the behaviour of the effective coupling of spin-1 mediator $g_{A}$ with the varying mass the spin-1 mediator $m_{A}$ 
for s-channel process with three distinct value of the tensor-to-scalar ratio $r=0.001$, $r=0.01$ and $r=0.1$ 
respectively in RSII membrane. We also consider two different values of the dark matter mass $m_{\chi}=100 \;{\rm GeV}$ and $m_{\chi}=1\;{\rm TeV}$ for the s-channel analysis. From fig.~(\ref{fig11a}) and fig.~(\ref{fig11b}), it is clearly observed
that the behaviour of the effective coupling of spin-1 mediator $g_{A}$ with the varying mass the spin-1 mediator $m_{A}$ are similar for both of the cases, where the coupling increases with mediator mass.
\subsubsection{\bf t/u- channel analysis}
\label{a5v2}
Next we consider the following localized interactions for a Majorana dark matter particle, $\chi$, and a spin-1 mediator, $V_\mu$,
within the framework of effective field theory written in RSII membrane as:
\begin{align}
\mathcal{L}_{membrane} &\supset \overline{\chi}\gamma^\mu(g_{\chi_V} + g_{\chi_a}\gamma_5)f V_\mu 
+ \overline{f}\gamma^\mu(g_{f_V} + g_{f_a})\chi V^\dag_\mu 
\end{align}
\begin{figure*}[!ht]
\centering
\subfigure[$g_{A}$ vs $m_{A}$ for t/u-channel with $m_{\chi}=100{\rm GeV}$.]{
    \includegraphics[width=14.5cm,height=8cm] {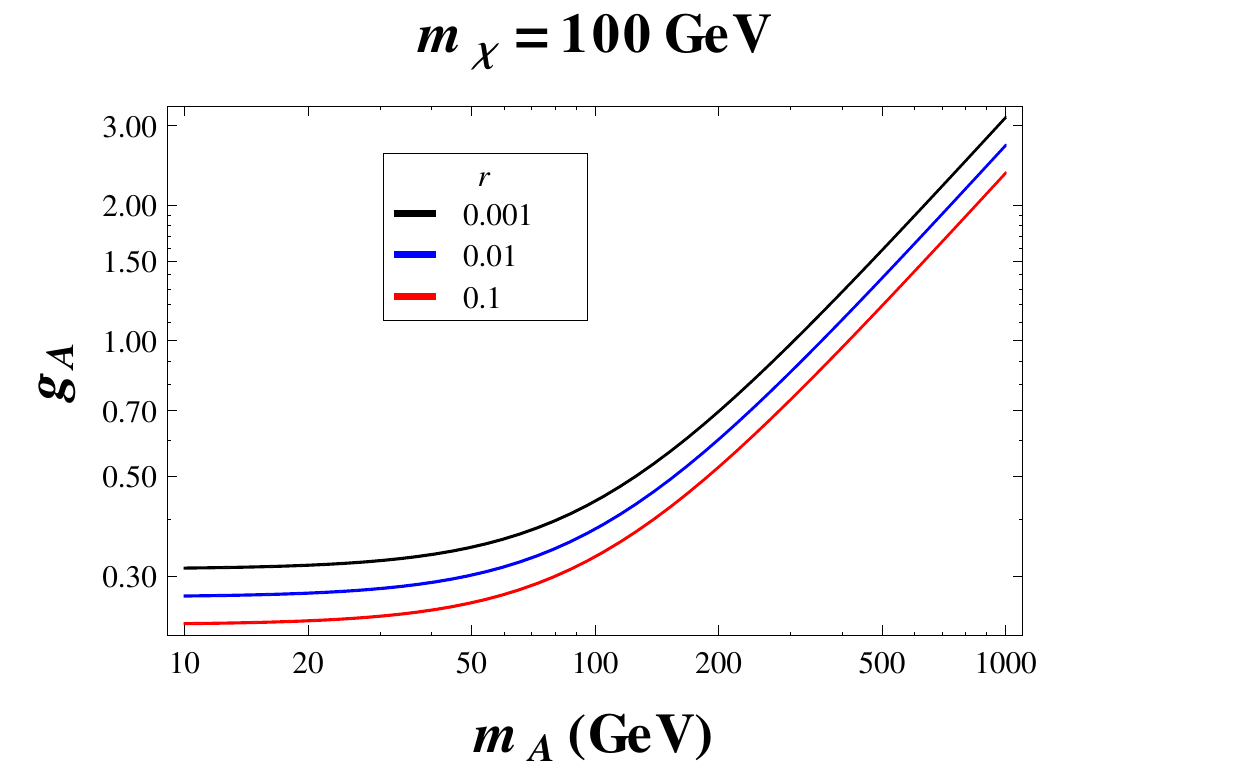}
    \label{fig12a}
}
\subfigure[$g_{A}$ vs $m_{A}$ for t/u-channel with $m_{\chi}=1{\rm TeV}$.]{
    \includegraphics[width=14.5cm,height=8cm] {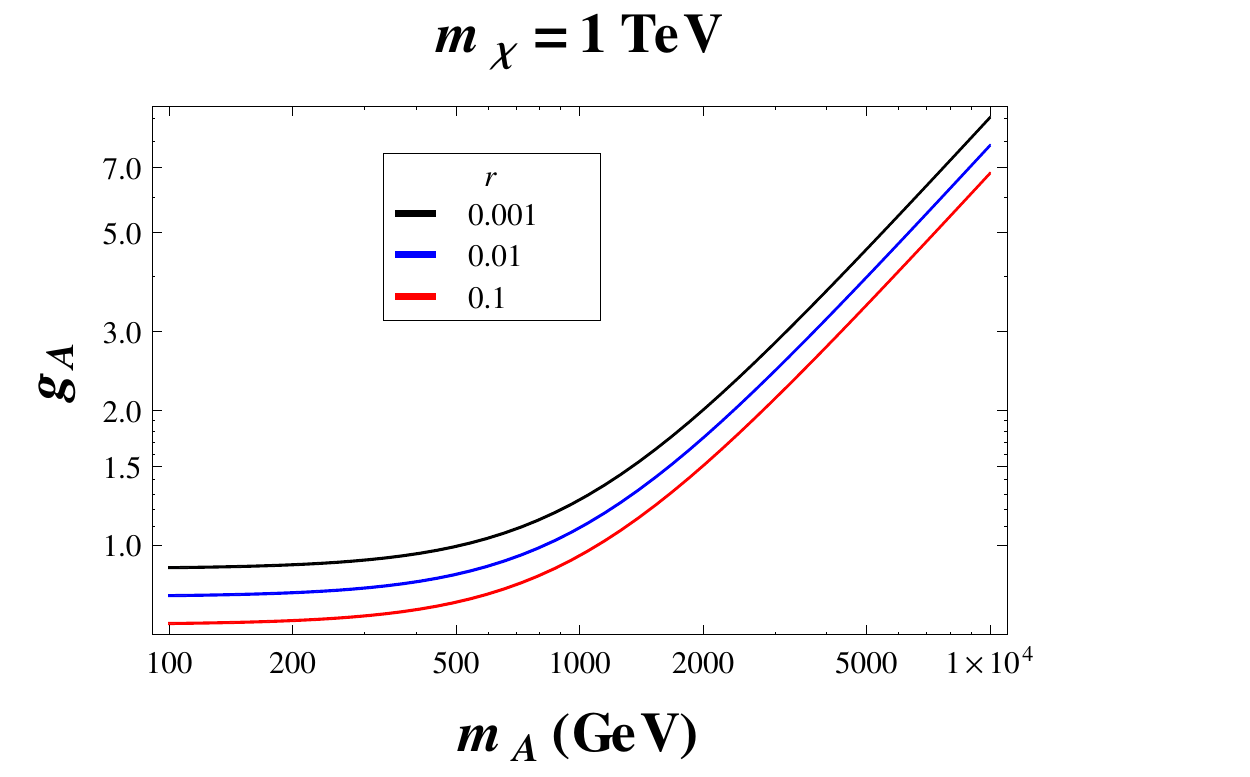}
    \label{fig12b}
}
\caption[Optional caption for list of figures]{In the above figure we have shown the allowed 
region of $g_A$ with respect to the mass of the mediator $m_A$, for the
t/u-channel processes.} 
\label{fig12}
\end{figure*}
Now taking equal coupling approximation for the couplings the cross-section from the above Lagrangian can be computed as:
\begin{align}
\sigma &= \frac{1}{8\pi (s-4m^2_\chi)}\int_{t_-}^{t_+}|\mathcal{M}|^2 dt
\end{align}
where the matrix element for the $S$ matrix and the symbol $t_{\pm}$ is defined as:
\begin{align}
 |\mathcal{M}|^2 &= 2g^2_A\left(\frac{s^2 + u^2 m^4_f m^2_\chi -2m^2_\chi(s+u) + m^2_f(6m^2_\chi-2(s+u)) }{(t-m^2_A)^2} \right. \nonumber \\
      &+\left. \frac{s^2 + t^2 m^4_f m^2_\chi -2m^2_\chi(s+t) + m^2_f(6m^2_\chi-2(s+t)) }{(u-m^2_A)^2} \right).
\end{align}
where $n_c=3$ for quarks and 1 for leptons, $g_A$ and $m_A$ are the respective coupling and the mass of the mediator. 
Now taking the following approximation: \be s=4m^2_\chi/\left(1+v^2/4\right)\ee we finally get the following simplified expression for the product of annihilation cross-section and velocity as:
\begin{align} \sigma v = a + \mathcal{O}(v^2) \end{align} where the factor $a$ is given by:
\begin{align}
    a &\approx \frac{n_c g^4_A\sqrt{1-m^2_f/m^2_\chi}}{8\pi m^4_A (m^2_A - m^2_f +  m^2_\chi)^2} 
\left[8m^6_f -8m^4_f\left(2m^2_A + m^2_\chi \right)  + 4 m^2_f \left(2m^2_A + m^2_\chi\right)^2\right]. 
\end{align}
 The relic abundance is given as:
\begin{align}
 \Omega_{DM} h^2 &= \frac{1.07\times 10^9}{J(x_f)g^{1/2}_*M_p}\end{align}
 where $J(x_f)$ for spin-1 mediator s-channel process is given by:
 \begin{align}J(x_f) &= \int_{x_f}^{\infty}\frac{n_c g^4_A\sqrt{1-m^2_f/m^2_\chi}}{8\pi m^4_A (m^2_A - m^2_f +  m^2_\chi)^2} 
\frac{\left[8m^6_f -8m^4_f\left(2m^2_A + m^2_\chi \right)  + 4 m^2_f \left(2m^2_A + m^2_\chi\right)^2\right]}{x^2f_{membrane}(x)} dx.
\end{align}
where the function $f_{membrane}(x)$ is the characteristic parameter for RS single braneworld and this can be expressed in terms of tensor-to-scalar ratio ($r$) which is given in Eq.\eqref{eq:ev1xxxxxc}. 
Now for the GR limiting case of the $f_{membrane}(x) \rightarrow 1$ and then the relic abundance will only depend on the mass of the Dark Matter ($m_\chi$), $g_A$ the 
coupling with the spin-0 mediator and the mass of the mediator ($m_A$). 
\begin{align}
 \Omega_{DM} h^2 &= \frac{1.07\times 10^9x_f}{g^{1/2}_*M_p}\left(\frac{n_c g^4_A\sqrt{1-m^2_f/m^2_\chi}}{8\pi m^4_A (m^2_A - m^2_f +  m^2_\chi)^2} \right. \nonumber \\
     & \left.\left[8m^6_f -8m^4_f\left(2m^2_A + m^2_\chi \right)  + 4 m^2_f \left(2m^2_A + m^2_\chi\right)^2\right]\right)^{-1}.
\end{align}
In order to constrain the coupling ($g_A$) and mass ($m_A$) we take the present data of the relic abundance ($\Omega_{DM}h^2 = 0.1199\pm0.0027$\cite{Ade:2015xua}) and constrain the function 
$J(x_f)$, which in turn constrain the coupling $g_A$ and $m_A$ for a particular tensor-to-scalar ratio ($r$). We have not shown the GR limiting case as it has been extensively been explored in 
\cite{Berlin:2014tja}.
In fig.~(\ref{fig12a}) and fig.~(\ref{fig12b}), we have depicted the behaviour of the effective coupling of spin-0 mediator $g_{A}$ with the varying mass the spin-0 mediator $m_{A}$ 
for t/u-channel process with three distinct value of the tensor-to-scalar ratio $r=0.001$, $r=0.01$and $r=0.1$ 
respectively in RSII membrane. We also consider three different values of the dark matter mass $m_{\chi}=100 {\rm GeV}$ and $m_{\chi}=1{\rm TeV}$
for the t/u-channel analysis. From fig.~(\ref{fig12a}) and fig.~(\ref{fig12b}), it is clearly observed
that the behaviour of the effective coupling of spin-0 mediator $g_{A}$ with the varying mass the spin-0 mediator $m_{A}$ are similar for all of the three cases, where the coupling increases with mediator mass.

 \subsection{\bf Complex scalar dark matter: spin-0 mediator}
 \label{a6}
\begin{figure*}[!htb]
\centering
\subfigure[$\chi{\chi}^{*}\rightarrow f\bar{f}$~process for s-channel.]{
    \includegraphics[width=6.5cm,height=4.2cm] {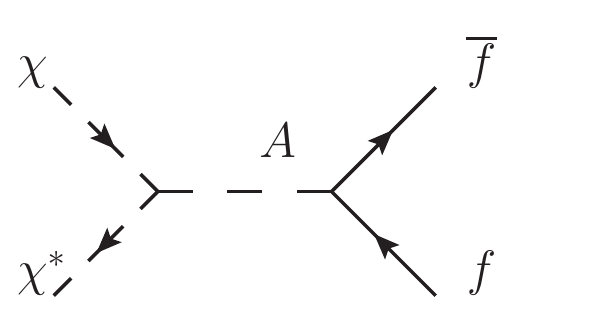}
    \label{fig13a}
}
\subfigure[$\chi{\chi}\rightarrow f\bar{f}$~process for s-channel.]{
    \includegraphics[width=6.5cm,height=4.2cm] {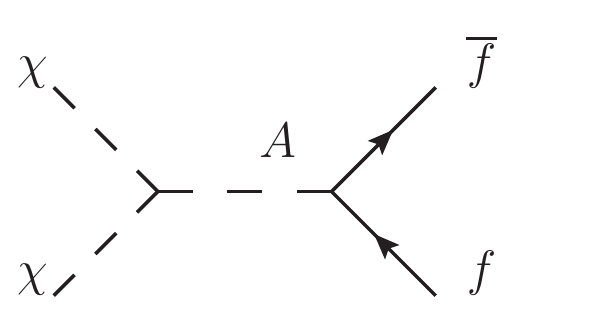}
    \label{fig13b}
}
\caption[Optional caption for list of figures]{Feynman diagrammatic representation of s-channel processes for complex and real scalar  dark matter with spin-0 mediator.} 
\label{fig13}
\end{figure*}
\begin{figure*}[!ht]
\centering
\subfigure[$g_{A}$ vs $m_{A}$ for s-channel with $m_{\chi}=100{\rm GeV}$.]{
    \includegraphics[width=14.5cm,height=8cm] {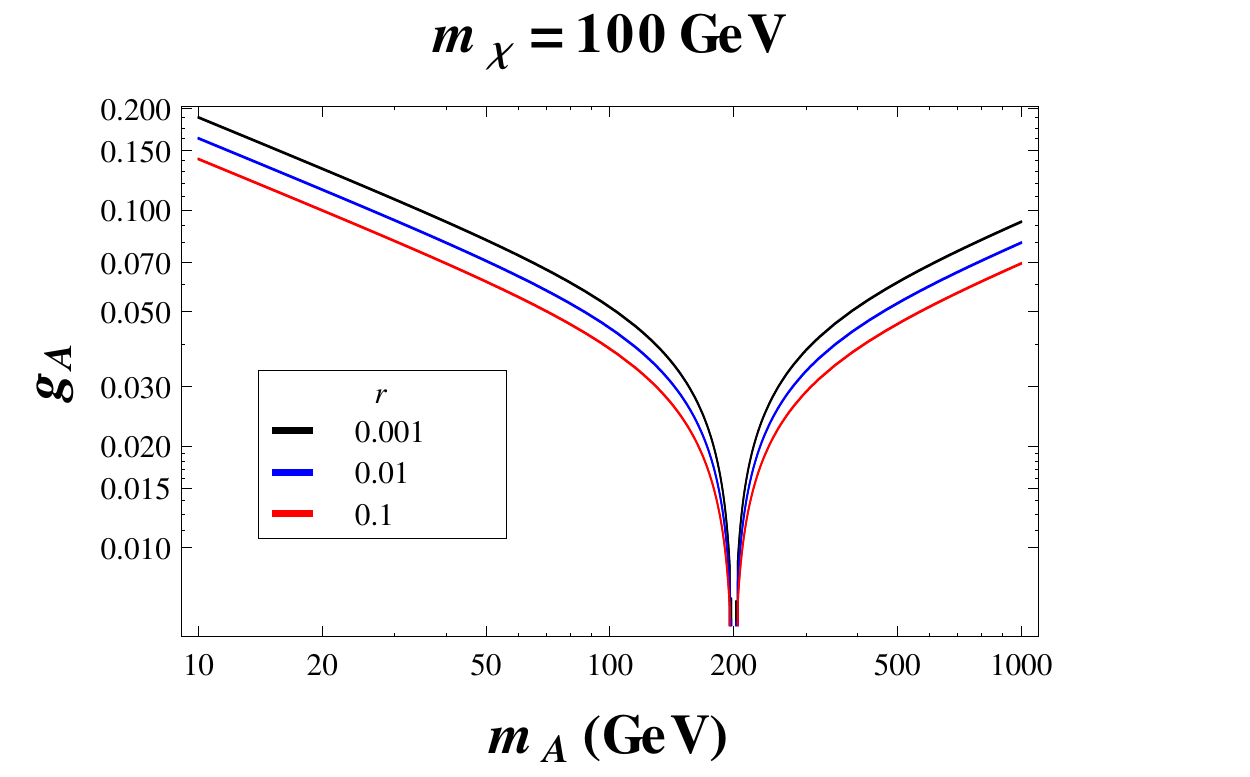}
    \label{fig14a}
}
\subfigure[$g_{A}$ vs $m_{A}$ for s-channel with $m_{\chi}=1{\rm TeV}$.]{
    \includegraphics[width=14.5cm,height=8cm] {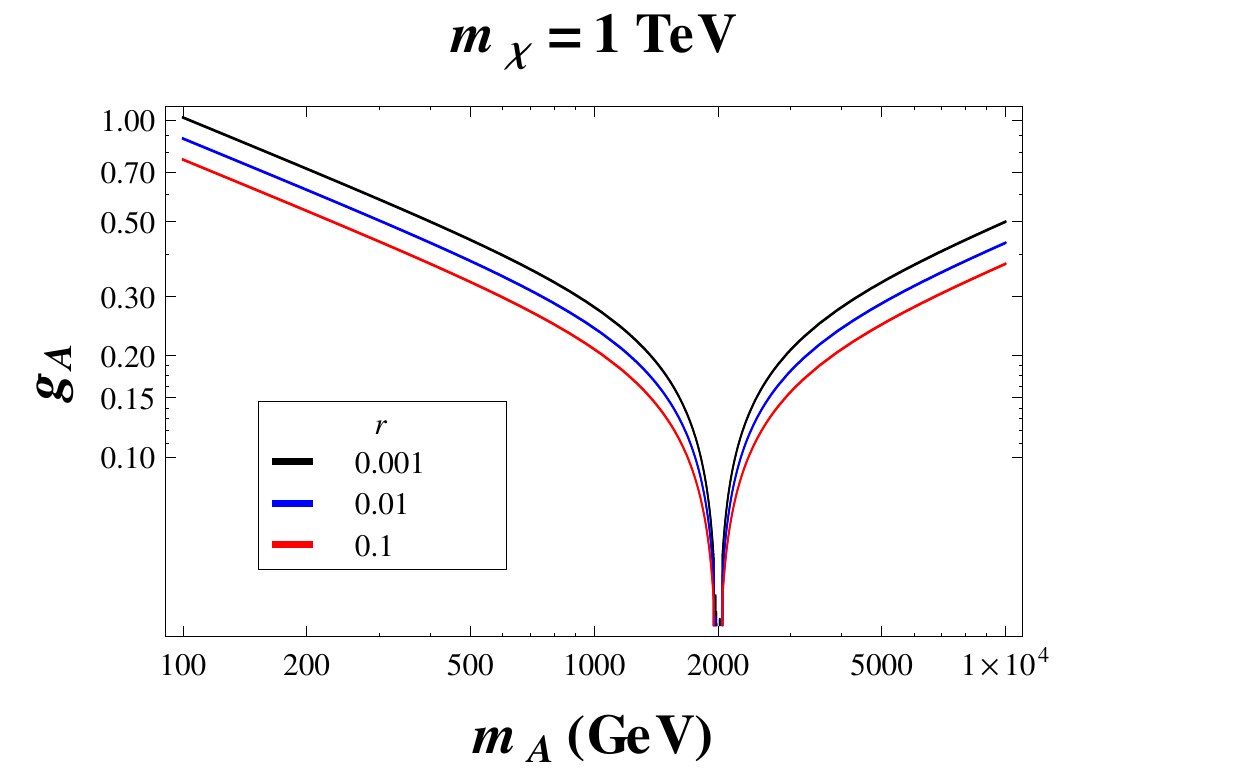}
    \label{fig14b}
}
\caption[Optional caption for list of figures]{In the above figure we have shown the allowed 
region of $g_A$ with respect to the mass of the mediator $m_A$, the upper panel is for the 
s-channel processes.} 
\label{fig14}
\end{figure*}
 Here we consider the following Lagrangian for for a complex scalar dark matter particle, $\phi$, and a spin-0
  mediator, $A$, within the framework of effective field theory written in RSII membrane as:
 \begin{align}
 \mathcal{L}_{membrane} &\supset \left[\mu_\phi |\phi|^2 + \overline{f}(\lambda_{f_s} 
 + \lambda_{f_p}i\gamma^5)f\right]A.
 \end{align}
 In fig.~(\ref{fig13a}), we have explicitly shown the
Feynman diagrammatic representation of possible s-channel process for complex scalar dark matter with spin-0 mediator.
Now, taking all the couplings to be the same the cross-section becomes: 
\begin{align}
\sigma &= \frac{1}{8\pi (s-4m^2_\chi)}\int_{t_-}^{t_+}|\mathcal{M}|^2 dt 
\end{align}
where the matrix element for the S-matrix and the symbol $t_{\pm}$ is given by:
\begin{align} |\mathcal{M}|^2 &= \frac{n_cg^4}{\pi s \left((s-m^2)^2 + m^2\Gamma^2 \right)}
\left\{s-2m^2_f\right\},\\
t_{\pm} &= (m^2_\chi + m^2_f-\frac{s}{2}) \pm \frac{\sqrt{(s-4m^2_\chi)(s-4m^2_f)}}{2}.\end{align}
where $n_c=3$ for quarks and 1 for leptons, $g_A$ and $m_A$ are the respective coupling and the mass of the mediator. 
Now taking the following approximation: \be s=4m^2_\chi/\left(1+v^2/4\right)\ee we finally get the following simplified expression for the product of annihilation cross-section and velocity as:
\begin{align} \sigma v = a + \mathcal{O}(v^2) \end{align} where the factor $a$ is given by:
\begin{align}
    a &= n_cg^4 \frac{m^2}{2\pi }\sqrt{1-\frac{m^2_f}{m^2_\chi}}\frac{1-\frac{m^2f}{2m^2_\chi}}{ (m^2-4m^2_\chi)^2 +m^2 \Gamma^2} \end{align}
   and the mediator's width to SM fermions is given by:
    \begin{align} 
 \Gamma &= \frac{g^4n_cm}{4\pi}\left[1-4\frac{m^2_f}{m^2}\right]^{1/2}\left[1-2\frac{m^2_f}{m^2}\right].
\end{align}
The dark matter relic abundance in this case is given as:
\begin{align}
 \Omega_{DM} h^2 &= \frac{1.07\times 10^9}{J(x_f)g^{1/2}_*M_p}\end{align}
 where $J(x_f)$ for complex dark matter with spin-0 mediator is given by:
 \begin{align}
 J(x_f) &= \int_{x_f}^{\infty}n_cg^4 \frac{m^2}{2\pi }\sqrt{1-\frac{m^2_f}{m^2_\chi}}\frac{1-\frac{m^2f}{2m^2_\chi}}{ \left((m^2-4m^2_\chi)^2 +m^2 \Gamma^2\right)x^2f_{membrane}(x)}
dx.
\end{align}
where the function $f_{membrane}(x)$ is the characteristic parameter for RS single braneworld and this can be expressed in terms of tensor-to-scalar ratio ($r$) which is given in Eq.\eqref{eq:ev1xxxxxc}. 
Now for the GR limiting case of the $f_{membrane}(x) \rightarrow 1$ and then the relic abundance will only depend on the mass of the Dark Matter ($m_\chi$), $g_A$ the 
coupling with the spin-0 mediator and the mass of the mediator ($m_A$). 
\begin{align}
 \Omega_{DM} h^2 &= \frac{1.07\times 10^9x_f}{g^{1/2}_*M_p}\left(n_cg^4 \frac{m^2}{2\pi }\sqrt{1-\frac{m^2_f}{m^2_\chi}}\frac{1-\frac{m^2f}{2m^2_\chi}}{ (m^2-4m^2_\chi)^2 +m^2 \Gamma^2}
     \right)^{-1}.
\end{align}
In order to constrain the coupling ($g_A$) and mass ($m_A$) we take the present data of the relic abundance ($\Omega_{DM}h^2 = 0.1199\pm0.0027$\cite{Ade:2015xua}) and constrain the function 
$J(x_F)$, which in turn constrain the coupling $g_A$ and $m_A$ for a particular tensor-to-scalar ratio ($r$). We have not shown the GR limiting case as it has been extensively been explored in 
\cite{Berlin:2014tja}.
In fig.~(\ref{fig14a}) and fig.~(\ref{fig14b}), we have depicted the behaviour of the effective coupling of spin-0 mediator $g_{A}$ with the varying mass the spin-0 mediator $m_{A}$ 
for s-channel process with three distinct value of the tensor-to-scalar ratio $r=0.001$, $r=0.01$and $r=0.1$ 
respectively in RSII membrane. We also consider two different values of the dark matter mass $m_{\chi}=100 \; {\rm GeV}$ and $m_{\chi}=1\;{\rm TeV}$ for the s-channel analysis. From fig.~(\ref{fig14a}) and fig.~(\ref{fig14b}), it is clearly observed
that the behaviour of the effective coupling of spin-0 mediator $g_{A}$ with the varying mass the spin-0 mediator $m_{A}$ are similar for all of the three cases and also 
sensitive in the vicinity of $m_{A}=2\times 10^{2}\;{\rm GeV}$ and $m_{A}=2\times 10^{3}\;{\rm GeV}$ respectively as it has a resonance 
(i.e $2m_X = m_A$).
Most importantly, in both the sides of $m_{A}=2\times 10^{2}{\rm GeV}$ and $m_{A}=2\times 10^{3}{\rm GeV}$
the coupling of spin-0 mediator $g_{A}$ behave in completely opposite manner. 

Now, for real scalar dark matter with spin-0 mediator we can also write down the the Lagrangian for a real scalar DM particle, $\phi$, that interacts with the SM via a 
 spin-0 meidator, A, within the framework of effective field theory written in RSII membrane as:
 \begin{align}
\mathcal{L}_{membrane} &\supset \left[\frac{1}{2}\mu_\phi\phi^2 + \overline{f}(\lambda_{f_S} + \lambda_{f_p}i\gamma^5)f \right]A.
\end{align}
In fig.~(\ref{fig13b}), we have explicitly shown the
Feynman diagrammatic representation of possible s-channel process for real scalar dark matter with spin-0 mediator.
Hence after taking all the coupling same and further 
taking the following approximation: \be s=4m^2_\chi/\left(1+v^2/4\right)\ee we finally get the following simplified expression for the product of annihilation cross-section and velocity as:
\begin{align} \sigma v = a + \mathcal{O}(v^2) \end{align} where the factor $a$ is given by:
\begin{align}
    a &= \frac{n_c g^4_A}{4 \pi (m^2_A - 4 m^2_\phi)^2}\sqrt{1-m^2_f/m^2_\chi} \left[2 -m^2_f/m^2_\phi \right]. \end{align}
\subsection{\bf Complex scalar dark matter: spin-1 mediator}
\label{a8}
\begin{figure*}[!htb]
\centering
\subfigure[$\chi{\chi}^{*}\rightarrow f\bar{f}$~process for s-channel.]{
    \includegraphics[width=6.5cm,height=4.2cm] {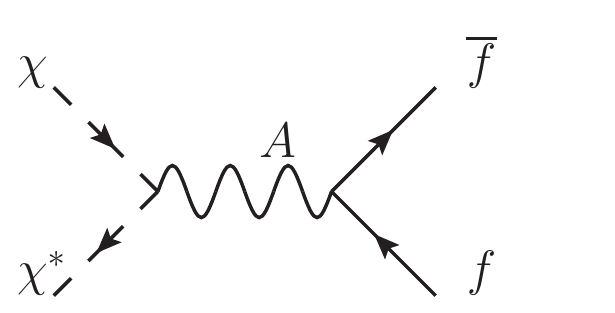}
    \label{fig15a}
}
\subfigure[$\chi{\chi}\rightarrow f\bar{f}$~process for s-channel.]{
    \includegraphics[width=6.5cm,height=4.2cm] {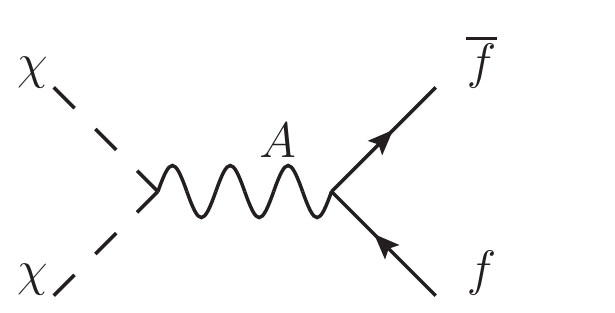}
    \label{fig15b}
}
\caption[Optional caption for list of figures]{Feynman diagrammatic representation of s-channel processes for complex and real scalar dark matter with spin-1 mediator.} 
\label{fig15}
\end{figure*}
\begin{figure*}[!htb]
\centering
\subfigure[$g_{A}$ vs $m_{A}$ for s-channel with $m_{\chi}=100{\rm GeV}$.]{
    \includegraphics[width=14.5cm,height=8cm] {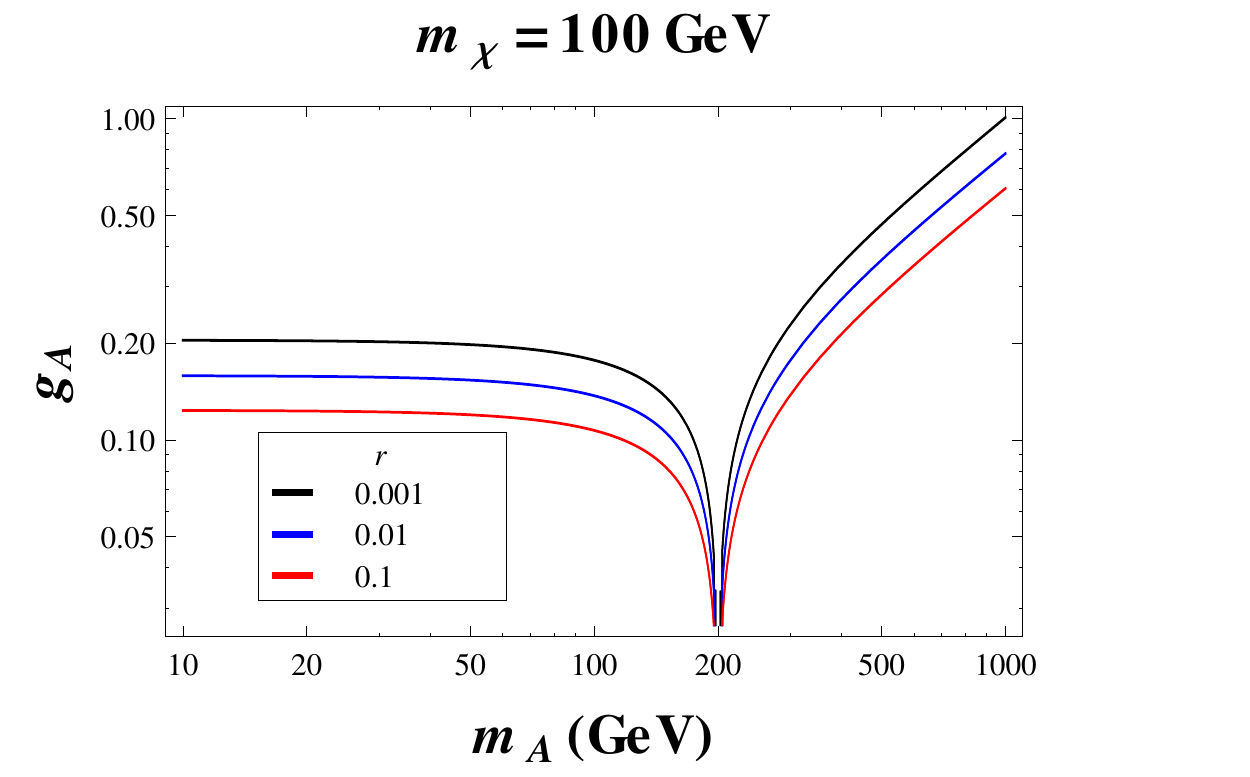}
    \label{fig16a}
}
\caption[Optional caption for list of figures]{In the above figure we have shown the allowed 
region of $g_A$ with respect to the mass of the mediator $m_A$, the upper panel is for the 
s-channel processes.} 
\label{fig16}
\end{figure*}

 Here we consider the following interactions for a complex scalar dark matter particle, $\phi$, and a spin-0
  mediator, $A$, within the framework of effective field theory written in RSII membrane as:
 \begin{align}
 \mathcal{L}_{membrane} &\supset \left[ig_\phi \phi^\dagger \overleftrightarrow{\partial_{\mu}}\phi + \overline{f}\gamma_\mu(g_{fv} + g_{fa}\gamma^5)f\right]V^\mu
 \end{align}
 In fig.~(\ref{fig15a}) and fig.~(\ref{fig15b}), we have explicitly shown the
Feynman diagrammatic representation of possible s-channel for complex and real scalar dark matter with spin-1 mediator.
The cross-section from the above Lagrangian after taking all the coupling equal is given as: 
\begin{align}
\sigma &= \frac{1}{8\pi (s-4m^2_\chi)}\int_{t_-}^{t_+}|\mathcal{M}|^2 dt 
\end{align}
where the matrix element for the S-matrix and the symbol $t_{\pm}$ is defined as:
\begin{align} |\mathcal{M}|^2 &= \frac{8n_cg^4}{3\pi \left((s-m^2)^2 + m^2\Gamma^2 \right)}\left(1-\frac{2m^2_f}{s}\right),\\
t_{\pm}& = (m^2_\chi + m^2_f-\frac{s}{2}) \pm \frac{\sqrt{(s-4m^2_\chi)(s-4m^2_f)}}{2}.
\end{align}
where $n_c=3$ for quarks and 1 for leptons, $g_A$ and $m_A$ are the respective coupling and the mass of the mediator. 
Further 
taking the following approximation: \be s=4m^2_\chi/\left(1+v^2/4\right),\ee we finally get the following simplified expression for the product of annihilation cross-section and velocity as:
\begin{align} \sigma v = bv^2 + \mathcal{O}(v^3) \end{align} where the factor $b$ is given by:
\begin{align}
b &= n_cg^4 \frac{m^2_\chi}{3\pi }\sqrt{1-\frac{m^2_f}{m^2_\chi}}\frac{1-\frac{m^2f}{2m^2_\chi}}{ (m^2-4m^2_\chi)^2+m^2 \Gamma^2},\end{align}
 and the mediator's width to SM fermions is given by:
    \begin{align} 
\Gamma &= \frac{g^2n_cm}{6\pi}\sqrt{1-\frac{4m^2_f}{m^2}}\left[1-\frac{m^2_f}{m^2}\right].
\end{align}
The dark matter relic abundance in this case is given as:
\begin{align}
 \Omega_{DM} h^2 &= \frac{1.07\times 10^9}{J(x_f)g^{1/2}_*M_p}\end{align}
 where $J(x_f)$ for complex scalar dark matter with spin-1 mediator is given by:
 \begin{align}
 J(x_f) &= \int_{x_f}^{\infty}n_cg^4 \frac{m^2_\chi}{3\pi }\sqrt{1-\frac{m^2_f}{m^2_\chi}}\frac{1-\frac{m^2f}{2m^2_\chi}}{ \left((m^2-4m^2_\chi)^2+m^2 \Gamma^2\right)x^3f_{membrane}(x)}
dx.
\end{align}
where the function $f_{membrane}(x)$ is the characteristic parameter for RS single braneworld and this can be expressed in terms of tensor-to-scalar ratio ($r$) which is given in Eq.\eqref{eq:ev1xxxxxc}. 
Now for the GR limiting case of the $f_{membrane}(x) \rightarrow 1$ and then the relic abundance will only depend on the mass of the Dark Matter ($m_\chi$), $g_A$ the 
coupling with the spin-0 mediator and the mass of the mediator ($m_A$). 
\begin{align}
 \Omega_{DM} h^2 &= \frac{1.07\times 10^9x^2_f}{g^{1/2}_*M_p}\left(n_cg^4 \frac{m^2_\chi}{3\pi }\sqrt{1-\frac{m^2_f}{m^2_\chi}}\frac{1-\frac{m^2f}{2m^2_\chi}}{ (m^2-4m^2_\chi)^2
                  +m^2 \Gamma^2}\right)^{-1}.
\end{align}
In order to constrain the coupling ($g_A$) and mass ($m_A$) we take the present data of the relic abundance ($\Omega_{DM}h^2 = 0.1199\pm0.0027$\cite{Ade:2015xua}) and constrain the function 
$J(x_f)$, which in turn constrain the coupling $g_A$ and $m_A$ for a particular tensor-to-scalar ratio ($r$). We have not shown the GR limiting case as it has been extensively been explored in 
\cite{Berlin:2014tja}.
In fig.~(\ref{fig16a}) , we have depicted the behaviour of the effective coupling of spin-0 mediator $g_{A}$ with the varying mass the spin-0 mediator $m_{A}$ 
for s-channel process with three distinct value of the tensor-to-scalar ratio $r=0.001$, $r=0.01$ and $r=0.1$ 
respectively in RSII membrane. We have only considered one value of the dark matter mass which is $m_{\chi}=100\; {\rm GeV}$,
cause as go to higher values like $\mathcal{O}({\rm TeV})$ the required coupling is going more than $\mathcal{O}$(1), for the s-channel analysis. 
From fig.~(\ref{fig16a}), it is clearly observed that the behaviour of the effective coupling of spin-0 mediator $g_{A}$ with the varying mass the spin-0 mediator $m_{A}$ is sensitive in the vicinity of $m_{A}=2\times 10^{2}{\rm GeV}$ as it is reaching a 
resonance (i.e $2 m_X = m_A$).
Most importantly, in both the sides of $m_{A}=2\times 10^{2}{\rm GeV}$
the coupling of spin-0 mediator $g_{A}$ behave in completely opposite manner. Now for the real vector dark matter with spin-1 mediator, 
the real scalar s-channel interaction vanishes identically, so that real scalar dark matter cannot couple to a vector at tree level of the effective field theory.
\subsection{\bf Complex vector dark matter: spin-0 mediator}
\label{a10}
Consider the following interaction for a complex vector dark matter, $\chi_\mu$, and a spin-0 mediator, A:
\begin{align}
\mathcal{L}_{membrane} &\supset \left[\mu_\chi \chi^\mu \chi^\dag_\mu  + \overline{f}(\lambda_{f_S} + \lambda_{f_p}i \gamma^5)f\right]A
\end{align}
\begin{figure*}[!htb]
\centering
\subfigure[$\chi_{\mu}{\chi}^{*}_{\nu}\rightarrow f\bar{f}$~process for s-channel.]{
    \includegraphics[width=6.5cm,height=4.2cm] {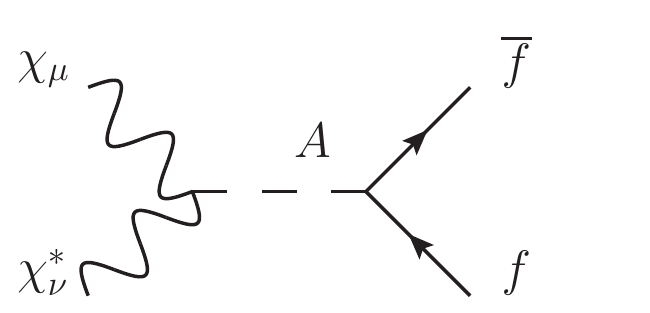}
    \label{fig17a}
}
\subfigure[$\chi_{\mu}{\chi}_{\nu}\rightarrow f\bar{f}$~process for s-channel.]{
    \includegraphics[width=6.5cm,height=4.2cm] {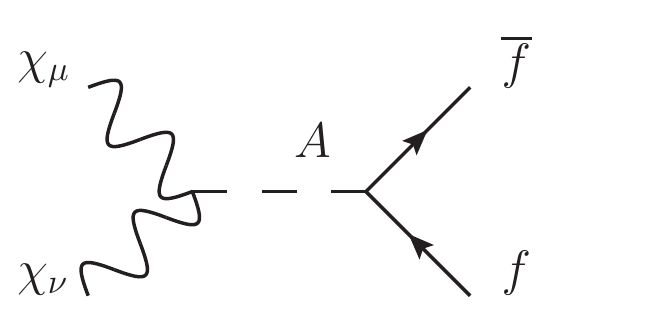}
    \label{fig17b}
}
\caption[Optional caption for list of figures]{Feynman diagrammatic representation of s-channel processes for complex and real vector dark matter with spin-0 mediator.} 
\label{fig17}
\end{figure*}
\begin{figure*}[!htb]
\centering
\subfigure[$g_{A}$ vs $m_{A}$ for s-channel with $m_{\chi}=100{\rm GeV}$.]{
    \includegraphics[width=14.5cm,height=8cm] {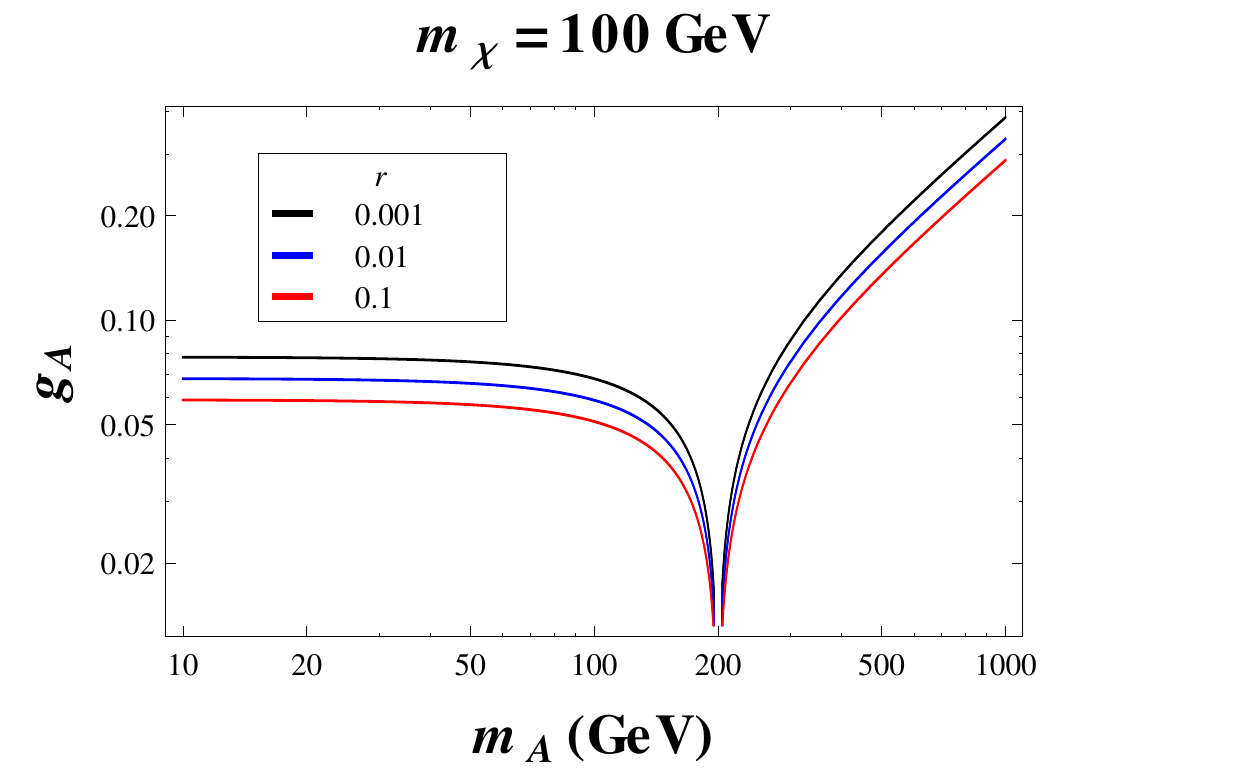}
    \label{fig18a}
}
\subfigure[$g_{A}$ vs $m_{A}$ for s-channel with $m_{\chi}=1{\rm TeV}$.]{
    \includegraphics[width=14.5cm,height=8cm] {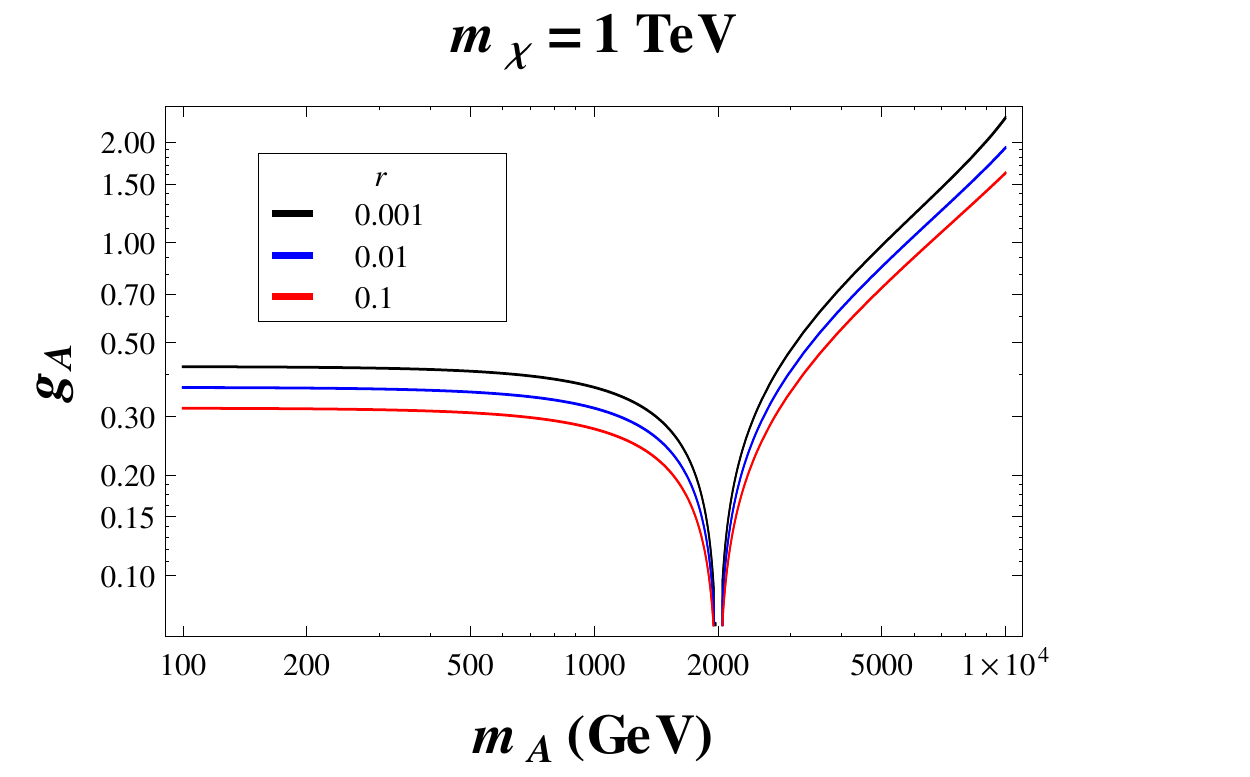}
    \label{fig18b}
}
\caption[Optional caption for list of figures]{In the above figure we have shown the allowed 
region of $g_A$ with respect to the mass of the mediator $m_A$, the upper panel is for the 
s-channel processes.} 
\label{fig18}
\end{figure*}
In fig.~(\ref{fig17a}), we have explicitly shown the
Feynman diagramatic representation of possible s-channel for complex vector dark matter with spin-0 mediator.

The cross-section from the above Lagrangian after taking all the coupling equal is given as:
\begin{align}
\sigma &= \frac{1}{8\pi (s-4m^2_\chi)}\int_{t_-}^{t_+}|\mathcal{M}|^2 dt 
\end{align}
where the matrix element for the S-matrix and the symbol $t_{\pm}$ is defined as:
\begin{align} |\mathcal{M}|^2 &= \frac{n_cg^4}{9\pi  \left((s-m^2)^2 + m^2\Gamma^2 \right)}\Big{[}\frac{s}{m^2_\chi}\Big{[}\frac{s}{4m^2_\chi}-1\Big{]}+3\Big{]},\\
t_{\pm} &= (m^2_\chi + m^2_f-\frac{s}{2}) \pm \frac{\sqrt{(s-4m^2_\chi)(s-4m^2_f)}}{2}.
\end{align}
where $n_c=3$ for quarks and 1 for leptons, $g_A$ and $m_A$ are the respective coupling and the mass of the mediator. 
Further 
taking the following approximation: \be s=4m^2_\chi/\left(1+v^2/4\right),\ee we finally get the following simplified expression for the product of annihilation cross-section and velocity as:
\begin{align} \sigma v = a + \mathcal{O}(v^2) \end{align} where the factor $a$ is given by:
\begin{align}
a &= n_cg^4 \frac{m^2_\chi}{6\pi }\sqrt{1-\frac{m^2_f}{m^2_\chi}}\frac{1-\frac{m^2f}{2m^2_\chi}}{ (m^2-4m^2_\chi)^2+m^2 \Gamma^2},
\end{align}
and the mediator's width to SM fermions is given by:
\begin{align}
    \Gamma &= \frac{g^2n_cm}{6\pi}\sqrt{1-\frac{4m^2_f}{m^2}}\left[1-\frac{m^2_f}{m^2}\right].
\end{align}
The dark matter relic abundance in this case is given by:
\begin{align}
 \Omega_{DM} h^2 &= \frac{1.07\times 10^9}{J(x_f)g^{1/2}_*M_p}\end{align}
 where $J(x_f)$ for complex vector dark matter with spin-0 mediator is given by:
 \begin{align} J(x_f) &= \int_{x_f}^{\infty}n_cg^4 \frac{m^2_\chi}{6\pi }\sqrt{1-\frac{m^2_f}{m^2_\chi}}\frac{1-\frac{m^2f}{2m^2_\chi}}{\left( (m^2-4m^2_\chi)^2+m^2 \Gamma^2\right)x^2f_{membrane}(x)}
dx.
\end{align}
where the function $f_{membrane}(x)$ is the characteristic parameter for RS single braneworld and this can be expressed in terms of tensor-to-scalar ratio ($r$) which is given in Eq.\eqref{eq:ev1xxxxxc}. 
Now for the GR limiting case of the $f_{membrane}(x) \rightarrow 1$ and then the relic abundance will only depend on the mass of the Dark Matter ($m_\chi$), $g_A$ the 
coupling with the spin-0 mediator and the mass of the mediator ($m_A$). 
\begin{align}
 \Omega_{DM} h^2 &= \frac{1.07\times 10^9x_f}{g^{1/2}_*M_p}\left(n_cg^4 \frac{m^2_\chi}{6\pi }\sqrt{1-\frac{m^2_f}{m^2_\chi}}\frac{1-\frac{m^2f}{2m^2_\chi}}{(m^2-4m^2_\chi)^2
                 +m^2 \Gamma^2}\right)^{-1}.
\end{align}
In order to constrain the coupling ($g_A$) and mass ($m_A$) we take the present data of the relic abundance ($\Omega_{DM}h^2 = 0.1199\pm0.0027$\cite{Ade:2015xua}) and constrain the function 
$J(x)$, which in turn constrain the coupling $g_A$ and $m_A$ for a particular tensor-to-scalar ratio ($r$). We have not shown the GR limiting case as it has been extensively been explored in 
\cite{Berlin:2014tja}.
In fig.~(\ref{fig18a}) and fig.~(\ref{fig18b}), we have depicted the behaviour of the effective coupling of spin-0 mediator $g_{A}$ with the varying mass the spin-0 mediator $m_{A}$ 
for s-channel process with three distinct value of the tensor-to-scalar ratio $r=0.001$, $r=0.01$ and $r=0.1$ 
respectively in RSII membrane. We also consider two different values of the dark matter mass $m_{\chi}=100\; {\rm GeV}$ and $m_{\chi}=1\;{\rm TeV}$
for the s-channel analysis. From fig.~(\ref{fig18a}) and fig.~(\ref{fig18b}), it is clearly observed
that the behaviour of the effective coupling of spin-0 mediator $g_{A}$ with the varying mass the spin-0 mediator $m_{A}$ are almost similar for both of the cases and also 
sensitive in the vicinity of $m_{A}=2\times 10^{2}\;{\rm GeV}$ and $m_{A}=2\times 10^{3}\;{\rm GeV}$ respectively as it has a resonance 
(i.e $2m_X = m_A$).
Most importantly, in both the sides of $m_{A}=2\times 10^{2}\;{\rm GeV}$ and $m_{A}=2\times 10^{3}\;{\rm GeV}$
the coupling of spin-0 mediator $g_{A}$ behave in completely opposite manner.

Now, similar to the earlier case of complex vector dark matter that interacts with SM through spin-0 mediator, 
a real vector matter Lagrangian will look like:
\begin{align}
\mathcal{L}_{membrane} &\supset \left[\frac{1}{2}\mu_\chi \chi^\mu \chi^\dag_\mu  + \overline{f}(\lambda_{f_S} + \lambda_{f_p}i \gamma^5)f\right]A
\label{eq:dmrvm0}
\end{align}
and after taking all the couplings equal the thermally averaged cross-section is same as mentioned in the previous subsection for the complex vector dark matter.
In fig.~(\ref{fig17b}), we have explicitly shown the
Feynman diagrammatic representation of possible s-channel for real vector dark matter with spin-0 mediator.
\subsection{\bf Complex vector dark matter: spin-1 mediator}
\label{a12}
\begin{figure*}[!htb]
\centering
\subfigure[$\chi_{\mu}{\chi}^{*}_{\nu}\rightarrow f\bar{f}$~process for s-channel.]{
    \includegraphics[width=6.5cm,height=4.2cm] {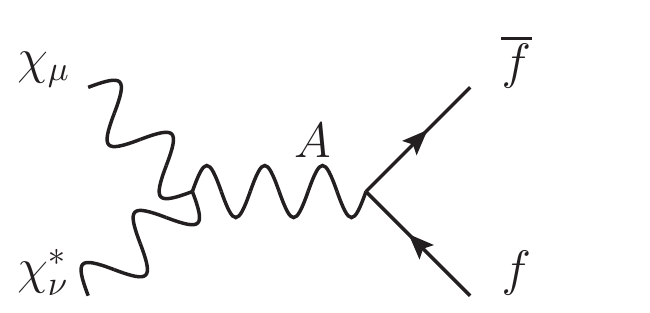}
    \label{fig19a}
}
\subfigure[$\chi_{\mu}{\chi}_{\nu}\rightarrow f\bar{f}$~process for s-channel.]{
    \includegraphics[width=6.5cm,height=4.2cm] {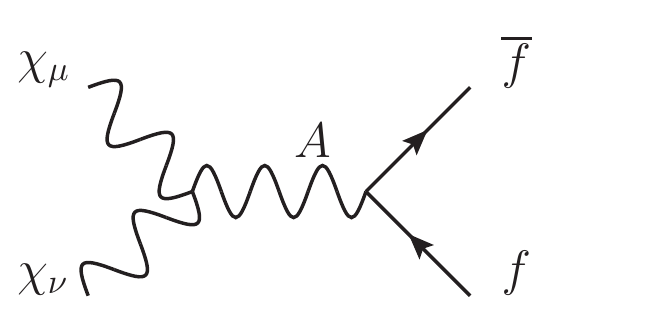}
    \label{fig19b}
}
\caption[Optional caption for list of figures]{Feynman diagrammatic representation of s-channel processes for complex and real vector dark matter with spin-1 mediator.} 
\label{fig19}
\end{figure*}
\begin{figure*}[!htb]
\centering
\subfigure[$g_{A}$ vs $m_{A}$ for s-channel with $m_{\chi}=100{\rm GeV}$.]{
    \includegraphics[width=14.5cm,height=8cm] {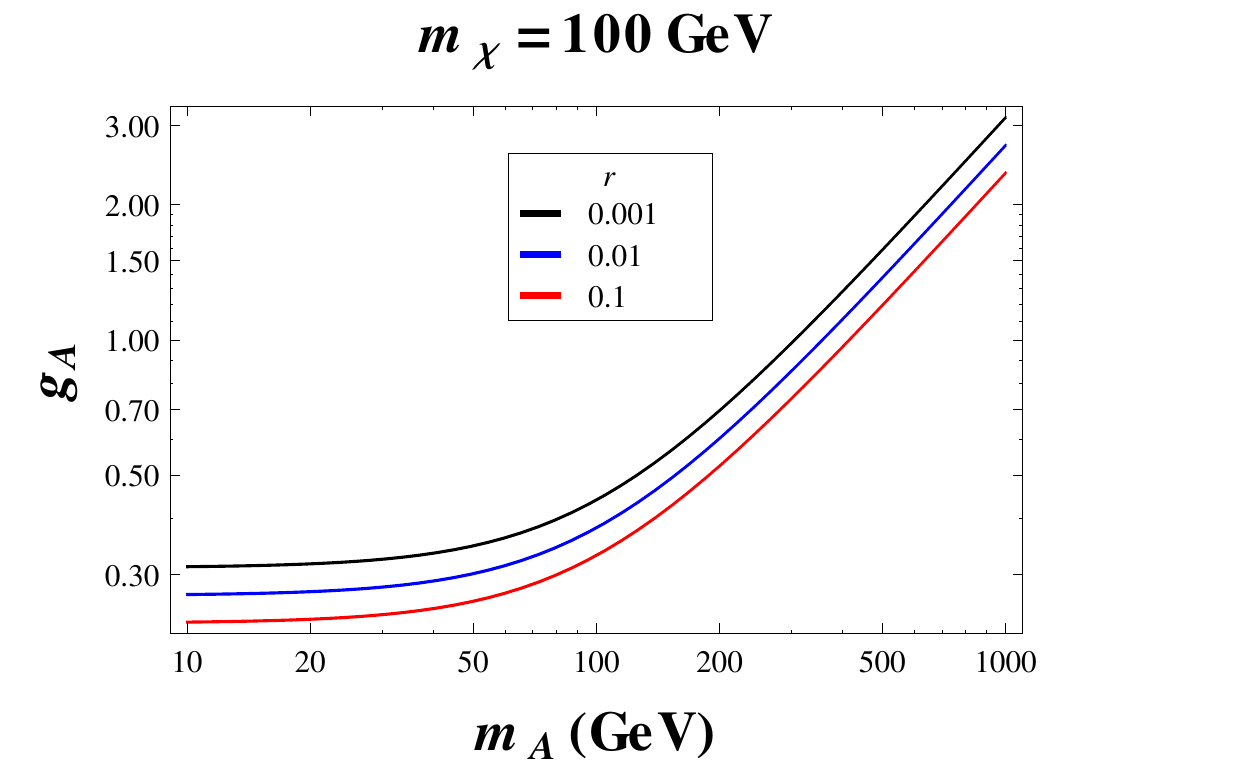}
    \label{fig20a}
}
\subfigure[$g_{A}$ vs $m_{A}$ for s-channel with $m_{\chi}=1{\rm TeV}$.]{
    \includegraphics[width=14.5cm,height=8cm] {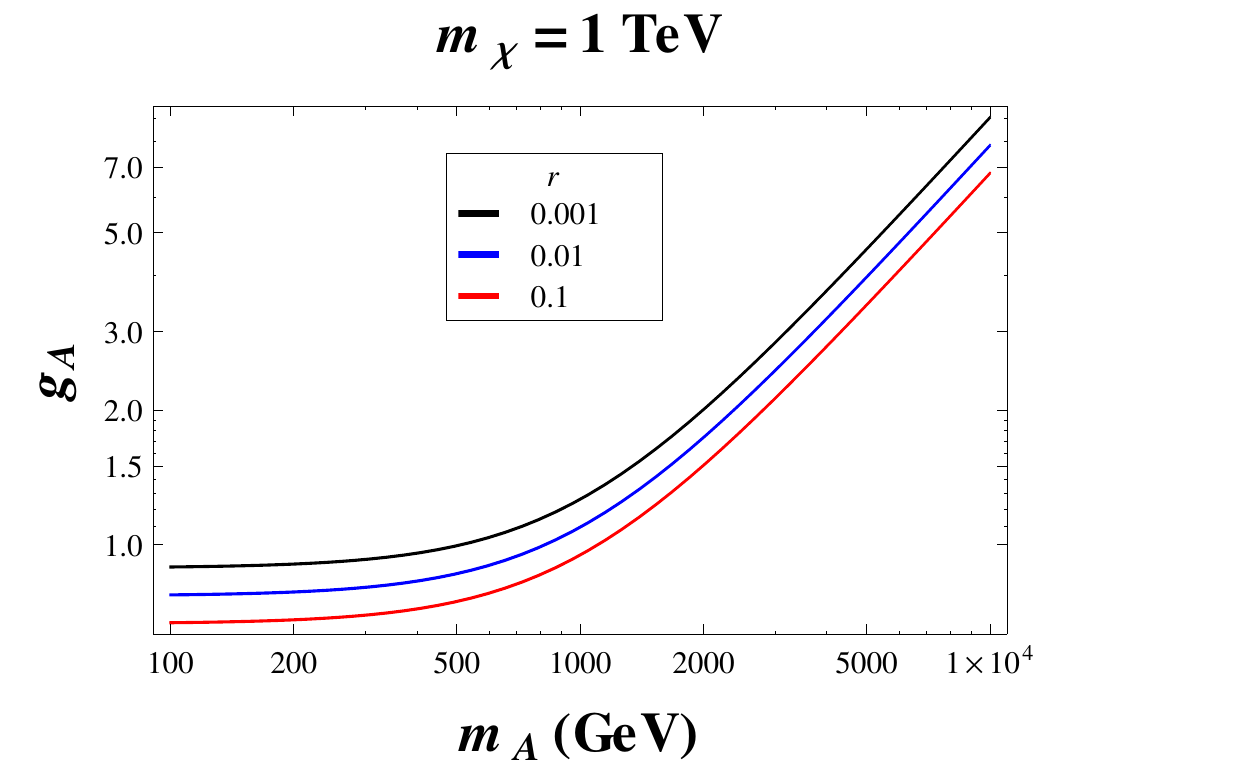}
    \label{fig20b}
}
\caption[Optional caption for list of figures]{In the above figure we have shown the allowed 
region of $g_A$ with respect to the mass of the mediator $m_A$, the above processes
are s-channel.} 
\label{fig20}
\end{figure*}
Here we consider the Lagrangian of complex scalar dark matter with spin-1 mediator is given as: 
\begin{align}
\mathcal{L}_{membrane} &\supset \left[ g_X\left(X^{\dagger \mu}\partial_\nu X^\nu + {\rm h.c}\right) + \overline{f}\gamma^\mu\left(g_{fv} 
    + g_{fa}\gamma^5\right)f\right] V_\mu
\end{align}
In fig.~(\ref{fig19a}) and fig.~(\ref{fig19b}), we have explicitly shown the
Feynman diagrammatic representation of possible s-channel for complex and real vector dark matter with spin-1 mediator.
The cross-section from the above Lagrangian after taking all the coupling equal is given as:
\begin{align}
\sigma &= \frac{1}{8\pi (s-4m^2_\chi)}\int_{t_-}^{t_+}|\mathcal{M}|^2 dt 
\end{align}
where the matrix element for S-matrix and the symbol $t_{\pm}$ is given by:
\begin{align}
 |\mathcal{M}|^2 &= \frac{n_cg^4 (s- 4 m^2_\chi)}{9\pi  m^4_{\chi}m^2\left((s-m^2)^2 + m^2\Gamma^2 \right)}\Big{[}2m^2_\chi m^4 (s+2m^2_f) \nonumber \\
    &+ 2m^2_\chi(s m^4 -2 m^2_f(5m^4-6m^2s +3s^2)) + 3 m^2_fs(s-m^2)^2\Big{]},\\
t_{\pm} &= (m^2_\chi + m^2_f-\frac{s}{2}) \pm \frac{\sqrt{(s-4m^2_\chi)(s-4m^2_f)}}{2}.
\end{align}
where $n_c=3$ for quarks and 1 for leptons, $g_A$ and $m_A$ are the respective coupling and the mass of the mediator. 
Further 
taking the following approximation: \be s=4m^2_\chi/\left(1+v^2/4\right),\ee we finally get the following simplified expression for the product of annihilation cross-section and velocity as:
\begin{align} \sigma v = bv^2 + \mathcal{O}(v^3) \end{align} where the factor $b$ is given by:
\begin{align}
b &= n_cg^4 \frac{m^2_\chi}{9\pi }\sqrt{1-\frac{m_f}{m^2_\chi}}\frac{4m^2_\chi}{ (m^2-4m^2_\chi)^2+m^2 \Gamma^2}, 
\end{align}
and the mediator's width to SM fermions is given by:
\begin{align}
\Gamma &= \frac{g^2n_cm}{6\pi}\sqrt{1-\frac{4m^2_f}{m^2}}\left[1-\frac{m^2_f}{m^2}\right].\end{align}
The dark matter relic abundance in this case is given by:
\begin{align}
 \Omega_{DM} h^2 &= \frac{1.07\times 10^9}{J(x_f)g^{1/2}_*M_p}\end{align}
 where $J(x_f)$ for complex vector dark matter with spin-1 mediator is given by:
 \begin{align}J(x_f) &= \int_{x_f}^{\infty}n_cg^4 \frac{m^2_\chi}{9\pi }\sqrt{1-\frac{m_f}{m^2_\chi}}\frac{4m^2_\chi}{\left((m^2-4m^2_\chi)^2+m^2 \Gamma^2\right)x^2f_{membrane}(x)}
dx.
\end{align}
where the function $f_{membrane}(x)$ is the characteristic parameter for RS single braneworld and this can be expressed in terms of tensor-to-scalar ratio ($r$) which is given in Eq.\eqref{eq:ev1xxxxxc}. 
Now for the GR limiting case of the $f_{membrane}(x) \rightarrow 1$ and then the relic abundance will only depend on the mass of the Dark Matter ($m_\chi$), $g_A$ the 
coupling with the spin-0 mediator and the mass of the mediator ($m_A$). 
\begin{align}
 \Omega_{DM} h^2 &= \frac{1.07\times 10^9x^2_f}{g^{1/2}_*M_p}\left(n_cg^4 \frac{m^2_\chi}{9\pi }\sqrt{1-\frac{m_f}{m^2_\chi}}\frac{4m^2_\chi}{ (m^2-4m^2_\chi)^2+m^2 \Gamma^2}\right)^{-1}.
\end{align}
In order to constrain the coupling ($g_A$) and mass ($m_A$) we take the present data of the relic abundance ($\Omega_{DM}h^2 = 0.1199\pm0.0027$\cite{Ade:2015xua}) and constrain the function 
$J(x)$, which in turn constrain the coupling $g_A$ and $m_A$ for a particular tensor-to-scalar ratio ($r$). We have not shown the GR limiting case as it has been extensively been explored in 
\cite{Berlin:2014tja}.
In fig.~(\ref{fig20a}) and fig.~(\ref{fig20b}), we have depicted the behaviour of the effective coupling of spin-1 mediator $g_{A}$ with the varying mass the spin-1 mediator $m_{A}$ 
for s-channel process with three distinct value of the tensor-to-scalar ratio $r=0.001$, $r=0.01$ and $r=0.1$ 
respectively in RSII membrane. We also consider two different values of the dark matter mass $m_{\chi}=100 {\rm GeV}$ and $m_{\chi}=1{\rm TeV}$
for the s-channel analysis. From fig.~(\ref{fig20a}) and fig.~(\ref{fig20b}), it is clearly observed
that the behaviour of the effective coupling of spin-1 mediator $g_{A}$ with the varying mass the spin-1 mediator $m_{A}$ are almost similar for both of the cases and also 
sensitive in the vicinity of $m_{A}=2\times 10^{2}{\rm GeV}$ and $m_{A}=2\times 10^{3}{\rm GeV}$ respectively as it has a resonance 
(i.e $2m_\chi = m_A$).
Most importantly, in both the sides of $m_{A}=1\times 10^{2}{\rm GeV}$ and $m_{A}=1\times 10^{3}{\rm GeV}$ 
the coupling of spin-1 mediator $g_{A}$ behave in completely opposite manner. 

Now for the real vector dark matter with spin-1 mediator the lagrangian is given as 
\begin{align}
\mathcal{L}_{membrane} &\supset \left[ \frac{1}{2}g_X\left(X^{\mu}\partial_\nu X^\nu + {\rm h.c}\right) + \overline{f}\gamma^\mu\left(g_{fv} 
    + g_{fa}\gamma^5\right)f\right] V_\mu.
\end{align}
The cross-section and the thermally averaged cross section is identical to the results obtained in the context of complex scalar dark matter with spin-1 mediator case.

\subsection{\bf Complex scalar dark matter: spin - 1/2 mediator}
\label{a14}
\begin{figure*}[!htb]
\centering
\subfigure[$\chi{\chi}^{*}\rightarrow f\bar{f}$~process for t/u-channel.]{
    \includegraphics[width=6.5cm,height=4.2cm] {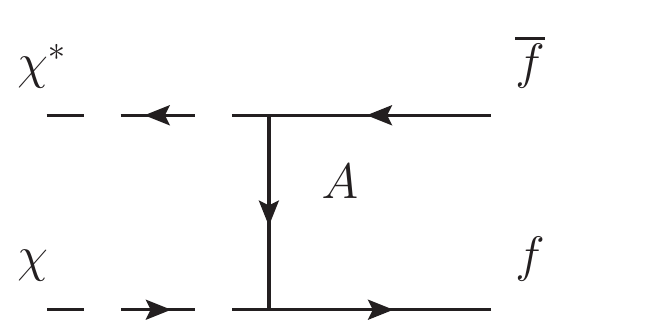}
    \label{fig22a}
}
\subfigure[$\chi_{\mu}{\chi}^{*}_{\nu}\rightarrow f\bar{f}$~process for t/u-channel.]{
    \includegraphics[width=6.5cm,height=4.2cm] {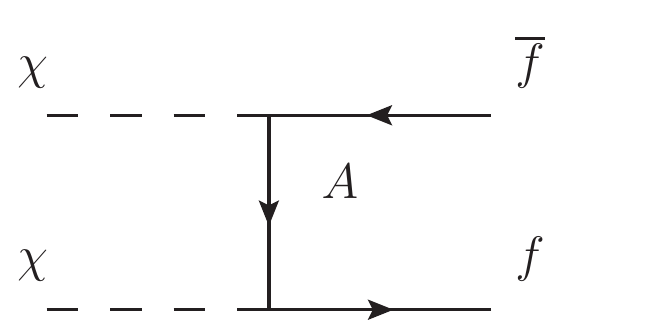}
    \label{fig22b}
}
\caption[Optional caption for list of figures]{Feynman diagrammatic representation of t/u-channel processes for complex and real scalar dark matter with spin-1/2 mediator.} 
\label{fig22}
\end{figure*}
The Lagrangian of complex scalar dark matter with spin -1/2 mediator is given as:
\begin{align}
\mathcal{L}_{membrane} &\supset \overline{\psi}(\lambda_s+\lambda_p\gamma^5)f\phi^\dagger 
+ \overline{f}(\lambda_s - \lambda_p \gamma^5)\psi \phi
\end{align}
In fig.~(\ref{fig22a}), we have explicitly shown the
Feynman diagramatic representation of possible t/u-channel for complex scalar dark matter with spin-1/2 mediator.

The cross-section from the above Lagrangian after taking all the coupling equal is given as:
\begin{align}
\sigma &= \frac{1}{8\pi (s-4m^2_\chi)}\int_{t_-}^{t_+}|\mathcal{M}|^2 dt
\end{align}
where the matrix element for the S-matrix and the symbol $t_{\pm}$ is given by:
\begin{align}
 |\mathcal{M}|^2 &= \frac{2}{\pi s}\frac{t(2m^2_f-s) - (m^2_\chi - m^2_f t)^2}{(t-m^2)^2}, \\
t_{\pm} &= (m^2_\chi + m^2_f-\frac{s}{2}) \pm \frac{\sqrt{(s-4m^2_\chi)(s-4m^2_f)}}{2}.
\end{align}
where $n_c=3$ for quarks and 1 for leptons, $g_A$ and $m_A$ are the respective coupling and the mass of the mediator. 
\begin{figure*}[!htb]
\centering
\subfigure[$g_{A}$ vs $m_{A}$ for t/u-channel with $m_{\chi}=100{\rm GeV}$.]{
    \includegraphics[width=14.5cm,height=8cm] {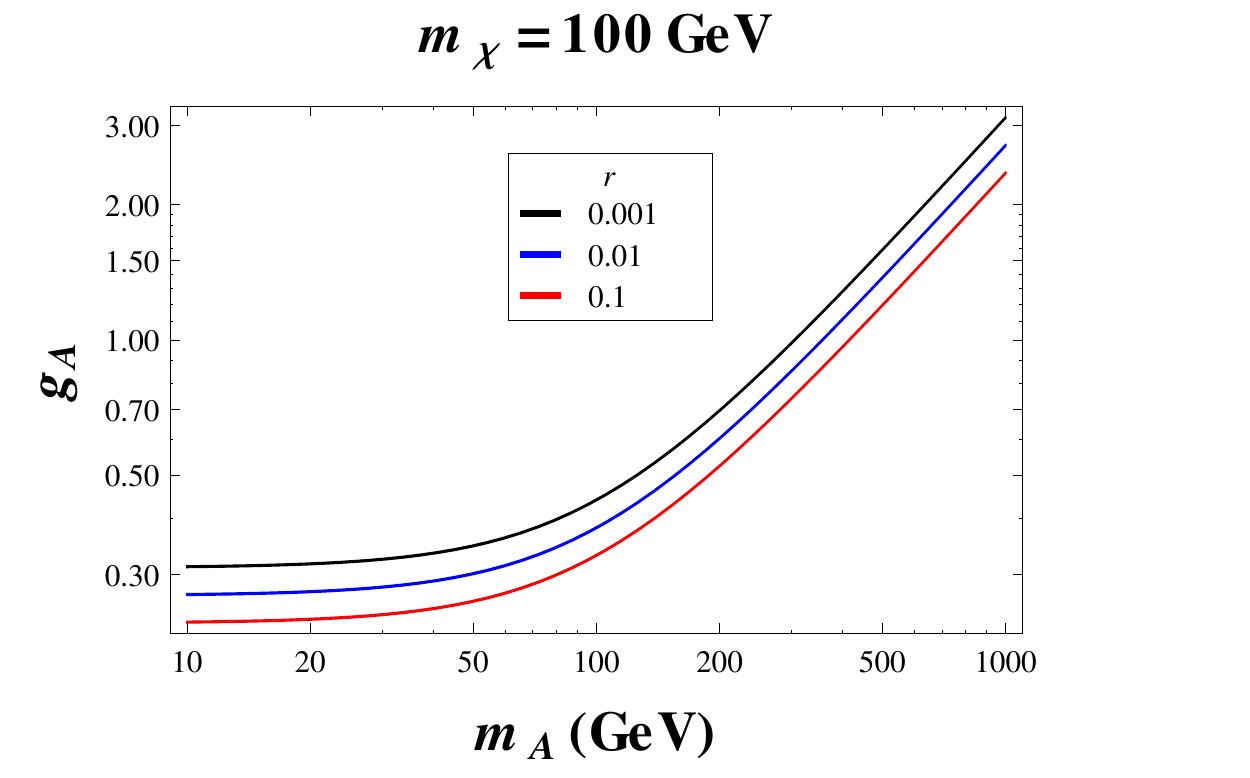}
    \label{fig21a}
}
\subfigure[$g_{A}$ vs $m_{A}$ for t/u-channel with $m_{\chi}=1{\rm TeV}$.]{
    \includegraphics[width=14.5cm,height=8cm] {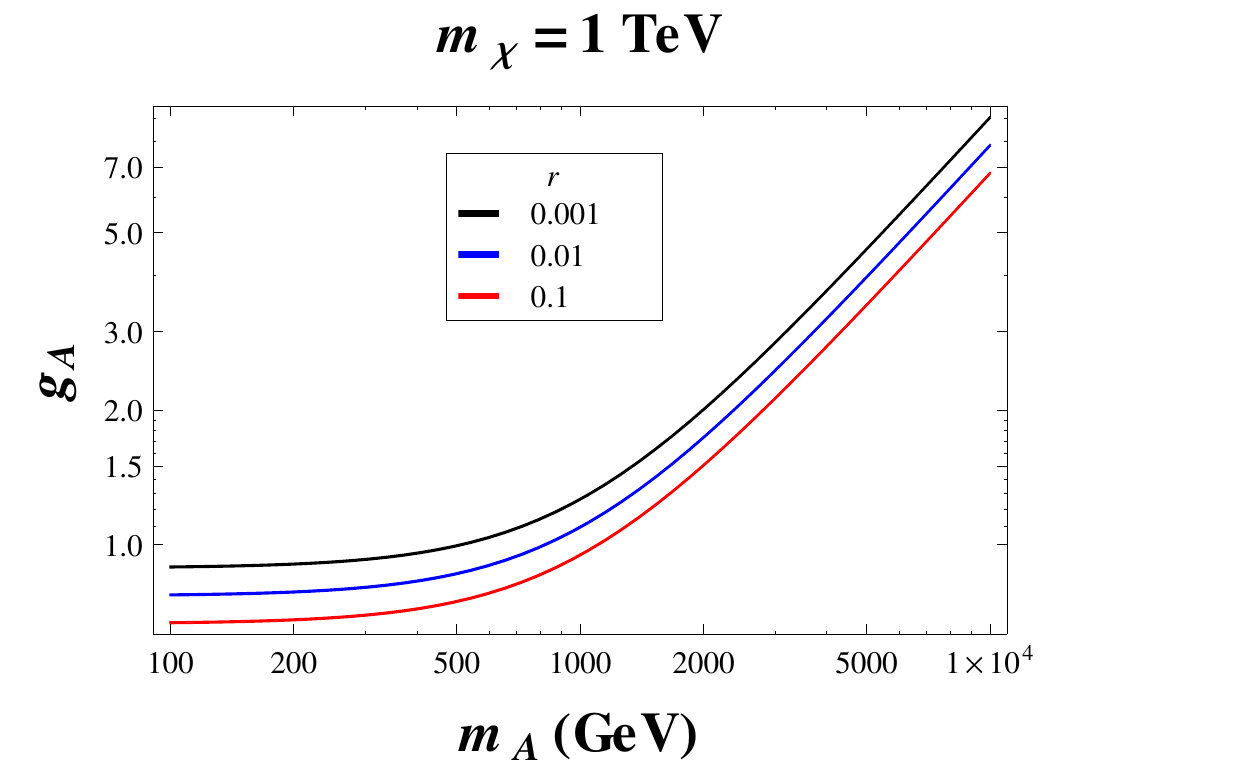}
    \label{fig21b}
}
\caption[Optional caption for list of figures]{In the above figure we have shown the allowed 
region of $g_A$ with respect to the mass of the mediator $m_A$, the above processes
are t/u-channel.} 
\label{fig21}
\end{figure*}
Further 
taking the following approximation: \be s=4m^2_\chi/\left(1+v^2/4\right),\ee we finally get the following simplified expression for the product of annihilation cross-section and velocity as:
\begin{align} \sigma v = a + \mathcal{O}(v^2), \end{align} where the factor $b$ is given by:
\begin{align}
a &= n_cg^4 \frac{m^2_f}{4\pi }\frac{\left[1-\frac{m_f}{m^2_\chi}\right]^{3/2}}{ (m^2 + m^2_\chi -m^2_f)}. 
\end{align}
The dark matter relic abundance in this case is given as:
\begin{align}
 \Omega_{DM} h^2 &= \frac{1.07\times 10^9}{J(x_f)g^{1/2}_*M_p}\end{align}
 where $J(x_f)$ for complex scalar dark matter with spin - 1/2 mediator is given by:
 \begin{align}
 J(x_f) &= \int_{x_f}^{\infty}n_cg^4 \frac{m^2_f}{4\pi }\frac{\left[1-\frac{m_f}{m^2_\chi}\right]^{3/2}}{\left((m^2 + m^2_\chi -m^2_f)\right)x^2f_{membrane}(x)}
dx.
\end{align}
where the function $f_{membrane}(x)$ is the characteristic parameter for RS single braneworld and this can be expressed in terms of tensor-to-scalar ratio ($r$) which is given in Eq.\eqref{eq:ev1xxxxxc}.
Now for the GR limiting case of the $f_{membrane}(x) \rightarrow 1$ and then the relic abundance will only depend on the mass of the Dark Matter ($m_\chi$), $g_A$ the 
coupling with the spin-0 mediator and the mass of the mediator ($m_A$). 
\begin{align}
 \Omega_{DM} h^2 &= \frac{1.07\times 10^9x^2_f}{g^{1/2}_*M_p}\left(n_cg^4 \frac{m^2_f}{4\pi }\frac{\left[1-\frac{m_f}{m^2_\chi}\right]^{3/2}}{ (m^2 + m^2_\chi -m^2_f)}\right)^{-1}.
\end{align}
In order to constrain the coupling ($g_A$) and mass ($m_A$) we take the present data of the relic abundance ($\Omega_{DM}h^2 = 0.1199\pm0.0027$\cite{Ade:2015xua}) and constrain the function 
$J(x_f)$, which in turn constrain the coupling $g_A$ and $m_A$ for a particular tensor-to-scalar ratio ($r$). We have not shown the GR limiting case as it has been extensively been explored in 
\cite{Berlin:2014tja}.
In fig.~(\ref{fig21a}) and fig.~(\ref{fig21b}), we have depicted the behaviour of the effective coupling of spin-1/2 mediator $g_{A}$ with the varying mass the spin-1/2 mediator $m_{A}$ 
for t/u-channel process with three distinct value of the tensor-to-scalar ratio $r=0.001$, $r=0.01$ and $r=0.1$ respectively in RSII membrane. We
also consider two different values of the dark matter mass 
$m_{\chi}=100\; {\rm GeV}$ and $m_{\chi}=1\;{\rm TeV}$ for the t/u-channel analysis. From fig.~(\ref{fig21a}) and fig.~(\ref{fig21b}), it is clearly observed
that the behaviour of the effective coupling of spin-1/2 mediator $g_{A}$ with the varying mass the spin-1/2 mediator $m_{A}$ are almost similar for both of the cases.
Most importantly, in both the sides of $m_{A}=1\times 10^{2}{\rm GeV}$ and $m_{A}=1\times 10^{3}{\rm GeV}$
the coupling of spin-1/2 mediator $g_{A}$ behave in completely opposite manner.

\subsection{\bf Real Scalar dark matter: spin-1/2 mediator}
\label{a15}
The Lagrangian of real scalar dark matter with spin -1/2 mediator is given as:
\begin{align}
\mathcal{L}_{membrane} &\supset \overline{\psi}(\lambda_s+\lambda_p\gamma^5)f\phi 
+ \overline{f}(\lambda_s - \lambda_p \gamma^5)\psi \phi.
\end{align}
\begin{figure*}[!htb]
\centering
\subfigure[$g_{A}$ vs $m_{A}$ for t/u-channel with $m_{\chi}=100{\rm GeV}$.]{
    \includegraphics[width=14.5cm,height=7cm] {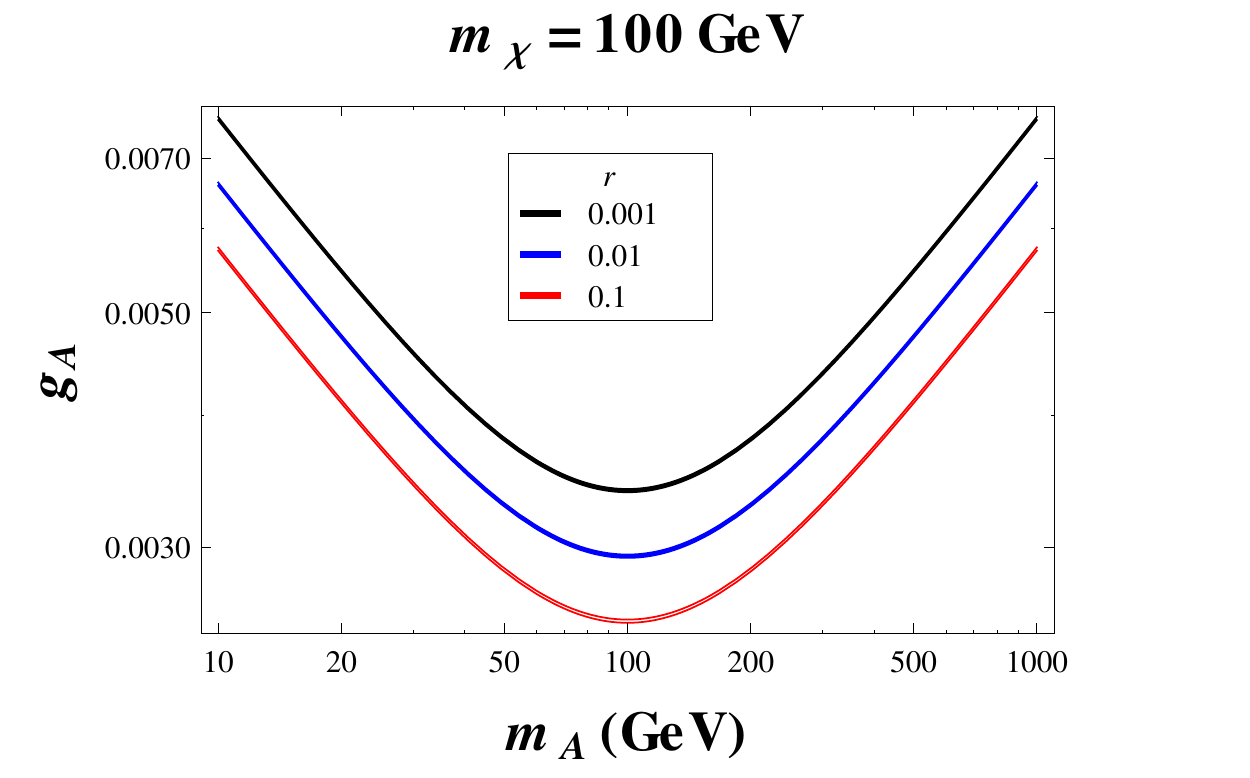}
    \label{fig23a}
}
\subfigure[$g_{A}$ vs $m_{A}$ for t/u-channel with $m_{\chi}=1{\rm TeV}$.]{
    \includegraphics[width=14.5cm,height=7cm] {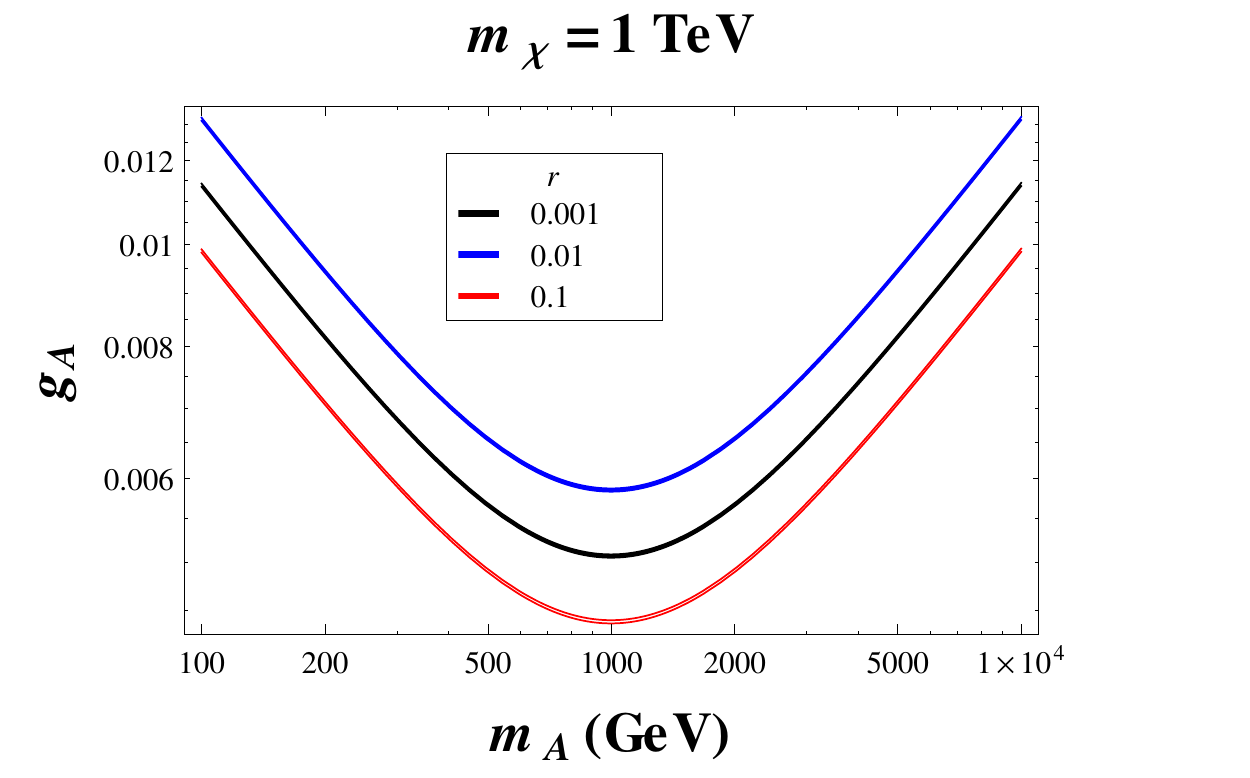}
    \label{fig23b}
}
\caption[Optional caption for list of figures]{In the above figure we have shown the allowed 
region of $g_A$ with respect to the mass of the mediator $m_A$, the above processes
are t/u-channel.} 
\label{fig23}
\end{figure*}
The cross-section from the above Lagrangian after taking all the coupling equal is given as:
\begin{align}
\sigma &= \frac{1}{8\pi (s-4m^2_\chi)}\int_{t_-}^{t_+}|\mathcal{M}|^2 dt
\end{align}
where the matrix element for the S-matrix and the symbol $t_{\pm}$ is given by:
\begin{align}
 |\mathcal{M}|^2 &= g^4 \left[\frac{t(2m^2_f-s)-(m^2_\chi - m^2_f-t)^2}{(t - m^2)^2} 
+ \frac{u(2m^2_f-s)-(m^2_\chi -m^2_f - u)^2}{(u - m^2)^2} \right. \nonumber \\
    &~~~~~~~~~~+ \left.2\frac{(u-m^2_\chi + m^2_f)(m^2_\chi - m^2_f - t) + (m^2_\chi - m^2_f)(s-2m^2_f)}{(u-m^2)(t-m^2)}\right],\\
t_{\pm} &= (m^2_\chi + m^2_f-\frac{s}{2}) \pm \frac{\sqrt{(s-4m^2_\chi)(s-4m^2_f)}}{2}.
\end{align}
where $n_c=3$ for quarks and 1 for leptons, $g_A$ and $m_A$ are the respective coupling and the mass of the mediator. 
Further 
taking the following approximation: \be s=4m^2_\chi/\left(1+v^2/4\right),\ee we finally get the following simplified expression for the product of annihilation cross-section and velocity as:
\begin{align} \sigma v = a + \mathcal{O}(v^2), \end{align} where the factor $a$ is given by:
\begin{align}
a &= 2n_cg^4 \frac{m^2_f}{\pi }\frac{\left[1-\frac{m_f}{m^2_\chi}\right]^{3/2}}{ (m^2 + m^2_\chi -m^2_f)}.
\end{align}
The dark matter relic abundance in this case is given as:
\begin{align}
 \Omega_{DM} h^2 &= \frac{1.07\times 10^9}{J(x_f)g^{1/2}_*M_p}\end{align}
 where $J(x_f)$ for real scalar dark matter with spin - 1/2 mediator is given by:
 \begin{align}
 J(x_f) &= \int_{x_f}^{\infty}2n_cg^4 \frac{m^2_f}{\pi }\frac{\left[1-\frac{m_f}{m^2_\chi}\right]^{3/2}}{\left((m^2 + m^2_\chi -m^2_f)\right)x^2f_{membrane}(x)}
dx.
\end{align}
where the function $f_{membrane}(x)$ is the characteristic parameter for RS single braneworld and this can be expressed in terms of tensor-to-scalar ratio ($r$) which is given in Eq.\eqref{eq:ev1xxxxxc}. 
Now for the GR limiting case of the $f_{membrane}(x) \rightarrow 1$ and then the relic abundance will only depend on the mass of the Dark Matter ($m_\chi$), $g_A$ the 
coupling with the spin-0 mediator and the mass of the mediator ($m_A$). 
\begin{align}
 \Omega_{DM} h^2 &= \frac{1.07\times 10^9x_f}{g^{1/2}_*M_p}\left(2n_cg^4 \frac{m^2_f}{\pi }\frac{\left[1-\frac{m_f}{m^2_\chi}\right]^{3/2}}{ (m^2 + m^2_\chi -m^2_f)}\right)^{-1}.
\end{align}
In order to constrain the coupling ($g_A$) and mass ($m_A$) we take the present data of the relic abundance ($\Omega_{DM}h^2 = 0.1199\pm0.0027$\cite{Ade:2015xua}) and constrain the function 
$J(x)$, which in turn constrain the coupling $g_A$ and $m_A$ for a particular tensor-to-scalar ratio ($r$). We have not shown the GR limiting case as it has been extensively been explored in 
\cite{Berlin:2014tja}.
In fig.~(\ref{fig23a}) and fig.~(\ref{fig23b}), we have depicted the behaviour of the effective coupling of spin-1/2 mediator $g_{A}$ with the varying mass the spin-1/2 mediator $m_{A}$ 
for t/u-channel process with three distinct value of the tensor-to-scalar ratio $r=0.001$, $r=0.01$and $r=0.1$ 
respectively in RSII membrane. We also consider three different values of the dark matter mass $m_{\chi}=100\; {\rm GeV}$ and $m_{\chi}=1\;{\rm TeV}$
for the t/u-channel analysis. From fig.~(\ref{fig23a}) and fig.~(\ref{fig23b}), it is clearly observed
that the behaviour of the effective coupling of spin-1/2 mediator $g_{A}$ with the varying mass the spin-1/2 mediator $m_{A}$ are almost similar for both of the cases.
Most importantly, in both the sides of $m_{A}=1\times 10^{2}{\rm GeV}$ and $m_{A}=1\times 10^{3}{\rm GeV}$ the coupling of spin-1/2 mediator $g_{A}$ behave in completely opposite manner.
\begin{figure*}[!ht]
\centering
\subfigure[$g_{A}$ vs $m_{A}$ for t/u-channel with $m_{\chi}=100{\rm GeV}$.]{
    \includegraphics[width=14.5cm,height=7cm] {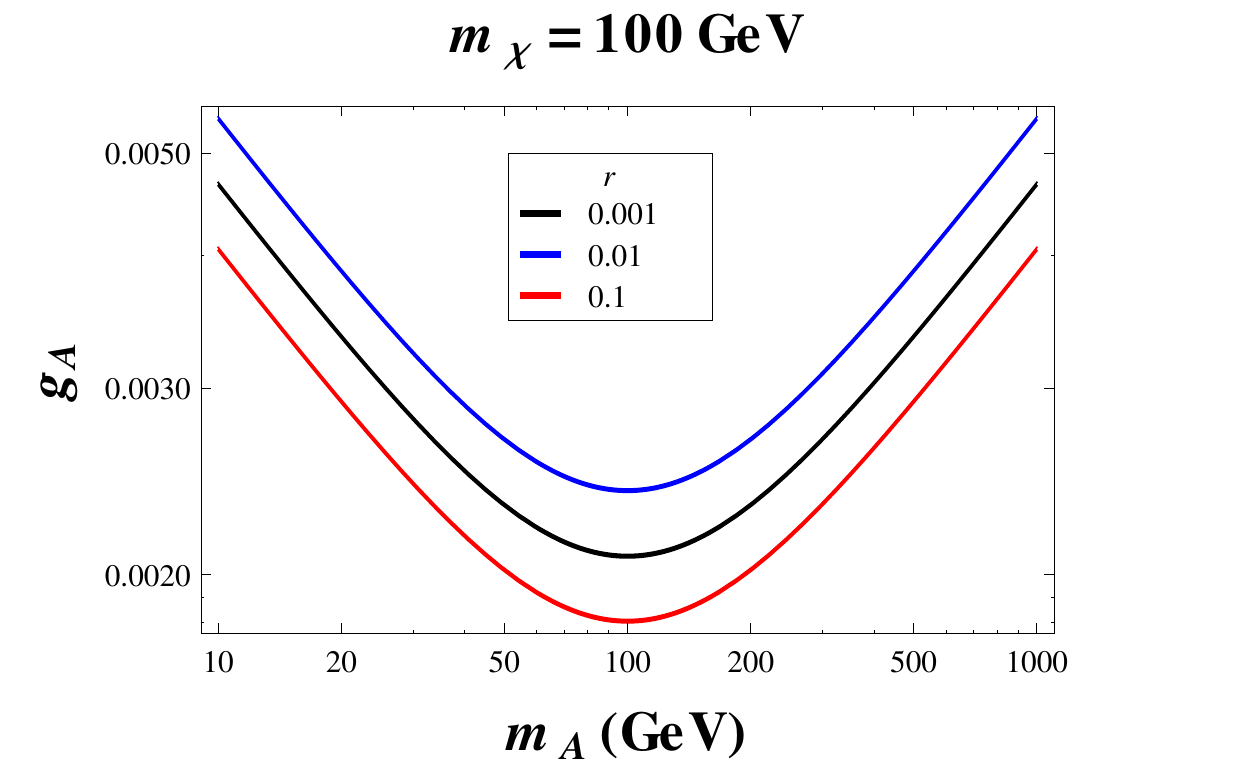}
    \label{fig24a}
}
\subfigure[$g_{A}$ vs $m_{A}$ for t/u-channel with $m_{\chi}=1{\rm TeV}$.]{
    \includegraphics[width=14.5cm,height=7cm] {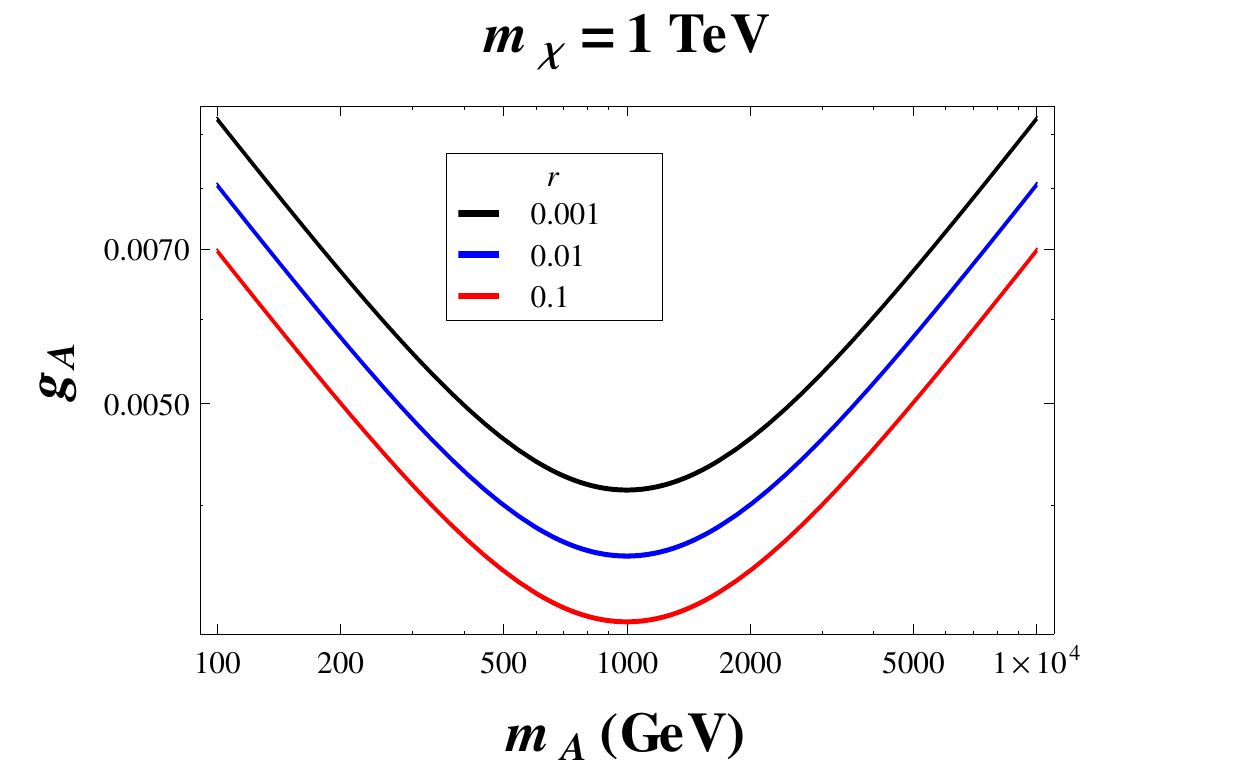}
    \label{fig24b}
}
\caption[Optional caption for list of figures]{In the above figure we have shown the allowed 
region of $g_A$ with respect to the mass of the mediator $m_A$, the above processes
are t/u-channel.} 
\label{fig24}
\end{figure*}
\subsection{\bf Complex Vector dark matter: spin-1/2 mediator}
\label{a16}
The Lagrangian of real vector dark matter with spin-1/2 mediator is given as 
\begin{align}
\mathcal{L}_{membrane} &\supset \left[\overline{\psi}\gamma^\mu \left(g_v + g_a\gamma^5 \right)f X^*_\mu + \overline{f}\gamma^\mu \left(g_v
- g_a\gamma^5 \right)\psi X_\mu \right].
\end{align}
The cross-section from the above Lagrangian after taking all the coupling equal is given as:
\begin{align}
\sigma &= \frac{1}{8\pi (s-4m^2_\chi)}\int_{t_-}^{t_+}|\mathcal{M}|^2 dt
\end{align}
where the matrix element for the S-matrix and the symbol $t_{\pm}$ is given by:
\begin{align}
 |\mathcal{M}|^2 &= g_A^4n_c\Big{[} \frac{1}{m_{\chi }^4 \left(t-m_A^2\right)^2}\left(m_{\chi }^4 \left(3 m_f^4-t (4 s+5 t)\right)+2 m_{\chi }^2 \left(t-m_f^2\right) \left(-2 t m_f^2+m_f^4
                  +t (2 s+t)\right) \right. \nonumber \\ 
                  -&\left.\left(t-m_f^2\right)^2 \left(-2 t m_f^2+m_f^4+t (s+t)\right)+4 m_{\chi }^6 \left(m_f^2+2 t\right)-4 m_{\chi }^8\right) \nonumber \\
                  &-\frac{1}{m_{\chi }^4 \left(m_A^2-2 m_f^2-2 m_{\chi}^2+s+t\right)^2}\left(-4 m_{\chi }^6 \left(5 (s+t)-7 m_f^2\right)+m_{\chi }^4 \left(-44 m_f^2 (s+t)+29 m_f^4 \right. \right.\nonumber \\
		  &+ \left. \left. 17 (s+t)^2\right)-2 m_{\chi }^2 \left(-3 m_f^2+2 s+3 t\right) \left(-m_f^2+s+t\right)^2 \right. \nonumber \\
		  &+ \left. \left(-m_f^2+s+t\right)^2 \left(-2 t m_f^2+m_f^4+t (s+t)\right)+8 m_{\chi }^8\right) \nonumber \\
		  &+ \frac{32 \left(s-2 m_f^2\right) \left(m_f^2-m_{\chi }^2\right)}{\left(t-m_A^2\right) \left(m_A^2-2 m_f^2-2 m_{\chi }^2+s+t\right)} \Big{]},\\
t_{\pm} &= (m^2_\chi + m^2_f-\frac{s}{2}) \pm \frac{\sqrt{(s-4m^2_\chi)(s-4m^2_f)}}{2}. 
\end{align}
where $n_c=3$ for quarks and 1 for leptons, $g_A$ and $m_A$ are the respective coupling and the mass of the mediator. 
Further 
taking the following approximation: \be s=4m^2_\chi/\left(1+v^2/4\right),\ee we finally get the following simplified expression for the product of annihilation cross-section and velocity as:
\begin{align} \sigma v = a + \mathcal{O}(v^2), \end{align} where the factor $a$ is given by:
\begin{align}
a &\approx \frac{n_c g^4 4\left(1-m^2_f/m^2_\chi \right)^{3/2}}{9 \pi (m^2_A - m^2_f + m^2_\chi)^2}\left(32m^2_\chi -4m^2_f\right).
\end{align}
The dark matter relic abundance in this case is given as:
\begin{align}
 \Omega_{DM} h^2 &= \frac{1.07\times 10^9}{J(x_f)g^{1/2}_*M_p}\end{align}
 where $J(x_f)$ for real scalar dark matter with spin - 1/2 mediator is given by:
 \begin{align}
 J(x_f) &= \int_{x_f}^{\infty}\frac{n_c g^4 4\left(1-m^2_f/m^2_\chi \right)^{3/2}}{9 \pi (m^2_A - m^2_f + m^2_\chi)^2}\frac{\left(32m^2_\chi -4m^2_f\right)}{x^2f_{membrane}(x)}
dx.
\end{align}
where the function $f_{membrane}(x)$ is the characteristic parameter for RS single braneworld and this can be expressed in terms of tensor-to-scalar ratio (r) which is given in Eq.\eqref{eq:ev1xxxxxc}. 
Now for the GR limiting case of the $f_{membrane}(x) \rightarrow 1$ and then the relic abundance will only depend on the mass of the Dark Matter ($m_\chi$), $g_A$ the 
coupling with the spin-0 mediator and the mass of the mediator ($m_A$). 
\begin{align}
 \Omega_{DM} h^2 &= \frac{1.07\times 10^9x_f}{g^{1/2}_*M_p}\left(\frac{n_c g^4 4\left(1-m^2_f/m^2_\chi \right)^{3/2}}{9 \pi (m^2_A - m^2_f + m^2_\chi)^2}\left(32m^2_\chi -4m^2_f\right)\right)^{-1}.
\end{align}
In order to constrain the coupling ($g_A$) and mass ($m_A$) we take the present data of the relic abundance ($\Omega_{DM}h^2 = 0.1199\pm0.0027$\cite{Ade:2015xua}) and constrain the function 
$J(x_f)$, which in turn constrain the coupling $g_A$ and $m_A$ for a particular tensor-to-scalar ratio ($r$). We have not shown the GR limiting case as it has been extensively been explored in 
\cite{Berlin:2014tja}.

In fig.~(\ref{fig24a}) and fig.~(\ref{fig24b}), we have depicted the behaviour of the effective coupling of spin-1/2 mediator $g_{A}$ with the varying mass the spin-1/2 mediator $m_{A}$ 
for t/u-channel process with three distinct value of the tensor-to-scalar ratio $r=0.001$, $r=0.01$and $r=0.1$ 
respectively in RSII membrane. We also consider two different values of the dark matter mass $m_{\chi}=100\; {\rm GeV}$ and $m_{\chi}=1\;{\rm TeV}$ for the t/u-channel analysis. From fig.~(\ref{fig24a}) and fig.~(\ref{fig24b}), it is clearly observed
that the behaviour of the effective coupling of spin-1/2 mediator $g_{A}$ with the varying mass the spin-1/2 mediator $m_{A}$ are almost similar for both of the cases.
Most importantly, in both the sides of $m_{A}=1\times 10^{2}{\rm GeV}$ and $m_{A}=1\times 10^{3}{\rm GeV}$ 
the coupling of spin-1/2 mediator $g_{A}$ behave in completely opposite manner. 

Now, for the real vector dark matter with spin-1/2 mediator the Lagrangian of is given as 
\begin{align}
\mathcal{L}_{membrane} &\supset \left[\overline{\psi}\gamma^\mu \left(g_v + g_a\gamma^5 \right)f X_\mu + \overline{f}\gamma^\mu \left(g_v
- g_a\gamma^5 \right)\psi X_\mu \right].
\end{align}
The cross-section and the thermally averaged cross section is almost identical to the results obtained in the context of complex scalar dark matter with spin-1/2 mediator case.

\section{Summary}
\label{a18}
To summarize, in the present article, we have addressed the following points:
\begin{itemize}
 \item We have established
a theoretical constraint relationship to explicitly show a direct connection between the dark matter relic abundance ($\Omega_{DM}h^2$) and primordial gravity waves
($r$), which establish a precise connection between inflation and generation of dark matter within the framework of effective field theory 
in RSII membrane paradigm. In fig.~(\ref{fig01}), we have shown the algorithm of the prescribed methodology proposed in this paper. Also 
in fig.~(\ref{fig1}), we have depicted the model independent inflationary constraint from the
primordial gravitational waves via tensor-to-scalar ratio on the dark matter relic abundance within the framework of 
effective field theory of RSII membrane paradigm. From fig.~(\ref{fig1}), it is clearly observed that for $r<0.01$, various inflationary models in 
membrane paradigm are in huge tension with the Planck 2015 constraint on dark matter relic abundance.
We have also shown the the dependence of the relic abundance with $\alpha$ (defined in eq \eqref{eq:alpha}) in fig.~(\ref{fig1a}) for 
various tensor-scalar ratio r. Obviously the result changes for different thermally averaged cross-section $\langle \sigma v\rangle$, 
but one may note the in direct relation of the tensor to scalar ratio to the model parameters which gives rise to the $\langle \sigma v\rangle$.
On the other hand, for $0.01\leq r\leq 0.12$, various inflationary models in 
membrane paradigm are consistent with $2\sigma$ constraint on dark matter relic abundance obtained
from Planck 2015 data. In fig.~(\ref{fig2a}) and fig.~(\ref{fig2b}), we have depicted
the model independent inflationary constraint from the primordial gravitational waves via tensor-to-scalar ratio on
the thermally averaged annihilation cross-section of dark matter content for the dark matter
mass, $m_{\chi}=100 {\rm GeV}$ and $m_{\chi}=1{\rm TeV}$ respectively, within the framework of 
effective field theory of RSII membrane paradigm. Here it is clearly observed that 
the value of the inflationary tensor-to-scalar ratio decreases with the increase in the thermally
averaged annihilation cross-section of dark matter content. Also it is important to note that for of these three cases
within the allowed range of tensor-to-scalar ratio, $0.01\leq r\leq 0.12$, thermally averaged annihilation cross-section of dark matter content
is constrained within the window, $\langle \sigma v\rangle\sim{\cal O}(10^{-28}-10^{-27}){\rm cm^3/s}$.

\item We have explicitly studied the details of 
      Effective Field Theory of dark matter from membrane paradigm paradigm. We have explicitly studied:
     \begin{itemize}\item the $s$ and $t/u$ channel interaction of the 
 Dirac dark matter with spin-0 and spin-1 mediator, Majorana dark matter with spin-1 mediator. 
\item After that, we have studied the consequences from 
s-channel interaction of complex and real scalar dark matter with spin-0 mediator, complex scalar and real vector dark matter with spin-1 mediator,
complex and real vector dark matter with spin-0 and spin-1 mediator respectively. 
\item Finally, we have studied the consequences from 
t/u-channel interaction of complex and real scalar and complex and real vector dark matter with spin-1/2 mediator.
\item In most of the case we have not gone to higher Dark Matter masses 
(i.e $\mathcal{O}$(10 TeV) in some cases $\mathcal{O}$(TeV)) as the couplings were exceeding $\mathcal{O}$(1) 
  violating perturbative limit.
\end{itemize}

\item Most significantly, once the signature of primordial
gravity waves will be predicted by in any near future observational probes, it will be possible to put further stringent constraint
on the dark matter abundance from our derived result.

\item  In this paper, we
have used important cosmological and particle physics constraints arising from Planck 2015 and Planck+BICEP2/Keck Array joint data on the 
 the upper bound on tenor to scalar ratio
and the bound on the dark matter abundance 
within $1.5\sigma-2\sigma$ statistical CL.

\end{itemize}
Further our
aim is to carry forward this work in a more broader sense,
where we will apply all the derived results to constrain the inflationary observables and cosmological parameters obtained from 
various models of membrane inflation. The other promising future prospects of this work are:-
\begin{enumerate}
\item One can follow the prescribed methodology to derive the cosmological constraints 
in the context of various modified gravity framework i.e. Dvali-Gabadadze-Porrati (DGP) braneworld \cite{Dvali:2000hr}, Einstein-Hilbert-Gauss-Bonnet
(EHGB) gravity \cite{Choudhury:2012yh,Choudhury:2015yna,Choudhury:2013yg,Choudhury:2013dia,Choudhury:2013aqa,Choudhury:2013eoa,Choudhury:2015wfa}, 
Einstein-Gauss-Bonnet-Dilaton (EGBD) gravity \cite{Choudhury:2013yg,Choudhury:2013dia,Choudhury:2013aqa,Choudhury:2013eoa,Choudhury:2015wfa}
and $f(R)$ theory of gravity \cite{DeFelice:2010aj,Sotiriou:2008rp} etc. 
\item Hence using the derived constraints one can further constrain various classes of (membrane) inflationary
models within the framework of other
modified theories of gravity.
\item One can explore the details of UV completion in the context of RSII membrane and for other versions of extended theories of gravity.
\item Detailed study of the collider constraints on the effective theory prescription from extended theories of gravity is one of the promising and unexplored areas 
in this context.
\item Generation of scalar dark matter and detailed study of the constraints from the UV complete extended theory of gravity is also an important 
issue, which we will explore very soon in our follow up work.
\item One can also implement the methodology for the alternative theories of inflation i.e. bouncing frameworks and related ideas. 
For an example one can investigate for the cosmological implications of cosmic hysteresis scenario \cite{Kanekar:2001qd,Sahni:2012er,Sahni:2015kga,Choudhury:2015baa,Choudhury:2015fzb,Choudhury:2016rtp} in the 
generation of dark matter.
\item Explaining the origin of dark matter in presence of non-standard/ non-canonical kinetic term, using non-minimal inflaton 
coupling to gravity sector \cite{Choudhury:2015zlc}, multi-field sector and also exploring the highly non-linear regime of effective field theory are 
 open issues in this literature. String theory originated DBI and tachyonic inflationary frameworks \cite{Choudhury:2015yna,Choudhury:2012yh,Choudhury:2015hvr} are the two prominent and well known
 examples of non-standard field theoretic setup through which one can explore various open questions in this area.
 \item One can also study the connecting relation between dark matter abundance and primordial gravity waves 
 with the inflationary magnetogenesis and standard 
 leptogenesis scenario \cite{Choudhury:2014hua} from the relevant effective field theory 
 operators. In the context of RSII single membrane we have recently studied some of these issues elaborately \cite{Choudhury:2015jaa}.
\end{enumerate}

\section*{Acknowledgments}
SC would like to thank Department of Theoretical Physics, Tata Institute of Fundamental
Research, Mumbai for providing me Visiting (Post-Doctoral) Research Fellowship. SC take this opportunity to thank
sincerely to Sandip P. Trivedi, Shiraz Minwalla, Soumitra SenGupta, Sudhakar Panda, Varun Sahni, Sayan Kar and Supratik Pal for their constant support
and inspiration. SC take this opportunity to thank all the active members and the regular participants of weekly
student discussion meet “COSMOMEET” from Department of Theoretical Physics and Department of Astronomy
and Astrophysics, Tata Institute of Fundamental Research for their strong support. SC also thank Sandip Trivedi and Shiraz Minwalla for giving the opportunity to be
the part of String Theory and Mathematical Physics Group. SC also thank the other
post-docs and doctoral students from String Theory and Mathematical Physics Group for
providing an excellant academic ambience during the research work. SC additionally take this
opportunity to thank the organizers of STRINGS, 2015,
International Centre for Theoretical Science, Tata Institute of Fundamental Research (ICTS,TIFR),
Indian Institute of Science (IISC) and specially Shiraz Minwalla for giving the opportunity
to participate in STRINGS, 2015 and also providing the local hospitality during the work. SC also thank 
the organizers of National String Meet 2015 and International Conference on Gravitation and Cosmology, IISER, Mohali 
and COSMOASTRO 2015, Institute of Physics for providing the local hospitality during the work. 
SC also thanks the organizers of School and Workshop on Large Scale Structure: From Galaxies to Cosmic Web, 
The Inter-University Centre for Astronomy and Astrophysics (IUCAA), Pune, India
and specially Aseem Paranjape and Varun Sahni for providing the academic visit during the work. 
Last but not the least, I would
all like to acknowledge our debt to the people of India for their generous and steady support for research in natural
sciences, especially for theoretical high energy physics, string theory and cosmology.

\section*{Appendix}
\subsection*{\bf A. Consistency relations in RSII membrane paradigm}

In the context of RSII the spectral tilts $(n_S, n_T)$, running of the tilts $(\alpha_S, \alpha_T)$ and running of the running of tilts $(\kappa_T,\kappa_S)$ at the momentum pivot scale $k_*$ 
can be expressed as:
\begin{eqnarray}
 n_S(k_*) -1&=& 2\eta_{b}(\phi_*)-6\epsilon_{b}(k_*),\\
\label{cv1}n_T(k_*) &=& -3\epsilon_{b}(k_*)=-\frac{r(k_*)}{8},\\
\alpha_S(k_*) &=& 16\eta_{b}(k_*)\epsilon_{b}(k_*)-18\epsilon^{2}_{b}(k_*)-2\xi^{2}_{b}(k_*),\\
\alpha_T(k_*) &=& 6\eta_{b}(k_*)\epsilon_{b}(k_*)-9\epsilon^{2}_{b}(k_*),\\
\kappa_S(k_*)&=&152\eta_{b}(k_*)\epsilon^{2}_{b}(k_*)-32\epsilon_{b}(k_*)\eta^{2}_{b}(k_*)-108\epsilon^{3}_{b}(k_*)\nonumber
\\ &&~~~~~-24\xi^{2}_{b}(k_*)\epsilon_{b}(k_*)+2\eta_{b}(k_*)\xi^{2}_{b}(k_*)+2\sigma^{3}_{b}(k_*),\\
\kappa_T(k_*)&=&66\eta_{b}(k_*)\epsilon^{2}_{b}(k_*)-12\epsilon_{b}(k_*)\eta^{2}_{b}(k_*)
-54\epsilon^{3}_{b}(k_*)-6\epsilon_{b}(k_*)\xi^{2}_{b}(k_*).
\end{eqnarray}
 In terms of slow-roll parameters in RSII setup one can also write the following sets of consistency conditions for membrane inflation:
\begin{eqnarray}
 \label{wqc1}n_T(k_*)-n_S(k_*)+1&=&\left(\frac{d\ln r(k)}{d\ln k}\right)_*=\left[\frac{r(k_*)}{8}-2\eta_{b}(k_*)\right],\\
\label{wqc2}\alpha_T(k_*)-\alpha_S(k_*)&=&\left(\frac{d^2\ln r(k)}{d\ln k^2}\right)_*=\left[\left(\frac{r(k_*)}{8}\right)^2-\frac{20}{3}\left(\frac{r(k_*)}{8}\right)+2\xi^2_{b}(k_*)\right],\\
\label{wqc3}\kappa_T(k_*)-\kappa_S(k_*)&=&\left(\frac{d^3\ln r(k)}{d\ln k^3}\right)_* \nonumber\\&=&
\left[2\left(\frac{r(k_*)}{8}\right)^3-\frac{86}{9}\left(\frac{r(k_*)}{8}\right)^2
\right.\\&&\left.~~+\frac{4}{3}\left(6\xi^2_{b}(k_*)+5\eta^{2}_{b}(k_*)\right)\left(\frac{r(k_*)}{8}\right)
+2\eta_{b}(k_*)\xi^2_{b}(k_*)+2\sigma^{3}_{b}(k_*)\right].\nonumber
\end{eqnarray}
Here Eq~(\ref{wqc1}-\ref{wqc3})) represent the running, running of the running and running of the double running of tensor-to-scalar ratio in RSII membrane inflationary setup. 
In this section let us In the high energy limit $\rho>>\sigma$,
Eq~(\ref{eq1}) is written using the slow-roll approximation as:
\begin{eqnarray}\label{eq2}
 H^{2}&\approx&\frac{\rho^{2}}{6M^{2}_{p}\sigma}\approx\frac{V^{2}(\phi)}{6M^{2}_{p}\sigma},
\end{eqnarray}
where $V(\phi)$ be the inflaton single field potential.
Within high energy limit $\rho>>\sigma$ the slow-roll parameters in the visible membrane can be expressed as:
\begin{eqnarray}
 \epsilon_{b}(\phi)&\approx& \frac{2M^{2}_{p}\sigma (V^{'}(\phi))^{2}}{V^{3}(\phi)},\\
\eta_{b}(\phi)&\approx& \frac{2M^{2}_{p}\sigma V^{''}(\phi)}{V^{2}(\phi)},\\
\xi^{2}_{b}(\phi)&\approx& \frac{4M^{4}_{p}\sigma^{2} V^{'}(\phi)V^{'''}(\phi)}{V^{4}(\phi)},\\
\sigma^{3}_{b}(\phi)&\approx& \frac{8M^{6}_{p}\sigma^{3} (V^{'}(\phi))^{2}V^{''''}(\phi)}{V^{6}(\phi)}.
\end{eqnarray}
and consequently the number of e-foldings can be written as:
\begin{eqnarray}\label{efolda}
\Delta {\cal N}_{b}={\cal N}_{b}(\phi_{cmb})-{\cal N}_{b}(\phi_{e})&\approx&  \frac{1}{2\sigma M^2_p}\int^{\phi_{cmb}}_{\phi_{e}} d\phi\frac{V^{2}(\phi)}{V^{'}(\phi)}
\end{eqnarray}
where $\phi_{e}$ corresponds to the field value at the end of inflation, which can be obtained from 
the following constraint equation:
\begin{eqnarray}
 \max_{\phi=\phi_{e}}\left[\epsilon_{b},|\eta_{b}|,|\xi^{2}_{b}|,|\sigma^{3}_{b}|\right]&=&1.
\end{eqnarray}
\subsection*{\bf B. Relic Abundance Results for GR limitng case}
\label{app:GR}
It is important to note that, for Standard
General Relativistic (GR) prescription the Boltzmann equation for the dark matter relic abundance is given as follows:
\begin{align}
\frac{d Y_{\rm DM}}{d\Theta} &= \frac{s\langle \sigma v \rangle}{H\Theta}\left[1+\frac{1}{3}\frac{d\ln g_s}{d\ln \Theta}\right]\left[(Y_{\rm DM}^{EQ})^2-Y_{\rm DM}^2 \right]
\label{eq:ev1}
\end{align}
where the solution for the dark matter relic abundance $Y^{EQ}_{\rm DM}(\Theta)$ is given as: 
\begin{align}
Y^{EQ}_{\rm DM}(\Theta) &= \frac{n^{EQ}_{\rm DM}}{s}=\frac{45}{2\pi^4}\left(\frac{\pi}{8}\right)^{1/2}\frac{g_{\rm DM}}{g_*}\Theta^{3/2}e^{-\Theta}.
\end{align}
Here $s$ is the entropy density as defined in Eq~(\ref{sen}), $g_{\rm DM}$ signifies the effective number of degrees of freedom corresponding to the dark matter content and for non-relativistic case the thermally averaged cross-section can be expressed as:
\begin{align}
\langle \sigma v \rangle &\approx \langle \sigma v \rangle_{NR}= a + \frac{3}{2}\frac{b}{\Theta} +\frac{15}{8}\frac{c}{\Theta^2}+\frac{35}{16}\frac{d}{\Theta^3}+\frac{315}{128}\frac{e}{\Theta^4}+\cdots.
\end{align}
For numerical estimation we will restrict ourselves up to the second term in the above series expansion of $\langle \sigma v \rangle_{NR}$. In more technical language this is commonly known as {\it s-wave approximation}. See 
the ref.~\cite{Gondolo:1990dk}, in which all the derivation have done in detail. Also it is important to mention here that, in a very specific physical situation when the energy density of the dark matter content is very very small compared to the 
membrane tension in RSII membrane paradigm ($\rho_{DM}=\rho_{\chi}<<\sigma$), then one can can reproduce all the known results for standard GR. On the other hand, the explicit role of RSII membarne paradigm can be clearly visualized 
when the energy density of the dark matter content is very very large compared to the membrane tension in RSII membrane paradigm ($\rho_{DM}=\rho_{\chi}>>\sigma$).

Now from the equation eq.\eqref{eq:sigtld},\eqref{eq:Sig} and \eqref{eq:fmem} for the standard GR case all three functions $\langle \widetilde{\sigma} v \rangle(\Theta)$, $\Sigma(\Theta)$ and characteristic function , $f_{brane}(\Theta) $ reduced to the following simplified expressions:\begin{align}
\langle \widetilde{\sigma} v \rangle(\Theta) &= \langle \sigma v \rangle, \\
\Sigma(\Theta)&=\sqrt{\frac{g_*}{90}}\frac{\pi m^2_{\chi}}{M_p \Theta^2}=H_{GR}(\Theta),  \\
f_{membrane}(\Theta) &= 1,
\label{eq:evqw}
\end{align}
where $H_{GR}(\Theta)$ is the Hubble parameter in GR.

where $\Theta_F$ is the freeze-out temperature which can be calculated by numerically solving the following transcendental equation as:
\begin{align}
\Theta_F &= \ln \left(\frac{0.038g_{\rm DM} m_{\chi} M_{p}\langle {\sigma} v\rangle (\Theta) }{g^{1/2}_*\Theta^{1/2}_F }\right).
\end{align}

Now, the dark matter relic abundance can be found out by
\begin{align}
\Omega_{\rm DM} h^2 &=\left(\frac{ m_{\chi}sY_{\rm DM}}{3H^2 M^2_p}\right)_{today}h^2\simeq \frac{1.07\times 10^9 \textrm{GeV}^{-1}}{J(\Theta_F)g^{1/2}_* M_{p}}
\end{align} 
where using s-wave approximation and for the context of standard GR the integral $J(\Theta_F)$ can be computed as:
\begin{align}\label{eqw:1}
J(\Theta_F) &= \int_{\Theta_F}^\infty d\Theta  \frac{\langle \sigma v \rangle(\Theta)}{\Theta^2}  \nonumber \\
 &\approx \int_{\Theta_F}^{\infty}d\Theta \int_{0}^{\infty}dv \frac{v^2 \left(a + b v^2\right)}{\sqrt{4\pi \Theta}}e^{-\Theta v^2/4}\nonumber \\
 &= \int_{\Theta_F}^{\infty} \left(a+\frac{3}{2}\frac{b}{\Theta}\right)\frac{d\Theta}{\Theta^2 }=\frac{1}{\Theta_F}\left(a+\frac{3b}{4\Theta_{F}}\right). 
\end{align}

In the context of GR, dark matter relic abundance can be expressed as:
\bea\label{omgr}
\Omega_{\rm DM} h^2 & \simeq &\frac{1.07\times 10^9 \Theta_{F}\textrm{GeV}^{-1}}{g^{1/2}_* M_{p}\left(a+\frac{3b}{4\Theta_{F}}\right)}.
\eea
It is important to note that, in the vicinity of resonance the Taylor series expansion breaks down and in such a case it is legitimate to consider 
the numerical solution of $J(\Theta_F)$, which is introduced in Eq~(\ref{eqwaaa:1}). Additionally, it is mention here that in our computation we 
treat the integral $J(\Theta_F)$ numerically to obtain the accurate result in the context of membrane paradigm. Here from Eq~(\ref{omgr}),
it is clearly observed that the dark matter relic abundance cannot be directly related to the tensor-to-scalar ratio ($r$) in GR limit using complete 
model independent prescription. If we specifically mention the model to describe the decay process of inflaton to dark matter then 
it is possible to obtain a connecting relationship between the dark matter relic abundance and tensor-to-scalar ratio in a model 
dependent fashion. See ref.~(\cite{Dev:2014tla}) for further details.

\end{document}